\documentclass[aps,prl,reprint,showpacs,superscriptaddress,longbibliography]{revtex4-2} %
\usepackage{mathtools,amssymb,graphicx,units}
\usepackage[usenames,dvipsnames]{color}
\usepackage[plainpages=false,pdfpagelabels,colorlinks=true,linkcolor=red,urlcolor=blue,citecolor=blue,pdftitle={Title},pdfauthor={},pdfdisplaydoctitle=true,pdfduplex=DuplexFlipLongEdge]{hyperref}
\usepackage{verbatim}
\usepackage[english]{babel}

\usepackage[normalem]{ulem}
\usepackage{appendix}
\usepackage{bbm}
\usepackage{makecell}
\usepackage{comment}
\usepackage{cases}
\usepackage{tikz}
\usepackage{xcolor}
\usetikzlibrary{shapes.geometric}

\newcommand{\beginsupplement}{%
	\setcounter{table}{0}
	\renewcommand{\thetable}{S\arabic{table}}%
	\setcounter{figure}{0}
	\renewcommand{\thefigure}{S\arabic{figure}}%
}
\graphicspath{{./fig/}}

\DeclareRobustCommand{\solidSquare}[2][black]{%
	\tikz[baseline=(S.south)] \node[rectangle, fill=#1, inner sep=0pt, minimum size=#2, anchor=south] (S) {};%
}
\DeclareRobustCommand{\solidCircle}[2][black]{%
	\tikz[baseline=(C.south)] \node[circle, fill=#1, inner sep=0pt, minimum size=#2, anchor=south] (C) {};%
}
\DeclareRobustCommand{\solidTriangle}[2][black]{%
	\tikz[baseline=(T.south)] \node[regular polygon, regular polygon sides=3, fill=#1, inner sep=0pt, minimum size=#2, anchor=south] (T) {};%
}

\DeclareRobustCommand{\solidStar}[2][black]{%
	\tikz[baseline=(base)]{%
		\pgfmathsetlengthmacro{\s}{#2}
		\pgfmathsetlengthmacro{\R}{0.5*\s}
		\pgfmathsetlengthmacro{\shift}{0.809016944*\R}
		
		\coordinate (base) at (0,0); 
		\coordinate (C) at (0,\shift); 
		
		\node[star, star points=5, star point ratio=2.25,
		fill=#1, inner sep=0pt, minimum size=\s, anchor=center] at (C) {};
	}%
}

\begin{document}
	\title{Distinct reentrant transitions in a quasi-periodic Raman lattice}
	\author{Xingbo Wei}
	\email{weixingbo@zstu.edu.cn}
	\affiliation{Zhejiang Key Laboratory of Quantum State Control and Optical Field Manipulation, Department of Physics, Zhejiang Sci-Tech University, 310018 Hangzhou, China}
	\author{Zhentian Xie}
	\affiliation{Zhejiang Key Laboratory of Quantum State Control and Optical Field Manipulation, Department of Physics, Zhejiang Sci-Tech University, 310018 Hangzhou, China}
	\author{Tong Liu}
	\affiliation{Department of Applied Physics, School of Science, Nanjing University of Posts and Telecommunications, Nanjing 210003, China}
	\author{Gao Xianlong}
	\affiliation{Department of Physics, Zhejiang Normal University, Jinhua 321004, China}
	\author{Yunbo Zhang}
	\email{ybzhang@zstu.edu.cn}
	\affiliation{Zhejiang Key Laboratory of Quantum State Control and Optical Field Manipulation, Department of Physics, Zhejiang Sci-Tech University, 310018 Hangzhou, China}

	\begin{abstract}	
		We investigate a one-dimensional lattice with spin-orbit coupling (SOC) and a Zeeman potential containing uniform and quasiperiodic components. By tuning SOC, anomalous mobility edges emerge that separate critical from non-critical states, yielding a reentrant transition between two mixed phases, M$_1$$\to$M$_2$$\to$M$_1$, where M$_1$ (M$_2$) lacks (hosts) anomalous mobility edges. A new \emph{reentrant criticality transition}, defined as multiple entries into the critical phase, is identified. As a counterpart to reentrant delocalization/localization transitions, it completes the basic framework of reentrant phenomena across extended, localized, and critical states. The uniform Zeeman potential drives a reentrant delocalization transition, arising from the splitting of the localized region induced by the shift of mobility edges. This reveals a distinct pathway for reentrant phenomena beyond hybridization mechanisms. 
	\end{abstract}
	
	\maketitle
	
	\paragraph{Introduction.---}In the field of Anderson localization~\cite{PhysRev.109.1492}, the competition between different types of quantum states has been extensively studied~\cite{zhou2025fundamentallocalizationphasesquasiperiodic,semeghini2015measurement,PhysRevLett.118.170403,PhysRevLett.132.236301,PhysRevLett.114.146601,PhysRevLett.125.196604,PhysRevLett.104.070601}, particularly in quasi-periodic systems~\cite{PhysRevLett.123.070405,10.21468/SciPostPhys.12.1.027,PhysRevB.108.174202,PhysRevLett.131.176401,wang2025exactnewmobilityedges}. 
	Many interesting phenomena are discovered, e.g., mobility edges~\cite{mott1967electrons}, well-defined critical energies that separate extended states from localized ones. Beyond the conventional picture of mobility edges, recent studies reveal the proliferation of critical states in the spectrum, inducing the emergence of anomalous mobility edges~\cite{10.21468/SciPostPhys.12.1.027,PhysRevB.108.174202}, also known as new mobility edges~\cite{PhysRevLett.131.176401,wang2025exactnewmobilityedges}. This phenomenon goes beyond the conventional mobility-edge paradigm and allows for rich coexistence manners among extended, localized, and critical states.
	These rich coexistence phenomena have been both theoretically predicted and experimentally confirmed across a variety of physical systems~\cite{PhysRevLett.120.160404,PhysRevLett.126.040603,PhysRevLett.129.103401,PhysRevLett.134.053601,xiao2021observation}.
	
	Recently, an intriguing phenomenon has been discovered in modified quasi-periodic systems, known as the reentrant delocalization/localization transition~\cite{goblot2020emergence,PhysRevLett.126.106803, PhysRevB.105.L220201}. In contrast to the general quasi-periodic systems, where the adjustment of a single parameter induces only a delocalization/localization transition~\cite{semeghini2015measurement,PhysRevLett.118.170403,PhysRevLett.132.236301,PhysRevLett.114.146601,PhysRevLett.125.196604,PhysRevLett.104.070601}, the modified quasi-periodic systems exhibit two or more distinct delocalization/localization transitions. This reentrant behavior generally originates from hybridization mechanisms, which can be driven by an enhanced density of states~\cite{goblot2020emergence,PhysRevResearch.3.033257,PhysRevB.111.134204,PhysRevB.105.L220201,xu2024observation,PhysRevLett.126.106803}. The relevant experimental verification and advances in photonic platforms further emphasize the practical relevance of reentrant transitions~\cite{goblot2020emergence, xu2024observation, PhysRevResearch.5.033170}, which has broad significance for exploring complex delocalization/localization transitions in superconducting simulators~\cite{guo2021observation,shi2024probing,PhysRevResearch.6.L042038}, cold atoms~\cite{PhysRevLett.120.160404,PhysRevLett.126.040603,PhysRevLett.129.103401,PhysRevLett.134.053601,PhysRevLett.125.073204,wu2016realization,song2019observation}, and other platforms. 
	These reentrant behaviors highlight a complex landscape of localization physics, which is crucial for understanding the interplay between disorder-induced localization and hybridization-driven delocalization.
	
	\begin{figure}[t]
		\includegraphics[width=1\columnwidth,height=0.6\columnwidth]{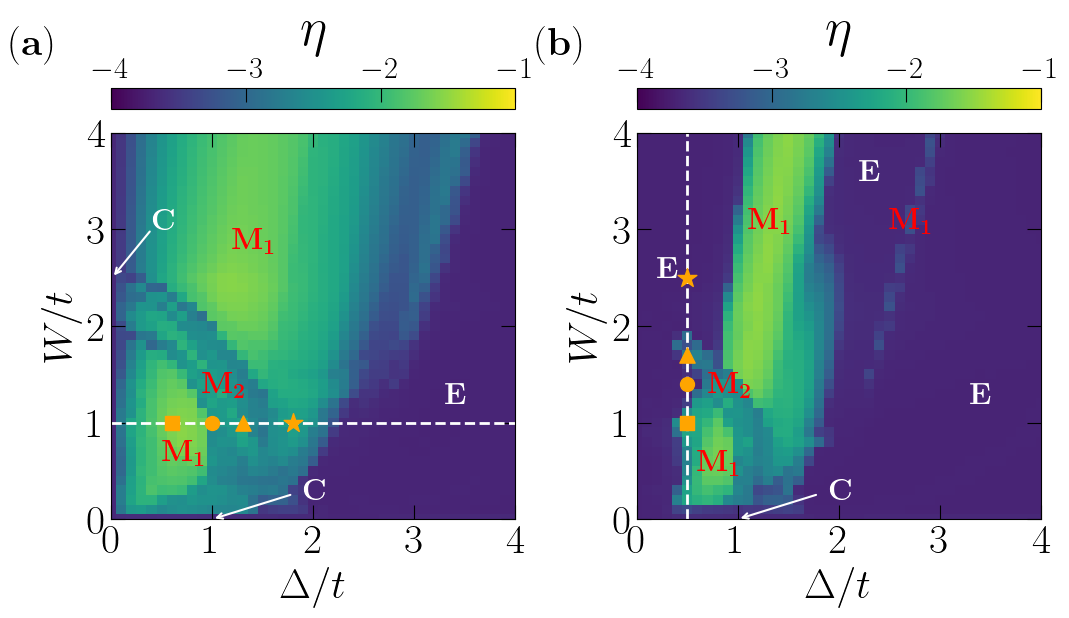}
		\vspace{-0.4cm}
		\caption{Phase diagrams in the $W\textendash\Delta$ plane for (a) $V/t=2.0$ and (b) $V/t=1.4$, respectively. Color indicates the value of $\eta$. White letter C (E) marks the purely critical phase (purely extended phase). M$_1$ and M$_2$ denote the mixed phases without and with anomalous mobility edges. White dotted lines mark $W/t=1$ in (a) and $\Delta/t=0.5$ in (b). Orange markers in (a) and (b) indicate points further analyzed in Figs.~\ref{fig2}\textendash\ref{fig3} and Fig.~\ref{fig5}, respectively. $\eta$ is used only to distinguish pure and mixed phases. The distinction between pure phases relies on fractal dimensions, and colored dot markers indicate transition points between different phases extracted from finite-size scaling analysis, see the Supplemental Material~\cite{SuppMateria}.
		}
		\label{fig1}
	\end{figure}

	Quasi-periodic systems host extended, critical, and localized states, yet reentrant phenomena involving critical states remain unexplored. The inclusion of critical states is expected to generate richer competition, leading to a more complex scenario for reentrant behavior. Furthermore, beyond the hybridization mechanism, it remains an open question whether other mechanisms can drive reentrant transitions.
	In this Letter, we study the reentrant transition in a quasi-periodic Raman lattice. In the presence of a Zeeman potential, spin-orbit coupling (SOC) induces a reentrant transition between mixed phases, distinguished by the absence or presence of anomalous mobility edges. Remarkably, the system repeatedly enters the purely critical phase, giving rise to a \emph{reentrant criticality transition}. By tuning the Zeeman potential, we demonstrate a reentrant delocalization transition via direct control of mobility edges, 
	revealing a distinct origin of reentrant behaviors. Since the model we adopted is feasible in optical Raman lattice experiments~\cite{PhysRevLett.125.073204,wu2016realization,song2019observation}, all phenomena can be realized experimentally.

	\paragraph{Model and quantities.--- }We investigate a 1D lattice model with SOC and a Zeeman potential, described by the Hamiltonian~\cite{PhysRevLett.110.076401}
	\begin{equation}\label{h1}
		\begin{aligned}
			\hat{H}=&-t\sum_{j}\left(\hat{c}_{j,\uparrow}^{\dagger}\hat{c}_{j+1,\uparrow}-\hat{c}_{j,\downarrow}^{\dagger} \hat{c}_{j+1,\downarrow}+\text{h.c.}\right) \\
			&+\Delta\sum_{j}\left(\hat{c}_{j,\uparrow}^{\dagger}\hat{c}_{j+1,\downarrow}-\hat{c}_{j,\uparrow}^{\dagger}\hat{c}_{j-1,\downarrow} +\text{h.c.}\right) \\
			&+ \sum_{j} V_j \left(\hat{n}_{j,\uparrow}-\hat{n}_{j,\downarrow}\right),
		\end{aligned}
	\end{equation}
	where $\hat{c}_{j,\sigma}^\dag(\hat{c}_{j,\sigma})$ is the creation (annihilation) operator with spin $\sigma=\uparrow,\downarrow$ at the site $j$, and $\hat{n}_{j,\sigma}=\hat{c}_{j,\sigma}^{\dagger}\hat{c}_{j,\sigma}$ is the number operator. $t$ and $\Delta$ denote the strengths of nearest-neighbor spin-conserved hopping and SOC induced spin-flip hopping, respectively. The Zeeman potential is given by 
	\begin{equation}
		\begin{aligned}
			V_j=V\cos(2\pi\alpha j+\phi)+W,\\
		\end{aligned}
	\end{equation}
	where $V$ ($W$) represents the amplitude of the quasi-periodic (uniform) component. $\alpha$ is taken as the ratio of adjacent Fibonacci numbers, $\alpha=F_{m-1}/F_{m}$~\cite{PhysRevLett.51.1198,PhysRevLett.125.073204,PhysRevB.108.054204}. For $m\rightarrow\infty$,  $\alpha$ approaches the inverse of the golden ratio, $\alpha=(\sqrt{5}-1)/2$. The random phase factor $\phi\in[0,2\pi)$ generates different examples of the quasi-periodic Zeeman potential. Unless otherwise specified, we set $t=1$ as the energy unit and adopt periodic boundary conditions (PBC). The system size is selected as a Fibonacci number $L=F_m$. Specifically, we take $L=2584$ with $\alpha=1597/2584$.
	
	For $\Delta=0$, Eq. \eqref{h1} reduces to two separated Aubry-Andr{\'e} (AA) models for each spin~\cite{aubry1980}, which exhibit a transition from a purely extended phase to a purely Anderson localized phase. The transition occurs at $V/t=2$~\cite{aubry1980}. In this regime, the properties of the system are independent of $W$. For $\Delta\neq0$ and $W=0$, the inclusion of SOC can induce a purely critical phase~\cite{PhysRevLett.125.073204} and the phase diagram is consistent with that of the non-Abelian AA model with $p$-wave superfluidity~\cite{PhysRevB.93.104504}. Upon the introduction of $W$, band splitting arises, weakening the correlation between different spin components and competing with SOC~\cite{SuppMateria}.

	To characterize different states, we employ multiple diagnostic quantities. The core one is the inverse participation ratio (IPR), defined as $\mathcal{I}_n=\sum_{j=0}^{L-1}(u_{j,n}^4+v_{j,n}^4)$~\cite{PhysRevB.93.104504,PhysRevB.100.064202,PhysRevB.94.125408}, where $n$ is the energy level index and $u_{j,n}$ ($v_{j,n}$) denotes the amplitude of the spin-$\uparrow$ (spin-$\downarrow$) component of the eigenstate $|\Psi^{(n)}\rangle=\sum_{j=0}^{L-1}\left(u_{j,n} \hat{c}_{j,\uparrow}^{\dagger}+v_{j,n} \hat{c}_{j, \downarrow}^{\dagger}\right)|\mathrm{vac}\rangle$. IPR is associated with the normalized participation ratio (NPR), $\mathcal{N}_n=(2L\mathcal{I}_n)^{-1}$~\cite{PhysRevLett.126.106803,PhysRevB.96.085119} and the fractal dimension, $D_n=-\ln(\mathcal{I}_n)/\ln(2L)$~\cite{k957-fcmh,PhysRevLett.112.057203,PhysRevLett.124.200602,ji2025fibonacci,zhang2025reentrant}, where $2L$ corresponds to the Hilbert space dimension. To distinguish pure phases from mixed phases,
	we calculate the composite participation ratio $\eta=\log_{10}[\langle \mathcal{I}_n \rangle \langle \mathcal{N}_n \rangle]$~\cite{PhysRevB.101.064203,PhysRevB.107.224201,PhysRevA.108.033305}, where $\langle \cdot \rangle$ denotes averaging over the entire spectrum.

	\paragraph{Phase diagrams.--- } Fig.~\ref{fig1} (a) presents a global phase diagram influenced by both $W$ and $\Delta$ for $V/t=2$. It contains purely critical phases (C) and purely extended phases (E), while no purely localized phases (L) are observed, which only appear for large $V$~\cite{SuppMateria}. 
	In addition, two mixed phases, denoted as M$_1$ and M$_2$, are identified. The M$_1$ (M$_2$) phase corresponds to the regions without (with) anomalous mobility edges. Specifically, M$_1$ consists of a mixture of localized and extended states, whereas M$_2$ mixes critical states with localized states, extended states, or both.
	Along the abscissa ($W/t = 0$), the system undergoes a transition from C to E as $\Delta$ increases, with the transition point located at $\Delta/t = 2$~\cite{PhysRevLett.125.073204,PhysRevB.93.104504}. In contrast, along the ordinate ($\Delta/t = 0$), Eq.~\eqref{h1} reduces to the AA model~\cite{aubry1980}, where $W$ does not change the properties of the system. Since $V/t=2$ is the critical point of the AA model, the entire ordinate is in C. In the limiting case of $\Delta/t\rightarrow\infty$, the system resides in E for finite $W$~\cite{SuppMateria}. Therefore, tuning $\Delta$ can induce a C\textendash{E} transition.
	In this process, conventional and anomalous mobility edges emerge. Interestingly, the M$_1$ phase is not contiguous, which is divided into two separate regions by the M$_2$ phase. This causes the system to experience a reentrant transition of M$_1$ as $\Delta$ increases. The fundamental reason for this phenomenon is analyzed below. Moreover, a \emph{reentrant criticality transition} is observed in the regime of perturbative $W$. In Fig.~\ref{fig1} (b), we show another phase diagram for $V/t=1.4$. In this scenario, the abscissa undergoes an E\textendash{C}\textendash{E} transition as $\Delta$ increases~\cite{PhysRevLett.125.073204,PhysRevB.93.104504}, whereas the ordinate is in E. Compared to Fig.~\ref{fig1} (a), E encroaches upon the survival region of mixed phases, resulting in a significant suppression of mixed phases. Note that the stability of mixed phases varies with $W$ and $\Delta$. It results in the M$_1$ phase being divided into three regions~\cite{SuppMateria}. More importantly, it causes the system to undergo a reentrant delocalization transition as $W$ increases for weak $\Delta$, e.g., $\Delta/t=0.5$, marked by the white dotted line in Fig.~\ref{fig1} (b). When perturbative random disorder and interactions are introduced, the phase diagram is not substantially modified, indicating the robustness of the reentrant behavior.
	
	\begin{figure}[t]
		\includegraphics[width=1.0\columnwidth,height=0.8\columnwidth]{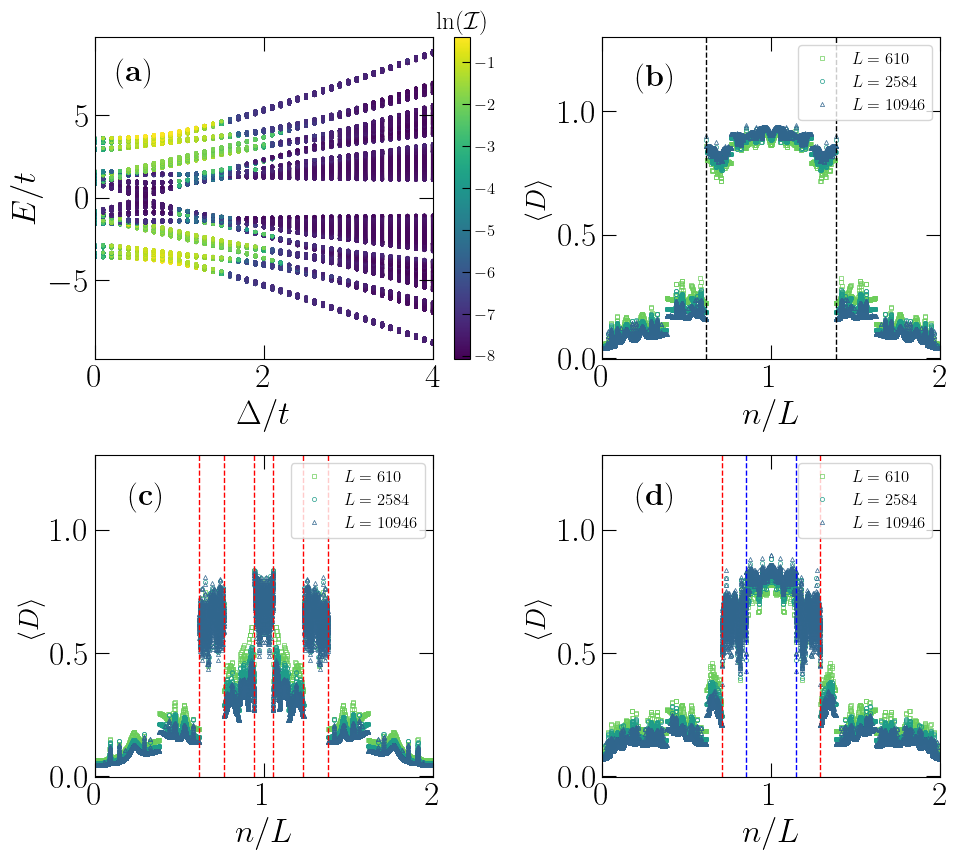}
		\vspace{-0.4cm}
		\caption{ (a) Spectrum as a function of $\Delta$, along the white dotted line in Fig.~\ref{fig1} (a). (b)\textendash(d) Mean fractal dimension $\langle D\rangle$ at various energy levels for $\Delta/t=0.6$, $1.0$, and $1.3$, as indicated by orange markers \protect\solidSquare[orange]{1.6ex}, \protect\solidCircle[orange]{1.6ex}, and \protect\solidTriangle[orange]{1.9ex} in Fig.~\ref{fig1} (a), respectively. Other cases for $\Delta/t=1.4$, $1.8$, and $3.0$ are provided in the Supplemental Material~\cite{SuppMateria}. Black dotted lines mark conventional mobility edges, whereas red and blue dotted lines denote two distinct classes of anomalous mobility edges. 
		}
		\label{fig2}
	\end{figure}
	
	\paragraph{Conventional and anomalous mobility edges.--- }
	In Fig.~\ref{fig2} (a), we show the spectrum as a function of $\Delta$ for $V/t=2$ and $W/t=1$. This demonstrates that localized states appear during the C\textendash{E} transition, leading to conventional and anomalous mobility edges. To distinguish two classes of anomalous mobility edges, we label them $aME_1$ and $aME_2$, defined as the critical energies separating critical states from localized and extended states in the spectrum, respectively. In Figs.~\ref{fig2} (b)\textendash(d), we use 1000 samples to calculate the mean fractal dimension $\langle D\rangle$ for various energy levels to show more details of mobility edges.
	In Fig.~\ref{fig2} (b), $\langle D \rangle$ approaches unity as the system size increases in the region $0.618 < n/L < 1.382$~\cite{SuppMateria}, indicating extended states~\cite{k957-fcmh,PhysRevLett.112.057203,PhysRevLett.124.200602,ji2025fibonacci,zhang2025reentrant}. Outside this region, $\langle D \rangle$ decreases with the system size and is expected to approach zero in the thermodynamic limit~\cite{SuppMateria}, signaling localized states~\cite{k957-fcmh,PhysRevLett.112.057203,PhysRevLett.124.200602,ji2025fibonacci,zhang2025reentrant}. The separation points, marked by black dotted lines, correspond to conventional mobility edges. Fig.~\ref{fig2} (c) illustrates anomalous mobility edges $aME_1$, where in the critical regions $0.618 < n/L < 0.766$, $0.944 < n/L < 1.056$, and $1.234 < n/L < 1.382$, $\langle D \rangle$ approaches a finite value within $0<\langle D \rangle<1$ as the system size increases~\cite{SuppMateria}, revealing the presence of critical states. The behavior of localized states outside the critical regions remains consistent with that in Fig.~\ref{fig2} (b). The anomalous mobility edges $aME_2$ are shown in Fig.~\ref{fig2} (d), marked by blue dotted lines. Here, extended states emerge within $0.854 < n/L < 1.146$, owing to the enhanced density of states near the middle of the spectrum at $\Delta/t = 1.3$, as illustrated in Fig.~\ref{fig2} (a). This is consistent with the re-emergence mechanism of extended states during reentrant localization transitions~\cite{PhysRevLett.126.106803,PhysRevB.111.134204}, where the increased density promotes overlap among non-extended states, facilitating the formation of extended states. When we increase $\Delta$ to reduce the density of states, these extended states are destroyed and reduced to critical states again~\cite{SuppMateria}.  As $\Delta$ further increases, the system reenters a M$_1$ phase before entering E. The coexistence behavior in this regime is similar to that in Fig.~\ref{fig2} (b)~\cite{SuppMateria}. It is worth noting that the rich coexistence phenomena remain robust against weak random disorder and interaction perturbations~\cite{SuppMateria}.

	\begin{figure}[t]
		\includegraphics[width=1.0\columnwidth,height=0.8\columnwidth]{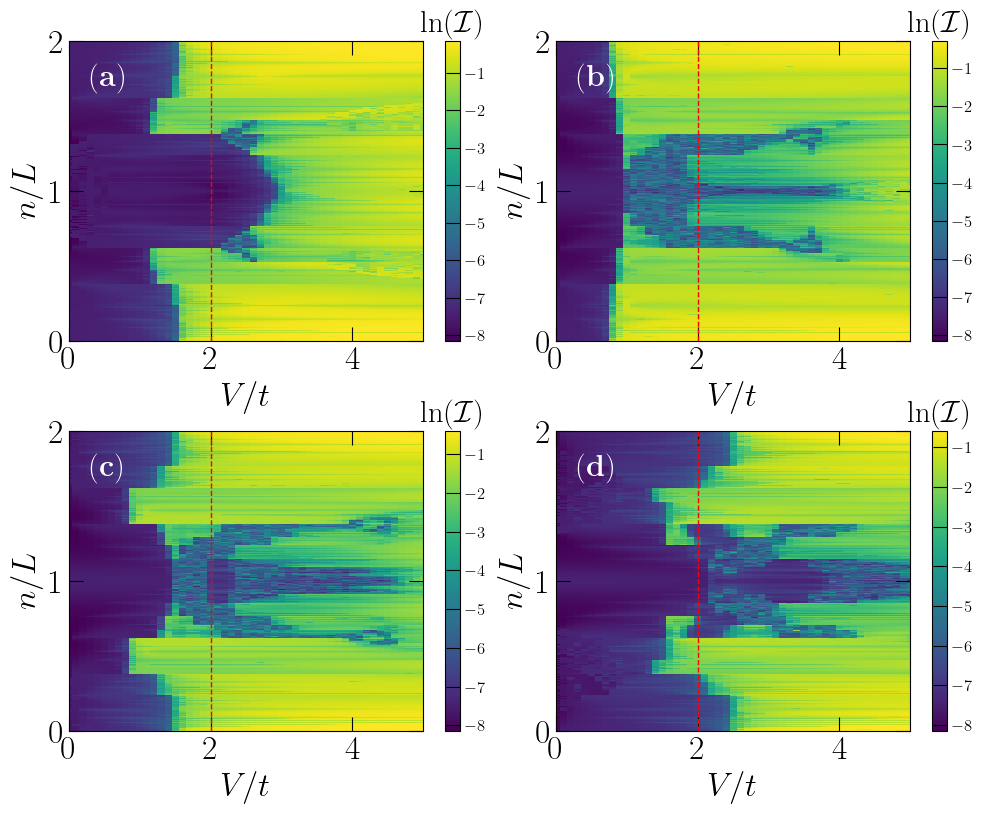}
		\vspace{-0.4cm}
		\caption{(a)\textendash(d) The IPR associated with the energy level index as a function of $V/t$ for $W/t=1$ and $\Delta/t=0.6$, $1.0$, $1.3$, and $1.8$, respectively. Red dotted lines mark $V/t=2$, corresponding to orange markers \protect\solidSquare[orange]{1.6ex}, \protect\solidCircle[orange]{1.6ex}, \protect\solidTriangle[orange]{1.9ex}, and \protect\solidStar[orange]{1.9ex} in Fig.~\ref{fig1} (a).
		}
		\label{fig3}
	\end{figure}
	
	To clarify the origin of the reentrant transition in Fig.~\ref{fig2} (a), we fix $W/t=1$ to investigate the localization transition induced by the quasi-periodic Zeeman potential for $\Delta/t=0.6$, $1.0$, $1.3$, and $1.8$, as shown in Figs.~\ref{fig3} (a)\textendash(d). For $\Delta/t=0.6$, the system undergoes an E\textendash{L} transition as $V/t$ increases. During this process, both conventional and anomalous mobility edges emerge in the spectrum. In comparison to the case of $W/t=0$ discussed in the Supplemental Material~\cite{SuppMateria}, the addition of the uniform Zeeman potential $W$ in Fig.~\ref{fig3} (a) causes extended and localized states to encroach upon regions originally occupied by critical states. This change results in the emergence of localized states at the specific amplitude of the quasi-periodic Zeeman potential, e.g., $V/t = 2$, indicated by the red dotted line in Fig.~\ref{fig3} (a).
	When $\Delta/t=1.0$ in Fig.~\ref{fig3} (b), E is suppressed relative to the case of $\Delta/t=0.6$, and the system forms localized states under a weaker quasi-periodic Zeeman potential. Moreover, the increase of $\Delta$ is more conducive to the existence of critical states~\cite{PhysRevLett.125.073204,PhysRevB.93.104504,SuppMateria}, promoting the formation of anomalous mobility edges. As $\Delta$ continues to increase, it produces the opposite effect in Fig.~\ref{fig3} (c), where a larger $\Delta$ facilitates the survival of E, requiring stronger $V$ to drive the localization transition. This trend is consistent with the case of $W/t=0$~\cite{SuppMateria}. Remarkably, as $V$ increases,  the extended states vanish completely at $V/t=1.4$ and then reappear near the middle of the spectrum within $1.9 < V/t < 2.4$, following the same mechanism discussed in Fig.~\ref{fig2} (d). For even larger $\Delta$, the localization transition with both conventional and anomalous mobility edges shifts to higher amplitudes of the quasi-periodic Zeeman potential in Fig.~\ref{fig3} (d). When we fix $V/t=2$ (red dotted lines) in Figs.~\ref{fig3} (a)\textendash(d), the system experiences mixed phases M$_1$$\to$M$_2$$\to$M$_2$$\to$M$_1$ as $\Delta$ increases. Fundamentally, the above phenomena, whether localized states emerge during the critical-extended transition or the reentrant transition of M$_1$, all originate from the nonmonotonic role of SOC. The enhancement of SOC initially suppresses extended states, thereby promoting critical and localized phases. In contrast, further increasing SOC favors the reemergence of extended states while suppressing critical and localized ones. Moreover, SOC regulates the density of states and induces hybridization-driven delocalization, which further contributes to the observed reentrant behavior.
	
	\begin{figure}[t]
		\includegraphics[width=1.0\columnwidth,height=0.7\columnwidth]{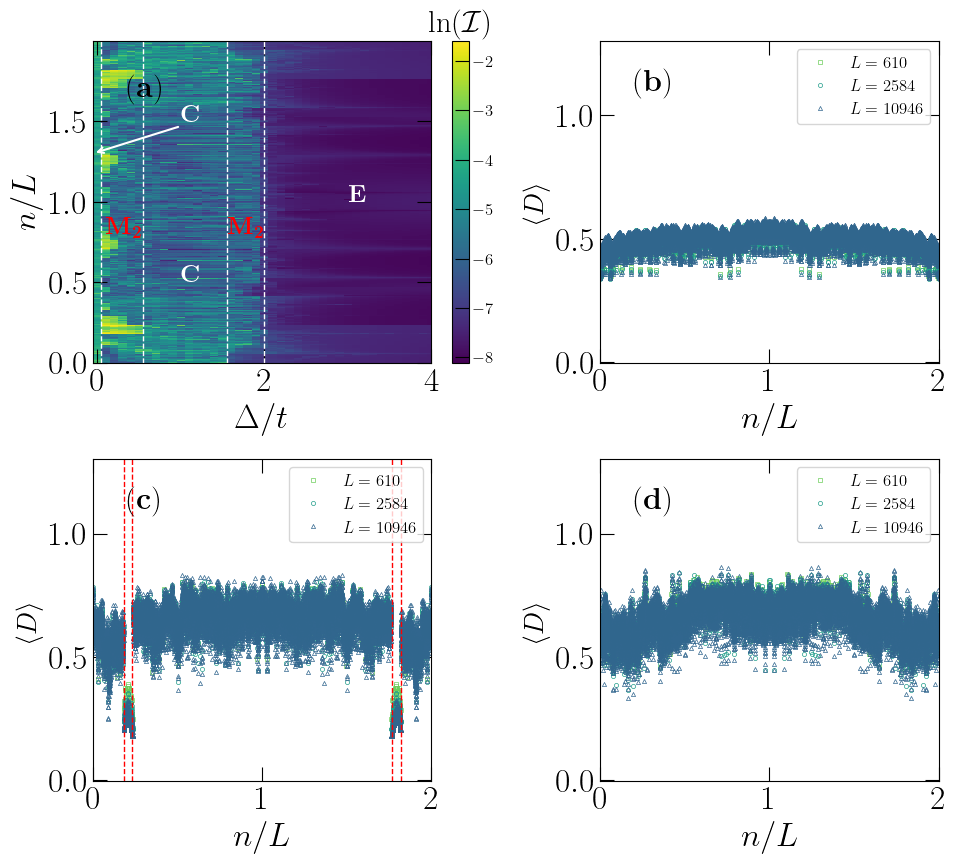}
		\vspace{-0.4cm}
		\caption{(a) IPR as a function of $\Delta$ for $V/t=2$ and $W/t=0.01$. (b)\textendash(d) $\langle D\rangle$ at various energy levels for $\Delta/t=0.0$, $0.4$, and $0.8$, respectively. Other parameters are the same as those in Fig.~\ref{fig1} (a). The white dotted lines in (a) mark boundaries between different phases, whereas the red dotted lines in (c) indicate anomalous mobility edges $aME_1$.
		}
		\label{fig4}
	\end{figure}

	\paragraph{Reentrant transitions.--- }Fig.~\ref{fig4} (a) displays the IPR associated with the energy level index $n$ as a function of $\Delta$ for $V/t=2$ and $W/t=0.01$, where $W$ is chosen sufficiently small to preserve the purely critical phase~\cite{SuppMateria}.
	As $\Delta$ decreases from $4$ to $0$, the system undergoes an E\textendash{M$_2$}\textendash{C}\textendash{M$_2$}\textendash{C} transition, entering the purely critical phase C twice. Notably, the two C originate from distinct mechanisms. For $\Delta/t=0$, the variation of $W$ does not affect the nature of the system, and the choice of $V/t=2$ ensures that the system remains in C regardless of $W$. In contrast, for $\Delta>0$, $W$ competes with $\Delta$, leading to the breakdown of part of C into the M$_{2}$ phase. The surviving C in the region $0.5<\Delta/t<1.6$, together with C at $\Delta/t=0$, gives rise to a \emph{reentrant criticality transition} as $\Delta$ decreases. Moreover, Fig.~\ref{fig4} (a) also shows that non-critical states penetrate C from multiple locations, resulting in two distinct M$_2$ phases.
	In Figs.~\ref{fig4} (b) and (d), we present $\langle D \rangle$ as a function of the energy level for $\Delta/t=0$ and $\Delta/t=0.8$, respectively. As the system size increases, the fractal dimensions of all states approach values within $0<\langle D \rangle<1$, indicating that the system resides in $C$. In contrast, for $\Delta/t=0.4$ in Fig.~\ref{fig4} (c), both localized and critical states coexist in the spectrum, producing anomalous mobility edges $aME_1$ (red dotted lines).
	
	\begin{figure}[htbp]
		\includegraphics[width=1.0\columnwidth,height=0.8\columnwidth]{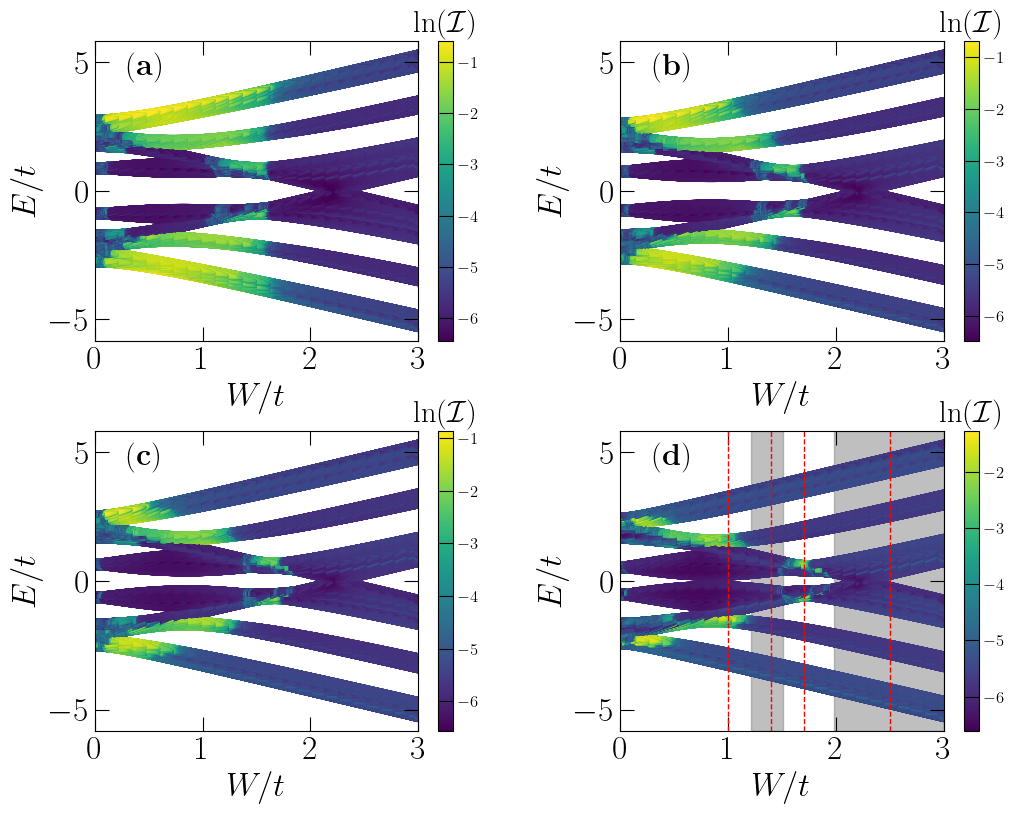}
		\vspace{-0.4cm}
		\caption{(a)\textendash(d) Spectra as a function of $W/t$ for $\Delta/t=0.8$, $0.7$, $0.6$ and $0.5$, sliced from Fig.~\ref{fig1} (b). In (d), the gray regions mark purely extended phases E and red dotted lines correspond to orange markers \protect\solidSquare[orange]{1.6ex}, \protect\solidCircle[orange]{1.6ex}, \protect\solidTriangle[orange]{1.9ex}, and \protect\solidStar[orange]{1.9ex} in Fig.~\ref{fig1} (b)
		}
		\label{fig5}
	\end{figure}
	
	In addition to the \emph{reentrant criticality transition}, we also observe a reentrant delocalization transition, as shown in Fig.~\ref{fig1} (b). To explore the mechanism behind this phenomenon, we vary $\Delta/t$ from $0.8$ to $0.5$ and plot the spectra as a function of $W$ in Fig.~\ref{fig5}. When $\Delta/t=0.8$, extended states appear near the middle of the spectrum for small $W/t$, in contrast to the case of $\Delta/t = 1.0$~\cite{SuppMateria}. This behavior can be attributed to two primary factors. First, the reduction of $\Delta/t$ facilitates the transition from critical states to extended states~\cite{PhysRevB.93.104504,PhysRevA.105.013315}. In the limit $\Delta/t \to 0$, Eq.~\eqref{h1} reduces to the AA model, where all states are extended for $V/t=1.4$~\cite{aubry1980}. Second, critical states in the middle of the spectrum possess larger fractal dimensions, indicating greater spatial extension, and are thus more likely to evolve into extended states as $\Delta/t$ decreases. This trend is also clearly illustrated in Figs.~\ref{fig5} (a)\textendash(d), where the region of extended states around the spectrum center expands systematically with decreasing $\Delta/t$. Crucially, Figs.~\ref{fig5} (a)\textendash(d) demonstrate that the region where localized states exist shrinks as $\Delta/t$ decreases. When $\Delta/t=0.5$ in Fig.~\ref{fig5} (d), localized states vanish entirely within the region $W/t \in [1.21, 1.51]$. 
	Although a finite $W$ makes the system violate the pure phase criterion~\cite{zhou2025fundamentallocalizationphasesquasiperiodic}, the absence of overlap between mobility edges and the spectrum restores a pure phase~\cite{SuppMateria}.
	As a result, the localized region is split into two parts, causing two distinct delocalization transitions. Specifically, the first transition into E occurs at $W/t = 1.21$, which is subsequently destroyed at $W/t = 1.51$. Upon further increasing $W/t$, the system reenters E at $W/t = 1.98$, thereby exhibiting the reentrant phenomenon. Experimentally, this reentrant delocalization can be probed via wave-packet dynamics and the retention of initial-state memory, which provide direct signatures of mobility-edge shifts, and the specific experimental steps and details are shown in the Supplementary Material~\cite{SuppMateria}.

	\paragraph{Conclusions.---}In this Letter, we investigate reentrant transitions in an optical Raman lattice. In the presence of the Zeeman potentials, SOC generates anomalous mobility edges that drive reentrant transitions between mixed phases. We also observe a brand new \emph{reentrant criticality transition}, supplementing the basic framework of reentrant phenomena in quasi-periodic systems. Moreover, we identify a reentrant delocalization transition induced by the uniform Zeeman potential. This reentrance occurs because the region supporting localized states enclosed by mobility edges fails to overlap with energy bands within a certain region, leading to the disappearance of localized states. This mechanism presents a novel way to induce reentrant transitions. 
	
	The underlying model studied here is supported by well-established experimental schemes~\cite{PhysRevLett.125.073204}, and its basic phase diagram has already been realized in ultracold atomic systems~\cite{xiao2021observation} and programmable superconducting simulators~\cite{PhysRevResearch.6.L042038}. The inclusion of a uniform Zeeman splitting introduces richer reentrant phenomena, and provides a promising platform for further theoretical and experimental research. It is also worthwhile to introduce this work into interacting systems and non-Hermitian systems to explore complex reentrant transitions.
	
	\paragraph{Acknowledgments.---}X. W. thanks useful discussions with Chao Gao. We acknowledge support from the Natural Science Foundation of China (Grant Nos. 12404322, 12474492, 12461160324), the Zhejiang Provincial Natural Science Foundation of China under Grant No. LQ24A040004, and the Science Foundation of Zhejiang Sci-Tech University (Grant No. 23062152-Y).
	
	\paragraph{Data availability.---}The data that support the findings of this Letter are openly available~\cite{data}
	
	\bibliography{reference}

\begin{thebibliography}{53}%
\makeatletter
\providecommand \@ifxundefined [1]{%
 \@ifx{#1\undefined}
}%
\providecommand \@ifnum [1]{%
 \ifnum #1\expandafter \@firstoftwo
 \else \expandafter \@secondoftwo
 \fi
}%
\providecommand \@ifx [1]{%
 \ifx #1\expandafter \@firstoftwo
 \else \expandafter \@secondoftwo
 \fi
}%
\providecommand \natexlab [1]{#1}%
\providecommand \enquote  [1]{``#1''}%
\providecommand \bibnamefont  [1]{#1}%
\providecommand \bibfnamefont [1]{#1}%
\providecommand \citenamefont [1]{#1}%
\providecommand \href@noop [0]{\@secondoftwo}%
\providecommand \href [0]{\begingroup \@sanitize@url \@href}%
\providecommand \@href[1]{\@@startlink{#1}\@@href}%
\providecommand \@@href[1]{\endgroup#1\@@endlink}%
\providecommand \@sanitize@url [0]{\catcode `\\12\catcode `\$12\catcode
  `\&12\catcode `\#12\catcode `\^12\catcode `\_12\catcode `\%12\relax}%
\providecommand \@@startlink[1]{}%
\providecommand \@@endlink[0]{}%
\providecommand \url  [0]{\begingroup\@sanitize@url \@url }%
\providecommand \@url [1]{\endgroup\@href {#1}{\urlprefix }}%
\providecommand \urlprefix  [0]{URL }%
\providecommand \Eprint [0]{\href }%
\providecommand \doibase [0]{https://doi.org/}%
\providecommand \selectlanguage [0]{\@gobble}%
\providecommand \bibinfo  [0]{\@secondoftwo}%
\providecommand \bibfield  [0]{\@secondoftwo}%
\providecommand \translation [1]{[#1]}%
\providecommand \BibitemOpen [0]{}%
\providecommand \bibitemStop [0]{}%
\providecommand \bibitemNoStop [0]{.\EOS\space}%
\providecommand \EOS [0]{\spacefactor3000\relax}%
\providecommand \BibitemShut  [1]{\csname bibitem#1\endcsname}%
\let\auto@bib@innerbib\@empty
\bibitem [{\citenamefont {Anderson}(1958)}]{PhysRev.109.1492}%
  \BibitemOpen
  \bibfield  {author} {\bibinfo {author} {\bibfnamefont {P.~W.}\ \bibnamefont
  {Anderson}},\ }\bibfield  {title} {\bibinfo {title} {{Absence of Diffusion in
  Certain Random Lattices}},\ }\href {https://doi.org/10.1103/PhysRev.109.1492}
  {\bibfield  {journal} {\bibinfo  {journal} {Phys. Rev.}\ }\textbf {\bibinfo
  {volume} {109}},\ \bibinfo {pages} {1492} (\bibinfo {year}
  {1958})}\BibitemShut {NoStop}%
\bibitem [{\citenamefont {Semeghini}\ \emph {et~al.}(2015)\citenamefont
  {Semeghini}, \citenamefont {Landini}, \citenamefont {Castilho}, \citenamefont
  {Roy}, \citenamefont {Spagnolli}, \citenamefont {Trenkwalder}, \citenamefont
  {Fattori}, \citenamefont {Inguscio},\ and\ \citenamefont
  {Modugno}}]{semeghini2015measurement}%
  \BibitemOpen
  \bibfield  {author} {\bibinfo {author} {\bibfnamefont {G.}~\bibnamefont
  {Semeghini}}, \bibinfo {author} {\bibfnamefont {M.}~\bibnamefont {Landini}},
  \bibinfo {author} {\bibfnamefont {P.}~\bibnamefont {Castilho}}, \bibinfo
  {author} {\bibfnamefont {S.}~\bibnamefont {Roy}}, \bibinfo {author}
  {\bibfnamefont {G.}~\bibnamefont {Spagnolli}}, \bibinfo {author}
  {\bibfnamefont {A.}~\bibnamefont {Trenkwalder}}, \bibinfo {author}
  {\bibfnamefont {M.}~\bibnamefont {Fattori}}, \bibinfo {author} {\bibfnamefont
  {M.}~\bibnamefont {Inguscio}},\ and\ \bibinfo {author} {\bibfnamefont
  {G.}~\bibnamefont {Modugno}},\ }\bibfield  {title} {\bibinfo {title}
  {{Measurement of the mobility edge for 3D Anderson localization}},\
  }\href@noop {} {\bibfield  {journal} {\bibinfo  {journal} {Nature Physics}\
  }\textbf {\bibinfo {volume} {11}},\ \bibinfo {pages} {554} (\bibinfo {year}
  {2015})}\BibitemShut {NoStop}%
\bibitem [{\citenamefont {Pasek}\ \emph {et~al.}(2017)\citenamefont {Pasek},
  \citenamefont {Orso},\ and\ \citenamefont
  {Delande}}]{PhysRevLett.118.170403}%
  \BibitemOpen
  \bibfield  {author} {\bibinfo {author} {\bibfnamefont {M.}~\bibnamefont
  {Pasek}}, \bibinfo {author} {\bibfnamefont {G.}~\bibnamefont {Orso}},\ and\
  \bibinfo {author} {\bibfnamefont {D.}~\bibnamefont {Delande}},\ }\bibfield
  {title} {\bibinfo {title} {{Anderson Localization of Ultracold Atoms: Where
  is the Mobility Edge?}},\ }\href
  {https://doi.org/10.1103/PhysRevLett.118.170403} {\bibfield  {journal}
  {\bibinfo  {journal} {Phys. Rev. Lett.}\ }\textbf {\bibinfo {volume} {118}},\
  \bibinfo {pages} {170403} (\bibinfo {year} {2017})}\BibitemShut {NoStop}%
\bibitem [{\citenamefont {Longhi}(2024)}]{PhysRevLett.132.236301}%
  \BibitemOpen
  \bibfield  {author} {\bibinfo {author} {\bibfnamefont {S.}~\bibnamefont
  {Longhi}},\ }\bibfield  {title} {\bibinfo {title} {{Dephasing-Induced
  Mobility Edges in Quasicrystals}},\ }\href
  {https://doi.org/10.1103/PhysRevLett.132.236301} {\bibfield  {journal}
  {\bibinfo  {journal} {Phys. Rev. Lett.}\ }\textbf {\bibinfo {volume} {132}},\
  \bibinfo {pages} {236301} (\bibinfo {year} {2024})}\BibitemShut {NoStop}%
\bibitem [{\citenamefont {Ganeshan}\ \emph {et~al.}(2015)\citenamefont
  {Ganeshan}, \citenamefont {Pixley},\ and\ \citenamefont
  {Das~Sarma}}]{PhysRevLett.114.146601}%
  \BibitemOpen
  \bibfield  {author} {\bibinfo {author} {\bibfnamefont {S.}~\bibnamefont
  {Ganeshan}}, \bibinfo {author} {\bibfnamefont {J.~H.}\ \bibnamefont
  {Pixley}},\ and\ \bibinfo {author} {\bibfnamefont {S.}~\bibnamefont
  {Das~Sarma}},\ }\bibfield  {title} {\bibinfo {title} {{Nearest Neighbor Tight
  Binding Models with an Exact Mobility Edge in One Dimension}},\ }\href
  {https://doi.org/10.1103/PhysRevLett.114.146601} {\bibfield  {journal}
  {\bibinfo  {journal} {Phys. Rev. Lett.}\ }\textbf {\bibinfo {volume} {114}},\
  \bibinfo {pages} {146601} (\bibinfo {year} {2015})}\BibitemShut {NoStop}%
\bibitem [{\citenamefont {Wang}\ \emph
  {et~al.}(2020{\natexlab{a}})\citenamefont {Wang}, \citenamefont {Xia},
  \citenamefont {Zhang}, \citenamefont {Yao}, \citenamefont {Chen},
  \citenamefont {You}, \citenamefont {Zhou},\ and\ \citenamefont
  {Liu}}]{PhysRevLett.125.196604}%
  \BibitemOpen
  \bibfield  {author} {\bibinfo {author} {\bibfnamefont {Y.}~\bibnamefont
  {Wang}}, \bibinfo {author} {\bibfnamefont {X.}~\bibnamefont {Xia}}, \bibinfo
  {author} {\bibfnamefont {L.}~\bibnamefont {Zhang}}, \bibinfo {author}
  {\bibfnamefont {H.}~\bibnamefont {Yao}}, \bibinfo {author} {\bibfnamefont
  {S.}~\bibnamefont {Chen}}, \bibinfo {author} {\bibfnamefont {J.}~\bibnamefont
  {You}}, \bibinfo {author} {\bibfnamefont {Q.}~\bibnamefont {Zhou}},\ and\
  \bibinfo {author} {\bibfnamefont {X.-J.}\ \bibnamefont {Liu}},\ }\bibfield
  {title} {\bibinfo {title} {{One-Dimensional Quasiperiodic Mosaic Lattice with
  Exact Mobility Edges}},\ }\href
  {https://doi.org/10.1103/PhysRevLett.125.196604} {\bibfield  {journal}
  {\bibinfo  {journal} {Phys. Rev. Lett.}\ }\textbf {\bibinfo {volume} {125}},\
  \bibinfo {pages} {196604} (\bibinfo {year} {2020}{\natexlab{a}})}\BibitemShut
  {NoStop}%
\bibitem [{\citenamefont {Biddle}\ and\ \citenamefont
  {Das~Sarma}(2010)}]{PhysRevLett.104.070601}%
  \BibitemOpen
  \bibfield  {author} {\bibinfo {author} {\bibfnamefont {J.}~\bibnamefont
  {Biddle}}\ and\ \bibinfo {author} {\bibfnamefont {S.}~\bibnamefont
  {Das~Sarma}},\ }\bibfield  {title} {\bibinfo {title} {{Predicted Mobility
  Edges in One-Dimensional Incommensurate Optical Lattices: An Exactly Solvable
  Model of Anderson Localization}},\ }\href
  {https://doi.org/10.1103/PhysRevLett.104.070601} {\bibfield  {journal}
  {\bibinfo  {journal} {Phys. Rev. Lett.}\ }\textbf {\bibinfo {volume} {104}},\
  \bibinfo {pages} {070601} (\bibinfo {year} {2010})}\BibitemShut {NoStop}%
\bibitem [{\citenamefont {Yao}\ \emph {et~al.}(2019)\citenamefont {Yao},
  \citenamefont {Khoudli}, \citenamefont {Bresque},\ and\ \citenamefont
  {Sanchez-Palencia}}]{PhysRevLett.123.070405}%
  \BibitemOpen
  \bibfield  {author} {\bibinfo {author} {\bibfnamefont {H.}~\bibnamefont
  {Yao}}, \bibinfo {author} {\bibfnamefont {A.}~\bibnamefont {Khoudli}},
  \bibinfo {author} {\bibfnamefont {L.}~\bibnamefont {Bresque}},\ and\ \bibinfo
  {author} {\bibfnamefont {L.}~\bibnamefont {Sanchez-Palencia}},\ }\bibfield
  {title} {\bibinfo {title} {{Critical Behavior and Fractality in Shallow
  One-Dimensional Quasiperiodic Potentials}},\ }\href
  {https://doi.org/10.1103/PhysRevLett.123.070405} {\bibfield  {journal}
  {\bibinfo  {journal} {Phys. Rev. Lett.}\ }\textbf {\bibinfo {volume} {123}},\
  \bibinfo {pages} {070405} (\bibinfo {year} {2019})}\BibitemShut {NoStop}%
\bibitem [{\citenamefont {Liu}\ \emph {et~al.}(2022)\citenamefont {Liu},
  \citenamefont {Xia}, \citenamefont {Longhi},\ and\ \citenamefont
  {Sanchez-Palencia}}]{10.21468/SciPostPhys.12.1.027}%
  \BibitemOpen
  \bibfield  {author} {\bibinfo {author} {\bibfnamefont {T.}~\bibnamefont
  {Liu}}, \bibinfo {author} {\bibfnamefont {X.}~\bibnamefont {Xia}}, \bibinfo
  {author} {\bibfnamefont {S.}~\bibnamefont {Longhi}},\ and\ \bibinfo {author}
  {\bibfnamefont {L.}~\bibnamefont {Sanchez-Palencia}},\ }\bibfield  {title}
  {\bibinfo {title} {{Anomalous mobility edges in one-dimensional quasiperiodic
  models}},\ }\href {https://doi.org/10.21468/SciPostPhys.12.1.027} {\bibfield
  {journal} {\bibinfo  {journal} {SciPost Phys.}\ }\textbf {\bibinfo {volume}
  {12}},\ \bibinfo {pages} {027} (\bibinfo {year} {2022})}\BibitemShut
  {NoStop}%
\bibitem [{\citenamefont {Wang}\ \emph {et~al.}(2023)\citenamefont {Wang},
  \citenamefont {Zhang}, \citenamefont {Wang},\ and\ \citenamefont
  {Chen}}]{PhysRevB.108.174202}%
  \BibitemOpen
  \bibfield  {author} {\bibinfo {author} {\bibfnamefont {Z.}~\bibnamefont
  {Wang}}, \bibinfo {author} {\bibfnamefont {Y.}~\bibnamefont {Zhang}},
  \bibinfo {author} {\bibfnamefont {L.}~\bibnamefont {Wang}},\ and\ \bibinfo
  {author} {\bibfnamefont {S.}~\bibnamefont {Chen}},\ }\bibfield  {title}
  {\bibinfo {title} {{Engineering mobility in quasiperiodic lattices with exact
  mobility edges}},\ }\href {https://doi.org/10.1103/PhysRevB.108.174202}
  {\bibfield  {journal} {\bibinfo  {journal} {Phys. Rev. B}\ }\textbf {\bibinfo
  {volume} {108}},\ \bibinfo {pages} {174202} (\bibinfo {year}
  {2023})}\BibitemShut {NoStop}%
\bibitem [{\citenamefont {Zhou}\ \emph {et~al.}(2023)\citenamefont {Zhou},
  \citenamefont {Wang}, \citenamefont {Poon}, \citenamefont {Zhou},\ and\
  \citenamefont {Liu}}]{PhysRevLett.131.176401}%
  \BibitemOpen
  \bibfield  {author} {\bibinfo {author} {\bibfnamefont {X.-C.}\ \bibnamefont
  {Zhou}}, \bibinfo {author} {\bibfnamefont {Y.}~\bibnamefont {Wang}}, \bibinfo
  {author} {\bibfnamefont {T.-F.~J.}\ \bibnamefont {Poon}}, \bibinfo {author}
  {\bibfnamefont {Q.}~\bibnamefont {Zhou}},\ and\ \bibinfo {author}
  {\bibfnamefont {X.-J.}\ \bibnamefont {Liu}},\ }\bibfield  {title} {\bibinfo
  {title} {{Exact New Mobility Edges between Critical and Localized States}},\
  }\href {https://doi.org/10.1103/PhysRevLett.131.176401} {\bibfield  {journal}
  {\bibinfo  {journal} {Phys. Rev. Lett.}\ }\textbf {\bibinfo {volume} {131}},\
  \bibinfo {pages} {176401} (\bibinfo {year} {2023})}\BibitemShut {NoStop}%
\bibitem [{\citenamefont {Wang}\ and\ \citenamefont
  {Zhou}(2025)}]{wang2025exactnewmobilityedges}%
  \BibitemOpen
  \bibfield  {author} {\bibinfo {author} {\bibfnamefont {Y.}~\bibnamefont
  {Wang}}\ and\ \bibinfo {author} {\bibfnamefont {Q.}~\bibnamefont {Zhou}},\
  }\href {https://arxiv.org/abs/2501.17523} {\bibinfo {title} {{Exact new
  mobility edges}}} (\bibinfo {year} {2025}),\ \Eprint
  {https://arxiv.org/abs/2501.17523} {arXiv:2501.17523 [math.DS]} \BibitemShut
  {NoStop}%
\bibitem [{\citenamefont {Mott}(1967)}]{mott1967electrons}%
  \BibitemOpen
  \bibfield  {author} {\bibinfo {author} {\bibfnamefont {N.}~\bibnamefont
  {Mott}},\ }\bibfield  {title} {\bibinfo {title} {{Electrons in disordered
  structures}},\ }\href@noop {} {\bibfield  {journal} {\bibinfo  {journal}
  {Advances in Physics}\ }\textbf {\bibinfo {volume} {16}},\ \bibinfo {pages}
  {49} (\bibinfo {year} {1967})}\BibitemShut {NoStop}%
\bibitem [{\citenamefont {L\"uschen}\ \emph {et~al.}(2018)\citenamefont
  {L\"uschen}, \citenamefont {Scherg}, \citenamefont {Kohlert}, \citenamefont
  {Schreiber}, \citenamefont {Bordia}, \citenamefont {Li}, \citenamefont
  {Das~Sarma},\ and\ \citenamefont {Bloch}}]{PhysRevLett.120.160404}%
  \BibitemOpen
  \bibfield  {author} {\bibinfo {author} {\bibfnamefont {H.~P.}\ \bibnamefont
  {L\"uschen}}, \bibinfo {author} {\bibfnamefont {S.}~\bibnamefont {Scherg}},
  \bibinfo {author} {\bibfnamefont {T.}~\bibnamefont {Kohlert}}, \bibinfo
  {author} {\bibfnamefont {M.}~\bibnamefont {Schreiber}}, \bibinfo {author}
  {\bibfnamefont {P.}~\bibnamefont {Bordia}}, \bibinfo {author} {\bibfnamefont
  {X.}~\bibnamefont {Li}}, \bibinfo {author} {\bibfnamefont {S.}~\bibnamefont
  {Das~Sarma}},\ and\ \bibinfo {author} {\bibfnamefont {I.}~\bibnamefont
  {Bloch}},\ }\bibfield  {title} {\bibinfo {title} {{Single-Particle Mobility
  Edge in a One-Dimensional Quasiperiodic Optical Lattice}},\ }\href
  {https://doi.org/10.1103/PhysRevLett.120.160404} {\bibfield  {journal}
  {\bibinfo  {journal} {Phys. Rev. Lett.}\ }\textbf {\bibinfo {volume} {120}},\
  \bibinfo {pages} {160404} (\bibinfo {year} {2018})}\BibitemShut {NoStop}%
\bibitem [{\citenamefont {An}\ \emph {et~al.}(2021)\citenamefont {An},
  \citenamefont {Padavi\ifmmode~\acute{c}\else \'{c}\fi{}}, \citenamefont
  {Meier}, \citenamefont {Hegde}, \citenamefont {Ganeshan}, \citenamefont
  {Pixley}, \citenamefont {Vishveshwara},\ and\ \citenamefont
  {Gadway}}]{PhysRevLett.126.040603}%
  \BibitemOpen
  \bibfield  {author} {\bibinfo {author} {\bibfnamefont {F.~A.}\ \bibnamefont
  {An}}, \bibinfo {author} {\bibfnamefont {K.}~\bibnamefont
  {Padavi\ifmmode~\acute{c}\else \'{c}\fi{}}}, \bibinfo {author} {\bibfnamefont
  {E.~J.}\ \bibnamefont {Meier}}, \bibinfo {author} {\bibfnamefont
  {S.}~\bibnamefont {Hegde}}, \bibinfo {author} {\bibfnamefont
  {S.}~\bibnamefont {Ganeshan}}, \bibinfo {author} {\bibfnamefont {J.~H.}\
  \bibnamefont {Pixley}}, \bibinfo {author} {\bibfnamefont {S.}~\bibnamefont
  {Vishveshwara}},\ and\ \bibinfo {author} {\bibfnamefont {B.}~\bibnamefont
  {Gadway}},\ }\bibfield  {title} {\bibinfo {title} {{Interactions and Mobility
  Edges: Observing the Generalized Aubry-Andr\'e Model}},\ }\href
  {https://doi.org/10.1103/PhysRevLett.126.040603} {\bibfield  {journal}
  {\bibinfo  {journal} {Phys. Rev. Lett.}\ }\textbf {\bibinfo {volume} {126}},\
  \bibinfo {pages} {040603} (\bibinfo {year} {2021})}\BibitemShut {NoStop}%
\bibitem [{\citenamefont {Wang}\ \emph {et~al.}(2022)\citenamefont {Wang},
  \citenamefont {Zhang}, \citenamefont {Li}, \citenamefont {Wu}, \citenamefont
  {Liu}, \citenamefont {Mei}, \citenamefont {Hu}, \citenamefont {Xiao},
  \citenamefont {Ma}, \citenamefont {Chin},\ and\ \citenamefont
  {Jia}}]{PhysRevLett.129.103401}%
  \BibitemOpen
  \bibfield  {author} {\bibinfo {author} {\bibfnamefont {Y.}~\bibnamefont
  {Wang}}, \bibinfo {author} {\bibfnamefont {J.-H.}\ \bibnamefont {Zhang}},
  \bibinfo {author} {\bibfnamefont {Y.}~\bibnamefont {Li}}, \bibinfo {author}
  {\bibfnamefont {J.}~\bibnamefont {Wu}}, \bibinfo {author} {\bibfnamefont
  {W.}~\bibnamefont {Liu}}, \bibinfo {author} {\bibfnamefont {F.}~\bibnamefont
  {Mei}}, \bibinfo {author} {\bibfnamefont {Y.}~\bibnamefont {Hu}}, \bibinfo
  {author} {\bibfnamefont {L.}~\bibnamefont {Xiao}}, \bibinfo {author}
  {\bibfnamefont {J.}~\bibnamefont {Ma}}, \bibinfo {author} {\bibfnamefont
  {C.}~\bibnamefont {Chin}},\ and\ \bibinfo {author} {\bibfnamefont
  {S.}~\bibnamefont {Jia}},\ }\bibfield  {title} {\bibinfo {title}
  {{Observation of Interaction-Induced Mobility Edge in an Atomic Aubry-Andr\'e
  Wire}},\ }\href {https://doi.org/10.1103/PhysRevLett.129.103401} {\bibfield
  {journal} {\bibinfo  {journal} {Phys. Rev. Lett.}\ }\textbf {\bibinfo
  {volume} {129}},\ \bibinfo {pages} {103401} (\bibinfo {year}
  {2022})}\BibitemShut {NoStop}%
\bibitem [{\citenamefont {Chang}\ \emph {et~al.}(2025)\citenamefont {Chang},
  \citenamefont {Zhang}, \citenamefont {Lu}, \citenamefont {Yang},
  \citenamefont {Mei}, \citenamefont {Ma}, \citenamefont {Jia},\ and\
  \citenamefont {Jin}}]{PhysRevLett.134.053601}%
  \BibitemOpen
  \bibfield  {author} {\bibinfo {author} {\bibfnamefont {Y.-J.}\ \bibnamefont
  {Chang}}, \bibinfo {author} {\bibfnamefont {J.-H.}\ \bibnamefont {Zhang}},
  \bibinfo {author} {\bibfnamefont {Y.-H.}\ \bibnamefont {Lu}}, \bibinfo
  {author} {\bibfnamefont {Y.-Y.}\ \bibnamefont {Yang}}, \bibinfo {author}
  {\bibfnamefont {F.}~\bibnamefont {Mei}}, \bibinfo {author} {\bibfnamefont
  {J.}~\bibnamefont {Ma}}, \bibinfo {author} {\bibfnamefont {S.}~\bibnamefont
  {Jia}},\ and\ \bibinfo {author} {\bibfnamefont {X.-M.}\ \bibnamefont {Jin}},\
  }\bibfield  {title} {\bibinfo {title} {{Observation of Photonic Mobility Edge
  Phases}},\ }\href {https://doi.org/10.1103/PhysRevLett.134.053601} {\bibfield
   {journal} {\bibinfo  {journal} {Phys. Rev. Lett.}\ }\textbf {\bibinfo
  {volume} {134}},\ \bibinfo {pages} {053601} (\bibinfo {year}
  {2025})}\BibitemShut {NoStop}%
\bibitem [{\citenamefont {Xiao}\ \emph {et~al.}(2021)\citenamefont {Xiao},
  \citenamefont {Xie}, \citenamefont {Dong}, \citenamefont {Chen},
  \citenamefont {Yi},\ and\ \citenamefont {Yan}}]{xiao2021observation}%
  \BibitemOpen
  \bibfield  {author} {\bibinfo {author} {\bibfnamefont {T.}~\bibnamefont
  {Xiao}}, \bibinfo {author} {\bibfnamefont {D.}~\bibnamefont {Xie}}, \bibinfo
  {author} {\bibfnamefont {Z.}~\bibnamefont {Dong}}, \bibinfo {author}
  {\bibfnamefont {T.}~\bibnamefont {Chen}}, \bibinfo {author} {\bibfnamefont
  {W.}~\bibnamefont {Yi}},\ and\ \bibinfo {author} {\bibfnamefont
  {B.}~\bibnamefont {Yan}},\ }\bibfield  {title} {\bibinfo {title}
  {{Observation of topological phase with critical localization in a
  quasi-periodic lattice}},\ }\href@noop {} {\bibfield  {journal} {\bibinfo
  {journal} {Science bulletin}\ }\textbf {\bibinfo {volume} {66}},\ \bibinfo
  {pages} {2175} (\bibinfo {year} {2021})}\BibitemShut {NoStop}%
\bibitem [{\citenamefont {Goblot}\ \emph {et~al.}(2020)\citenamefont {Goblot},
  \citenamefont {{\v{S}}trkalj}, \citenamefont {Pernet}, \citenamefont {Lado},
  \citenamefont {Dorow}, \citenamefont {Lema{\^\i}tre}, \citenamefont
  {Le~Gratiet}, \citenamefont {Harouri}, \citenamefont {Sagnes}, \citenamefont
  {Ravets} \emph {et~al.}}]{goblot2020emergence}%
  \BibitemOpen
  \bibfield  {author} {\bibinfo {author} {\bibfnamefont {V.}~\bibnamefont
  {Goblot}}, \bibinfo {author} {\bibfnamefont {A.}~\bibnamefont
  {{\v{S}}trkalj}}, \bibinfo {author} {\bibfnamefont {N.}~\bibnamefont
  {Pernet}}, \bibinfo {author} {\bibfnamefont {J.~L.}\ \bibnamefont {Lado}},
  \bibinfo {author} {\bibfnamefont {C.}~\bibnamefont {Dorow}}, \bibinfo
  {author} {\bibfnamefont {A.}~\bibnamefont {Lema{\^\i}tre}}, \bibinfo {author}
  {\bibfnamefont {L.}~\bibnamefont {Le~Gratiet}}, \bibinfo {author}
  {\bibfnamefont {A.}~\bibnamefont {Harouri}}, \bibinfo {author} {\bibfnamefont
  {I.}~\bibnamefont {Sagnes}}, \bibinfo {author} {\bibfnamefont
  {S.}~\bibnamefont {Ravets}}, \emph {et~al.},\ }\bibfield  {title} {\bibinfo
  {title} {{Emergence of criticality through a cascade of delocalization
  transitions in quasiperiodic chains}},\ }\href@noop {} {\bibfield  {journal}
  {\bibinfo  {journal} {Nature Physics}\ }\textbf {\bibinfo {volume} {16}},\
  \bibinfo {pages} {832} (\bibinfo {year} {2020})}\BibitemShut {NoStop}%
\bibitem [{\citenamefont {Roy}\ \emph {et~al.}(2021)\citenamefont {Roy},
  \citenamefont {Mishra}, \citenamefont {Tanatar},\ and\ \citenamefont
  {Basu}}]{PhysRevLett.126.106803}%
  \BibitemOpen
  \bibfield  {author} {\bibinfo {author} {\bibfnamefont {S.}~\bibnamefont
  {Roy}}, \bibinfo {author} {\bibfnamefont {T.}~\bibnamefont {Mishra}},
  \bibinfo {author} {\bibfnamefont {B.}~\bibnamefont {Tanatar}},\ and\ \bibinfo
  {author} {\bibfnamefont {S.}~\bibnamefont {Basu}},\ }\bibfield  {title}
  {\bibinfo {title} {{Reentrant Localization Transition in a Quasiperiodic
  Chain}},\ }\href {https://doi.org/10.1103/PhysRevLett.126.106803} {\bibfield
  {journal} {\bibinfo  {journal} {Phys. Rev. Lett.}\ }\textbf {\bibinfo
  {volume} {126}},\ \bibinfo {pages} {106803} (\bibinfo {year}
  {2021})}\BibitemShut {NoStop}%
\bibitem [{\citenamefont {Padhan}\ \emph {et~al.}(2022)\citenamefont {Padhan},
  \citenamefont {Giri}, \citenamefont {Mondal},\ and\ \citenamefont
  {Mishra}}]{PhysRevB.105.L220201}%
  \BibitemOpen
  \bibfield  {author} {\bibinfo {author} {\bibfnamefont {A.}~\bibnamefont
  {Padhan}}, \bibinfo {author} {\bibfnamefont {M.~K.}\ \bibnamefont {Giri}},
  \bibinfo {author} {\bibfnamefont {S.}~\bibnamefont {Mondal}},\ and\ \bibinfo
  {author} {\bibfnamefont {T.}~\bibnamefont {Mishra}},\ }\bibfield  {title}
  {\bibinfo {title} {{Emergence of multiple localization transitions in a
  one-dimensional quasiperiodic lattice}},\ }\href
  {https://doi.org/10.1103/PhysRevB.105.L220201} {\bibfield  {journal}
  {\bibinfo  {journal} {Phys. Rev. B}\ }\textbf {\bibinfo {volume} {105}},\
  \bibinfo {pages} {L220201} (\bibinfo {year} {2022})}\BibitemShut {NoStop}%
\bibitem [{\citenamefont {\ifmmode~\check{S}\else \v{S}\fi{}trkalj}\ \emph
  {et~al.}(2021)\citenamefont {\ifmmode~\check{S}\else \v{S}\fi{}trkalj},
  \citenamefont {Doggen}, \citenamefont {Gornyi},\ and\ \citenamefont
  {Zilberberg}}]{PhysRevResearch.3.033257}%
  \BibitemOpen
  \bibfield  {author} {\bibinfo {author} {\bibfnamefont {A.}~\bibnamefont
  {\ifmmode~\check{S}\else \v{S}\fi{}trkalj}}, \bibinfo {author} {\bibfnamefont
  {E.~V.~H.}\ \bibnamefont {Doggen}}, \bibinfo {author} {\bibfnamefont {I.~V.}\
  \bibnamefont {Gornyi}},\ and\ \bibinfo {author} {\bibfnamefont
  {O.}~\bibnamefont {Zilberberg}},\ }\bibfield  {title} {\bibinfo {title}
  {{Many-body localization in the interpolating Aubry-Andr\'e-Fibonacci
  model}},\ }\href {https://doi.org/10.1103/PhysRevResearch.3.033257}
  {\bibfield  {journal} {\bibinfo  {journal} {Phys. Rev. Res.}\ }\textbf
  {\bibinfo {volume} {3}},\ \bibinfo {pages} {033257} (\bibinfo {year}
  {2021})}\BibitemShut {NoStop}%
\bibitem [{\citenamefont {Wei}(2025)}]{PhysRevB.111.134204}%
  \BibitemOpen
  \bibfield  {author} {\bibinfo {author} {\bibfnamefont {X.}~\bibnamefont
  {Wei}},\ }\bibfield  {title} {\bibinfo {title} {{Reentrant localization
  transition induced by a composite potential}},\ }\href
  {https://doi.org/10.1103/PhysRevB.111.134204} {\bibfield  {journal} {\bibinfo
   {journal} {Phys. Rev. B}\ }\textbf {\bibinfo {volume} {111}},\ \bibinfo
  {pages} {134204} (\bibinfo {year} {2025})}\BibitemShut {NoStop}%
\bibitem [{\citenamefont {Xu}\ \emph {et~al.}(2024)\citenamefont {Xu},
  \citenamefont {Gao}, \citenamefont {Iovan}, \citenamefont {Khaymovich},
  \citenamefont {Zwiller},\ and\ \citenamefont {Elshaari}}]{xu2024observation}%
  \BibitemOpen
  \bibfield  {author} {\bibinfo {author} {\bibfnamefont {Z.-S.}\ \bibnamefont
  {Xu}}, \bibinfo {author} {\bibfnamefont {J.}~\bibnamefont {Gao}}, \bibinfo
  {author} {\bibfnamefont {A.}~\bibnamefont {Iovan}}, \bibinfo {author}
  {\bibfnamefont {I.~M.}\ \bibnamefont {Khaymovich}}, \bibinfo {author}
  {\bibfnamefont {V.}~\bibnamefont {Zwiller}},\ and\ \bibinfo {author}
  {\bibfnamefont {A.~W.}\ \bibnamefont {Elshaari}},\ }\bibfield  {title}
  {\bibinfo {title} {{Observation of reentrant metal-insulator transition in a
  random-dimer disordered SSH lattice}},\ }\href@noop {} {\bibfield  {journal}
  {\bibinfo  {journal} {npj Nanophotonics}\ }\textbf {\bibinfo {volume} {1}},\
  \bibinfo {pages} {8} (\bibinfo {year} {2024})}\BibitemShut {NoStop}%
\bibitem [{\citenamefont {Vaidya}\ \emph {et~al.}(2023)\citenamefont {Vaidya},
  \citenamefont {J\"org}, \citenamefont {Linn}, \citenamefont {Goh},\ and\
  \citenamefont {Rechtsman}}]{PhysRevResearch.5.033170}%
  \BibitemOpen
  \bibfield  {author} {\bibinfo {author} {\bibfnamefont {S.}~\bibnamefont
  {Vaidya}}, \bibinfo {author} {\bibfnamefont {C.}~\bibnamefont {J\"org}},
  \bibinfo {author} {\bibfnamefont {K.}~\bibnamefont {Linn}}, \bibinfo {author}
  {\bibfnamefont {M.}~\bibnamefont {Goh}},\ and\ \bibinfo {author}
  {\bibfnamefont {M.~C.}\ \bibnamefont {Rechtsman}},\ }\bibfield  {title}
  {\bibinfo {title} {{Reentrant delocalization transition in one-dimensional
  photonic quasicrystals}},\ }\href
  {https://doi.org/10.1103/PhysRevResearch.5.033170} {\bibfield  {journal}
  {\bibinfo  {journal} {Phys. Rev. Res.}\ }\textbf {\bibinfo {volume} {5}},\
  \bibinfo {pages} {033170} (\bibinfo {year} {2023})}\BibitemShut {NoStop}%
\bibitem [{\citenamefont {Guo}\ \emph {et~al.}(2021)\citenamefont {Guo},
  \citenamefont {Cheng}, \citenamefont {Sun}, \citenamefont {Song},
  \citenamefont {Li}, \citenamefont {Wang}, \citenamefont {Ren}, \citenamefont
  {Dong}, \citenamefont {Zheng}, \citenamefont {Zhang} \emph
  {et~al.}}]{guo2021observation}%
  \BibitemOpen
  \bibfield  {author} {\bibinfo {author} {\bibfnamefont {Q.}~\bibnamefont
  {Guo}}, \bibinfo {author} {\bibfnamefont {C.}~\bibnamefont {Cheng}}, \bibinfo
  {author} {\bibfnamefont {Z.-H.}\ \bibnamefont {Sun}}, \bibinfo {author}
  {\bibfnamefont {Z.}~\bibnamefont {Song}}, \bibinfo {author} {\bibfnamefont
  {H.}~\bibnamefont {Li}}, \bibinfo {author} {\bibfnamefont {Z.}~\bibnamefont
  {Wang}}, \bibinfo {author} {\bibfnamefont {W.}~\bibnamefont {Ren}}, \bibinfo
  {author} {\bibfnamefont {H.}~\bibnamefont {Dong}}, \bibinfo {author}
  {\bibfnamefont {D.}~\bibnamefont {Zheng}}, \bibinfo {author} {\bibfnamefont
  {Y.-R.}\ \bibnamefont {Zhang}}, \emph {et~al.},\ }\bibfield  {title}
  {\bibinfo {title} {{Observation of energy-resolved many-body localization}},\
  }\href@noop {} {\bibfield  {journal} {\bibinfo  {journal} {Nature Physics}\
  }\textbf {\bibinfo {volume} {17}},\ \bibinfo {pages} {234} (\bibinfo {year}
  {2021})}\BibitemShut {NoStop}%
\bibitem [{\citenamefont {Shi}\ \emph {et~al.}(2024)\citenamefont {Shi},
  \citenamefont {Sun}, \citenamefont {Wang}, \citenamefont {Wang},
  \citenamefont {Zhang}, \citenamefont {Ma}, \citenamefont {Liu}, \citenamefont
  {Zhao}, \citenamefont {Song}, \citenamefont {Liang} \emph
  {et~al.}}]{shi2024probing}%
  \BibitemOpen
  \bibfield  {author} {\bibinfo {author} {\bibfnamefont {Y.-H.}\ \bibnamefont
  {Shi}}, \bibinfo {author} {\bibfnamefont {Z.-H.}\ \bibnamefont {Sun}},
  \bibinfo {author} {\bibfnamefont {Y.-Y.}\ \bibnamefont {Wang}}, \bibinfo
  {author} {\bibfnamefont {Z.-A.}\ \bibnamefont {Wang}}, \bibinfo {author}
  {\bibfnamefont {Y.-R.}\ \bibnamefont {Zhang}}, \bibinfo {author}
  {\bibfnamefont {W.-G.}\ \bibnamefont {Ma}}, \bibinfo {author} {\bibfnamefont
  {H.-T.}\ \bibnamefont {Liu}}, \bibinfo {author} {\bibfnamefont
  {K.}~\bibnamefont {Zhao}}, \bibinfo {author} {\bibfnamefont {J.-C.}\
  \bibnamefont {Song}}, \bibinfo {author} {\bibfnamefont {G.-H.}\ \bibnamefont
  {Liang}}, \emph {et~al.},\ }\bibfield  {title} {\bibinfo {title} {{Probing
  spin hydrodynamics on a superconducting quantum simulator}},\ }\href@noop {}
  {\bibfield  {journal} {\bibinfo  {journal} {Nature Communications}\ }\textbf
  {\bibinfo {volume} {15}},\ \bibinfo {pages} {7573} (\bibinfo {year}
  {2024})}\BibitemShut {NoStop}%
\bibitem [{\citenamefont {Li}\ \emph {et~al.}(2024)\citenamefont {Li},
  \citenamefont {Xu}, \citenamefont {Wang}, \citenamefont {Tang}, \citenamefont
  {Zhang}, \citenamefont {Yang}, \citenamefont {Su}, \citenamefont {Wang},
  \citenamefont {Mi}, \citenamefont {Sun}, \citenamefont {Liang}, \citenamefont
  {Chen}, \citenamefont {Li}, \citenamefont {Zhang}, \citenamefont {Linghu},
  \citenamefont {Han}, \citenamefont {Liu}, \citenamefont {Feng}, \citenamefont
  {Liu}, \citenamefont {Xue}, \citenamefont {Zhang}, \citenamefont {Jin},
  \citenamefont {Zhu}, \citenamefont {Yu}, \citenamefont {Zhao},\ and\
  \citenamefont {Xue}}]{PhysRevResearch.6.L042038}%
  \BibitemOpen
  \bibfield  {author} {\bibinfo {author} {\bibfnamefont {X.}~\bibnamefont
  {Li}}, \bibinfo {author} {\bibfnamefont {H.}~\bibnamefont {Xu}}, \bibinfo
  {author} {\bibfnamefont {J.}~\bibnamefont {Wang}}, \bibinfo {author}
  {\bibfnamefont {L.-Z.}\ \bibnamefont {Tang}}, \bibinfo {author}
  {\bibfnamefont {D.-W.}\ \bibnamefont {Zhang}}, \bibinfo {author}
  {\bibfnamefont {C.}~\bibnamefont {Yang}}, \bibinfo {author} {\bibfnamefont
  {T.}~\bibnamefont {Su}}, \bibinfo {author} {\bibfnamefont {C.}~\bibnamefont
  {Wang}}, \bibinfo {author} {\bibfnamefont {Z.}~\bibnamefont {Mi}}, \bibinfo
  {author} {\bibfnamefont {W.}~\bibnamefont {Sun}}, \bibinfo {author}
  {\bibfnamefont {X.}~\bibnamefont {Liang}}, \bibinfo {author} {\bibfnamefont
  {M.}~\bibnamefont {Chen}}, \bibinfo {author} {\bibfnamefont {C.}~\bibnamefont
  {Li}}, \bibinfo {author} {\bibfnamefont {Y.}~\bibnamefont {Zhang}}, \bibinfo
  {author} {\bibfnamefont {K.}~\bibnamefont {Linghu}}, \bibinfo {author}
  {\bibfnamefont {J.}~\bibnamefont {Han}}, \bibinfo {author} {\bibfnamefont
  {W.}~\bibnamefont {Liu}}, \bibinfo {author} {\bibfnamefont {Y.}~\bibnamefont
  {Feng}}, \bibinfo {author} {\bibfnamefont {P.}~\bibnamefont {Liu}}, \bibinfo
  {author} {\bibfnamefont {G.}~\bibnamefont {Xue}}, \bibinfo {author}
  {\bibfnamefont {J.}~\bibnamefont {Zhang}}, \bibinfo {author} {\bibfnamefont
  {Y.}~\bibnamefont {Jin}}, \bibinfo {author} {\bibfnamefont {S.-L.}\
  \bibnamefont {Zhu}}, \bibinfo {author} {\bibfnamefont {H.}~\bibnamefont
  {Yu}}, \bibinfo {author} {\bibfnamefont {S.~P.}\ \bibnamefont {Zhao}},\ and\
  \bibinfo {author} {\bibfnamefont {Q.-K.}\ \bibnamefont {Xue}},\ }\bibfield
  {title} {\bibinfo {title} {{Mapping the topology-localization phase diagram
  with quasiperiodic disorder using a programmable superconducting
  simulator}},\ }\href {https://doi.org/10.1103/PhysRevResearch.6.L042038}
  {\bibfield  {journal} {\bibinfo  {journal} {Phys. Rev. Res.}\ }\textbf
  {\bibinfo {volume} {6}},\ \bibinfo {pages} {L042038} (\bibinfo {year}
  {2024})}\BibitemShut {NoStop}%
\bibitem [{\citenamefont {Wang}\ \emph
  {et~al.}(2020{\natexlab{b}})\citenamefont {Wang}, \citenamefont {Zhang},
  \citenamefont {Niu}, \citenamefont {Yu},\ and\ \citenamefont
  {Liu}}]{PhysRevLett.125.073204}%
  \BibitemOpen
  \bibfield  {author} {\bibinfo {author} {\bibfnamefont {Y.}~\bibnamefont
  {Wang}}, \bibinfo {author} {\bibfnamefont {L.}~\bibnamefont {Zhang}},
  \bibinfo {author} {\bibfnamefont {S.}~\bibnamefont {Niu}}, \bibinfo {author}
  {\bibfnamefont {D.}~\bibnamefont {Yu}},\ and\ \bibinfo {author}
  {\bibfnamefont {X.-J.}\ \bibnamefont {Liu}},\ }\bibfield  {title} {\bibinfo
  {title} {{Realization and Detection of Nonergodic Critical Phases in an
  Optical Raman Lattice}},\ }\href
  {https://doi.org/10.1103/PhysRevLett.125.073204} {\bibfield  {journal}
  {\bibinfo  {journal} {Phys. Rev. Lett.}\ }\textbf {\bibinfo {volume} {125}},\
  \bibinfo {pages} {073204} (\bibinfo {year} {2020}{\natexlab{b}})}\BibitemShut
  {NoStop}%
\bibitem [{\citenamefont {Wu}\ \emph {et~al.}(2016)\citenamefont {Wu},
  \citenamefont {Zhang}, \citenamefont {Sun}, \citenamefont {Xu}, \citenamefont
  {Wang}, \citenamefont {Ji}, \citenamefont {Deng}, \citenamefont {Chen},
  \citenamefont {Liu},\ and\ \citenamefont {Pan}}]{wu2016realization}%
  \BibitemOpen
  \bibfield  {author} {\bibinfo {author} {\bibfnamefont {Z.}~\bibnamefont
  {Wu}}, \bibinfo {author} {\bibfnamefont {L.}~\bibnamefont {Zhang}}, \bibinfo
  {author} {\bibfnamefont {W.}~\bibnamefont {Sun}}, \bibinfo {author}
  {\bibfnamefont {X.-T.}\ \bibnamefont {Xu}}, \bibinfo {author} {\bibfnamefont
  {B.-Z.}\ \bibnamefont {Wang}}, \bibinfo {author} {\bibfnamefont {S.-C.}\
  \bibnamefont {Ji}}, \bibinfo {author} {\bibfnamefont {Y.}~\bibnamefont
  {Deng}}, \bibinfo {author} {\bibfnamefont {S.}~\bibnamefont {Chen}}, \bibinfo
  {author} {\bibfnamefont {X.-J.}\ \bibnamefont {Liu}},\ and\ \bibinfo {author}
  {\bibfnamefont {J.-W.}\ \bibnamefont {Pan}},\ }\bibfield  {title} {\bibinfo
  {title} {{Realization of two-dimensional spin-orbit coupling for
  Bose-Einstein condensates}},\ }\href@noop {} {\bibfield  {journal} {\bibinfo
  {journal} {Science}\ }\textbf {\bibinfo {volume} {354}},\ \bibinfo {pages}
  {83} (\bibinfo {year} {2016})}\BibitemShut {NoStop}%
\bibitem [{\citenamefont {Song}\ \emph {et~al.}(2019)\citenamefont {Song},
  \citenamefont {He}, \citenamefont {Niu}, \citenamefont {Zhang}, \citenamefont
  {Ren}, \citenamefont {Liu},\ and\ \citenamefont {Jo}}]{song2019observation}%
  \BibitemOpen
  \bibfield  {author} {\bibinfo {author} {\bibfnamefont {B.}~\bibnamefont
  {Song}}, \bibinfo {author} {\bibfnamefont {C.}~\bibnamefont {He}}, \bibinfo
  {author} {\bibfnamefont {S.}~\bibnamefont {Niu}}, \bibinfo {author}
  {\bibfnamefont {L.}~\bibnamefont {Zhang}}, \bibinfo {author} {\bibfnamefont
  {Z.}~\bibnamefont {Ren}}, \bibinfo {author} {\bibfnamefont {X.-J.}\
  \bibnamefont {Liu}},\ and\ \bibinfo {author} {\bibfnamefont {G.-B.}\
  \bibnamefont {Jo}},\ }\bibfield  {title} {\bibinfo {title} {{Observation of
  nodal-line semimetal with ultracold fermions in an optical lattice}},\
  }\href@noop {} {\bibfield  {journal} {\bibinfo  {journal} {Nature Physics}\
  }\textbf {\bibinfo {volume} {15}},\ \bibinfo {pages} {911} (\bibinfo {year}
  {2019})}\BibitemShut {NoStop}%
\bibitem [{\citenamefont {Liu}\ \emph {et~al.}(2013)\citenamefont {Liu},
  \citenamefont {Liu},\ and\ \citenamefont {Cheng}}]{PhysRevLett.110.076401}%
  \BibitemOpen
  \bibfield  {author} {\bibinfo {author} {\bibfnamefont {X.-J.}\ \bibnamefont
  {Liu}}, \bibinfo {author} {\bibfnamefont {Z.-X.}\ \bibnamefont {Liu}},\ and\
  \bibinfo {author} {\bibfnamefont {M.}~\bibnamefont {Cheng}},\ }\bibfield
  {title} {\bibinfo {title} {{Manipulating Topological Edge Spins in a
  One-Dimensional Optical Lattice}},\ }\href
  {https://doi.org/10.1103/PhysRevLett.110.076401} {\bibfield  {journal}
  {\bibinfo  {journal} {Phys. Rev. Lett.}\ }\textbf {\bibinfo {volume} {110}},\
  \bibinfo {pages} {076401} (\bibinfo {year} {2013})}\BibitemShut {NoStop}%
\bibitem [{\citenamefont {Kohmoto}(1983)}]{PhysRevLett.51.1198}%
  \BibitemOpen
  \bibfield  {author} {\bibinfo {author} {\bibfnamefont {M.}~\bibnamefont
  {Kohmoto}},\ }\bibfield  {title} {\bibinfo {title} {{Metal-Insulator
  Transition and Scaling for Incommensurate Systems}},\ }\href
  {https://doi.org/10.1103/PhysRevLett.51.1198} {\bibfield  {journal} {\bibinfo
   {journal} {Phys. Rev. Lett.}\ }\textbf {\bibinfo {volume} {51}},\ \bibinfo
  {pages} {1198} (\bibinfo {year} {1983})}\BibitemShut {NoStop}%
\bibitem [{\citenamefont {Zhao}\ \emph {et~al.}(2023)\citenamefont {Zhao},
  \citenamefont {Zhao}, \citenamefont {Wang}, \citenamefont {Li},\ and\
  \citenamefont {Bai}}]{PhysRevB.108.054204}%
  \BibitemOpen
  \bibfield  {author} {\bibinfo {author} {\bibfnamefont {J.}~\bibnamefont
  {Zhao}}, \bibinfo {author} {\bibfnamefont {Y.}~\bibnamefont {Zhao}}, \bibinfo
  {author} {\bibfnamefont {J.-G.}\ \bibnamefont {Wang}}, \bibinfo {author}
  {\bibfnamefont {Y.}~\bibnamefont {Li}},\ and\ \bibinfo {author}
  {\bibfnamefont {X.-D.}\ \bibnamefont {Bai}},\ }\bibfield  {title} {\bibinfo
  {title} {{Phase transition of a non-Abelian quasiperiodic mosaic lattice
  model with $p$-wave superfluidity}},\ }\href
  {https://doi.org/10.1103/PhysRevB.108.054204} {\bibfield  {journal} {\bibinfo
   {journal} {Phys. Rev. B}\ }\textbf {\bibinfo {volume} {108}},\ \bibinfo
  {pages} {054204} (\bibinfo {year} {2023})}\BibitemShut {NoStop}%
\bibitem [{\citenamefont {Aubry}\ and\ \citenamefont
  {Andr{\'e}}(1980)}]{aubry1980}%
  \BibitemOpen
  \bibfield  {author} {\bibinfo {author} {\bibfnamefont {S.}~\bibnamefont
  {Aubry}}\ and\ \bibinfo {author} {\bibfnamefont {G.}~\bibnamefont
  {Andr{\'e}}},\ }\bibfield  {title} {\bibinfo {title} {{Analyticity breaking
  and Anderson localization in incommensurate lattices}},\ }\href@noop {}
  {\bibfield  {journal} {\bibinfo  {journal} {Ann. Isr. Phys. Soc}\ }\textbf
  {\bibinfo {volume} {3}},\ \bibinfo {pages} {18} (\bibinfo {year}
  {1980})}\BibitemShut {NoStop}%
\bibitem [{\citenamefont {Wang}\ \emph {et~al.}(2016)\citenamefont {Wang},
  \citenamefont {Liu}, \citenamefont {Xianlong},\ and\ \citenamefont
  {Hu}}]{PhysRevB.93.104504}%
  \BibitemOpen
  \bibfield  {author} {\bibinfo {author} {\bibfnamefont {J.}~\bibnamefont
  {Wang}}, \bibinfo {author} {\bibfnamefont {X.-J.}\ \bibnamefont {Liu}},
  \bibinfo {author} {\bibfnamefont {G.}~\bibnamefont {Xianlong}},\ and\
  \bibinfo {author} {\bibfnamefont {H.}~\bibnamefont {Hu}},\ }\bibfield
  {title} {\bibinfo {title} {{Phase diagram of a non-Abelian
  Aubry-Andr\'e-Harper model with $p$-wave superfluidity}},\ }\href
  {https://doi.org/10.1103/PhysRevB.93.104504} {\bibfield  {journal} {\bibinfo
  {journal} {Phys. Rev. B}\ }\textbf {\bibinfo {volume} {93}},\ \bibinfo
  {pages} {104504} (\bibinfo {year} {2016})}\BibitemShut {NoStop}%
\bibitem [{Sup()}]{SuppMateria}%
  \BibitemOpen
  \href@noop {} {}\bibinfo {note} {See the Supplemental Material for details on
  (i) Matrix form of the Hamiltonian, (ii) In the limiting cases, (iii)
  Destruction of the re-emerged extended states, (iv) The transition in the
  case of $W/t=0$, (v) Spectrum for $\Delta/t=1$ and typical states (vi)
  $\Delta$ induced transition for large $W$, (vii)The phase diagram for
  $V/t=2.5$, (viii)Multifractal analysis of different states, (ix)Reentrant
  delocalization transitions}\BibitemShut {NoStop}%
\bibitem [{\citenamefont {Yahyavi}\ \emph {et~al.}(2019)\citenamefont
  {Yahyavi}, \citenamefont {Het\'enyi},\ and\ \citenamefont
  {Tanatar}}]{PhysRevB.100.064202}%
  \BibitemOpen
  \bibfield  {author} {\bibinfo {author} {\bibfnamefont {M.}~\bibnamefont
  {Yahyavi}}, \bibinfo {author} {\bibfnamefont {B.}~\bibnamefont {Het\'enyi}},\
  and\ \bibinfo {author} {\bibfnamefont {B.}~\bibnamefont {Tanatar}},\
  }\bibfield  {title} {\bibinfo {title} {{Generalized Aubry-Andr\'e-Harper
  model with modulated hopping and $p$-wave pairing}},\ }\href
  {https://doi.org/10.1103/PhysRevB.100.064202} {\bibfield  {journal} {\bibinfo
   {journal} {Phys. Rev. B}\ }\textbf {\bibinfo {volume} {100}},\ \bibinfo
  {pages} {064202} (\bibinfo {year} {2019})}\BibitemShut {NoStop}%
\bibitem [{\citenamefont {Zeng}\ \emph {et~al.}(2016)\citenamefont {Zeng},
  \citenamefont {Chen},\ and\ \citenamefont {L\"u}}]{PhysRevB.94.125408}%
  \BibitemOpen
  \bibfield  {author} {\bibinfo {author} {\bibfnamefont {Q.-B.}\ \bibnamefont
  {Zeng}}, \bibinfo {author} {\bibfnamefont {S.}~\bibnamefont {Chen}},\ and\
  \bibinfo {author} {\bibfnamefont {R.}~\bibnamefont {L\"u}},\ }\bibfield
  {title} {\bibinfo {title} {{Generalized Aubry-Andr\'e-Harper model with
  $p$-wave superconducting pairing}},\ }\href
  {https://doi.org/10.1103/PhysRevB.94.125408} {\bibfield  {journal} {\bibinfo
  {journal} {Phys. Rev. B}\ }\textbf {\bibinfo {volume} {94}},\ \bibinfo
  {pages} {125408} (\bibinfo {year} {2016})}\BibitemShut {NoStop}%
\bibitem [{\citenamefont {Li}\ \emph {et~al.}(2017)\citenamefont {Li},
  \citenamefont {Li},\ and\ \citenamefont {Das~Sarma}}]{PhysRevB.96.085119}%
  \BibitemOpen
  \bibfield  {author} {\bibinfo {author} {\bibfnamefont {X.}~\bibnamefont
  {Li}}, \bibinfo {author} {\bibfnamefont {X.}~\bibnamefont {Li}},\ and\
  \bibinfo {author} {\bibfnamefont {S.}~\bibnamefont {Das~Sarma}},\ }\bibfield
  {title} {\bibinfo {title} {{Mobility edges in one-dimensional bichromatic
  incommensurate potentials}},\ }\href
  {https://doi.org/10.1103/PhysRevB.96.085119} {\bibfield  {journal} {\bibinfo
  {journal} {Phys. Rev. B}\ }\textbf {\bibinfo {volume} {96}},\ \bibinfo
  {pages} {085119} (\bibinfo {year} {2017})}\BibitemShut {NoStop}%
\bibitem [{\citenamefont {Ghosh}\ \emph {et~al.}(2025)\citenamefont {Ghosh},
  \citenamefont {Sarkar},\ and\ \citenamefont {Khaymovich}}]{k957-fcmh}%
  \BibitemOpen
  \bibfield  {author} {\bibinfo {author} {\bibfnamefont {R.}~\bibnamefont
  {Ghosh}}, \bibinfo {author} {\bibfnamefont {M.}~\bibnamefont {Sarkar}},\ and\
  \bibinfo {author} {\bibfnamefont {I.~M.}\ \bibnamefont {Khaymovich}},\
  }\bibfield  {title} {\bibinfo {title} {{Reentrant localization induced by
  short-range hopping in the fractal Rosenzweig-Porter model}},\ }\href
  {https://doi.org/10.1103/k957-fcmh} {\bibfield  {journal} {\bibinfo
  {journal} {Phys. Rev. B}\ }\textbf {\bibinfo {volume} {111}},\ \bibinfo
  {pages} {L220102} (\bibinfo {year} {2025})}\BibitemShut {NoStop}%
\bibitem [{\citenamefont {Luitz}\ \emph {et~al.}(2014)\citenamefont {Luitz},
  \citenamefont {Alet},\ and\ \citenamefont
  {Laflorencie}}]{PhysRevLett.112.057203}%
  \BibitemOpen
  \bibfield  {author} {\bibinfo {author} {\bibfnamefont {D.~J.}\ \bibnamefont
  {Luitz}}, \bibinfo {author} {\bibfnamefont {F.}~\bibnamefont {Alet}},\ and\
  \bibinfo {author} {\bibfnamefont {N.}~\bibnamefont {Laflorencie}},\
  }\bibfield  {title} {\bibinfo {title} {{Universal Behavior beyond
  Multifractality in Quantum Many-Body Systems}},\ }\href
  {https://doi.org/10.1103/PhysRevLett.112.057203} {\bibfield  {journal}
  {\bibinfo  {journal} {Phys. Rev. Lett.}\ }\textbf {\bibinfo {volume} {112}},\
  \bibinfo {pages} {057203} (\bibinfo {year} {2014})}\BibitemShut {NoStop}%
\bibitem [{\citenamefont {De~Tomasi}\ and\ \citenamefont
  {Khaymovich}(2020)}]{PhysRevLett.124.200602}%
  \BibitemOpen
  \bibfield  {author} {\bibinfo {author} {\bibfnamefont {G.}~\bibnamefont
  {De~Tomasi}}\ and\ \bibinfo {author} {\bibfnamefont {I.~M.}\ \bibnamefont
  {Khaymovich}},\ }\bibfield  {title} {\bibinfo {title} {{Multifractality Meets
  Entanglement: Relation for Nonergodic Extended States}},\ }\href
  {https://doi.org/10.1103/PhysRevLett.124.200602} {\bibfield  {journal}
  {\bibinfo  {journal} {Phys. Rev. Lett.}\ }\textbf {\bibinfo {volume} {124}},\
  \bibinfo {pages} {200602} (\bibinfo {year} {2020})}\BibitemShut {NoStop}%
\bibitem [{\citenamefont {Ji}\ and\ \citenamefont
  {Xu}(2025)}]{ji2025fibonacci}%
  \BibitemOpen
  \bibfield  {author} {\bibinfo {author} {\bibfnamefont {R.}~\bibnamefont
  {Ji}}\ and\ \bibinfo {author} {\bibfnamefont {Z.}~\bibnamefont {Xu}},\
  }\bibfield  {title} {\bibinfo {title} {{Fibonacci-modulation-induced multiple
  topological Anderson insulators}},\ }\href@noop {} {\bibfield  {journal}
  {\bibinfo  {journal} {Communications Physics}\ }\textbf {\bibinfo {volume}
  {8}},\ \bibinfo {pages} {336} (\bibinfo {year} {2025})}\BibitemShut {NoStop}%
\bibitem [{\citenamefont {Zhang}\ \emph {et~al.}(2025)\citenamefont {Zhang},
  \citenamefont {Tang}, \citenamefont {Quezada}, \citenamefont {Dong},\ and\
  \citenamefont {Zhang}}]{zhang2025reentrant}%
  \BibitemOpen
  \bibfield  {author} {\bibinfo {author} {\bibfnamefont {G.-Q.}\ \bibnamefont
  {Zhang}}, \bibinfo {author} {\bibfnamefont {L.-Z.}\ \bibnamefont {Tang}},
  \bibinfo {author} {\bibfnamefont {L.}~\bibnamefont {Quezada}}, \bibinfo
  {author} {\bibfnamefont {S.-H.}\ \bibnamefont {Dong}},\ and\ \bibinfo
  {author} {\bibfnamefont {D.-W.}\ \bibnamefont {Zhang}},\ }\bibfield  {title}
  {\bibinfo {title} {{Reentrant topological phases and spin density wave
  induced by 1D moir{\'e} potentials}},\ }\href@noop {} {\bibfield  {journal}
  {\bibinfo  {journal} {Communications Physics}\ }\textbf {\bibinfo {volume}
  {8}},\ \bibinfo {pages} {275} (\bibinfo {year} {2025})}\BibitemShut {NoStop}%
\bibitem [{\citenamefont {Li}\ and\ \citenamefont
  {Das~Sarma}(2020)}]{PhysRevB.101.064203}%
  \BibitemOpen
  \bibfield  {author} {\bibinfo {author} {\bibfnamefont {X.}~\bibnamefont
  {Li}}\ and\ \bibinfo {author} {\bibfnamefont {S.}~\bibnamefont {Das~Sarma}},\
  }\bibfield  {title} {\bibinfo {title} {{Mobility edge and intermediate phase
  in one-dimensional incommensurate lattice potentials}},\ }\href
  {https://doi.org/10.1103/PhysRevB.101.064203} {\bibfield  {journal} {\bibinfo
   {journal} {Phys. Rev. B}\ }\textbf {\bibinfo {volume} {101}},\ \bibinfo
  {pages} {064203} (\bibinfo {year} {2020})}\BibitemShut {NoStop}%
\bibitem [{\citenamefont {Qi}\ \emph {et~al.}(2023)\citenamefont {Qi},
  \citenamefont {Cao},\ and\ \citenamefont {Jiang}}]{PhysRevB.107.224201}%
  \BibitemOpen
  \bibfield  {author} {\bibinfo {author} {\bibfnamefont {R.}~\bibnamefont
  {Qi}}, \bibinfo {author} {\bibfnamefont {J.}~\bibnamefont {Cao}},\ and\
  \bibinfo {author} {\bibfnamefont {X.-P.}\ \bibnamefont {Jiang}},\ }\bibfield
  {title} {\bibinfo {title} {{Multiple localization transitions and novel
  quantum phases induced by a staggered on-site potential}},\ }\href
  {https://doi.org/10.1103/PhysRevB.107.224201} {\bibfield  {journal} {\bibinfo
   {journal} {Phys. Rev. B}\ }\textbf {\bibinfo {volume} {107}},\ \bibinfo
  {pages} {224201} (\bibinfo {year} {2023})}\BibitemShut {NoStop}%
\bibitem [{\citenamefont {Guan}\ \emph {et~al.}(2023)\citenamefont {Guan},
  \citenamefont {Wang}, \citenamefont {Guan},\ and\ \citenamefont
  {Cai}}]{PhysRevA.108.033305}%
  \BibitemOpen
  \bibfield  {author} {\bibinfo {author} {\bibfnamefont {E.}~\bibnamefont
  {Guan}}, \bibinfo {author} {\bibfnamefont {G.}~\bibnamefont {Wang}}, \bibinfo
  {author} {\bibfnamefont {X.-W.}\ \bibnamefont {Guan}},\ and\ \bibinfo
  {author} {\bibfnamefont {X.}~\bibnamefont {Cai}},\ }\bibfield  {title}
  {\bibinfo {title} {{Reentrant localization and mobility edges in a spinful
  Aubry-Andr\'e-Harper model with a non-Abelian potential}},\ }\href
  {https://doi.org/10.1103/PhysRevA.108.033305} {\bibfield  {journal} {\bibinfo
   {journal} {Phys. Rev. A}\ }\textbf {\bibinfo {volume} {108}},\ \bibinfo
  {pages} {033305} (\bibinfo {year} {2023})}\BibitemShut {NoStop}%
\bibitem [{\citenamefont {Lv}\ \emph {et~al.}(2022)\citenamefont {Lv},
  \citenamefont {Yi}, \citenamefont {Li}, \citenamefont {Sun},\ and\
  \citenamefont {You}}]{PhysRevA.105.013315}%
  \BibitemOpen
  \bibfield  {author} {\bibinfo {author} {\bibfnamefont {T.}~\bibnamefont
  {Lv}}, \bibinfo {author} {\bibfnamefont {T.-C.}\ \bibnamefont {Yi}}, \bibinfo
  {author} {\bibfnamefont {L.}~\bibnamefont {Li}}, \bibinfo {author}
  {\bibfnamefont {G.}~\bibnamefont {Sun}},\ and\ \bibinfo {author}
  {\bibfnamefont {W.-L.}\ \bibnamefont {You}},\ }\bibfield  {title} {\bibinfo
  {title} {{Quantum criticality and universality in the $p$-wave-paired
  Aubry-Andr\'e-Harper model}},\ }\href
  {https://doi.org/10.1103/PhysRevA.105.013315} {\bibfield  {journal} {\bibinfo
   {journal} {Phys. Rev. A}\ }\textbf {\bibinfo {volume} {105}},\ \bibinfo
  {pages} {013315} (\bibinfo {year} {2022})}\BibitemShut {NoStop}%
\bibitem [{\citenamefont {Xingbo~Wei}\ and\ \citenamefont
  {Zhang}(2025)}]{data}%
  \BibitemOpen
  \bibfield  {author} {\bibinfo {author} {\bibfnamefont {T.~L. G.~X.}\
  \bibnamefont {Xingbo~Wei}, \bibfnamefont {Zhentian~Xie}}\ and\ \bibinfo
  {author} {\bibfnamefont {Y.}~\bibnamefont {Zhang}},\ }\bibfield  {title}
  {\bibinfo {title} {{Distinct reentrant transitions in a quasi-periodic Raman
  lattice}}\ }\href {https://doi.org/10.5281/zenodo.17082777}
  {10.5281/zenodo.17082777} (\bibinfo {year} {2025})\BibitemShut {NoStop}%
\bibitem [{\citenamefont {Pietracaprina}\ \emph {et~al.}(2018)\citenamefont
  {Pietracaprina}, \citenamefont {Mac{\'e}}, \citenamefont {Luitz},\ and\
  \citenamefont {Alet}}]{pietracaprina2018shift}%
  \BibitemOpen
  \bibfield  {author} {\bibinfo {author} {\bibfnamefont {F.}~\bibnamefont
  {Pietracaprina}}, \bibinfo {author} {\bibfnamefont {N.}~\bibnamefont
  {Mac{\'e}}}, \bibinfo {author} {\bibfnamefont {D.~J.}\ \bibnamefont
  {Luitz}},\ and\ \bibinfo {author} {\bibfnamefont {F.}~\bibnamefont {Alet}},\
  }\bibfield  {title} {\bibinfo {title} {{Shift-invert diagonalization of large
  many-body localizing spin chains}},\ }\href@noop {} {\bibfield  {journal}
  {\bibinfo  {journal} {SciPost Physics}\ }\textbf {\bibinfo {volume} {5}},\
  \bibinfo {pages} {045} (\bibinfo {year} {2018})}\BibitemShut {NoStop}%
\bibitem [{\citenamefont {Wei}\ \emph {et~al.}(2019)\citenamefont {Wei},
  \citenamefont {Cheng}, \citenamefont {Xianlong},\ and\ \citenamefont
  {Mondaini}}]{PhysRevB.99.165137}%
  \BibitemOpen
  \bibfield  {author} {\bibinfo {author} {\bibfnamefont {X.}~\bibnamefont
  {Wei}}, \bibinfo {author} {\bibfnamefont {C.}~\bibnamefont {Cheng}}, \bibinfo
  {author} {\bibfnamefont {G.}~\bibnamefont {Xianlong}},\ and\ \bibinfo
  {author} {\bibfnamefont {R.}~\bibnamefont {Mondaini}},\ }\bibfield  {title}
  {\bibinfo {title} {{Investigating many-body mobility edges in isolated
  quantum systems}},\ }\href {https://doi.org/10.1103/PhysRevB.99.165137}
  {\bibfield  {journal} {\bibinfo  {journal} {Phys. Rev. B}\ }\textbf {\bibinfo
  {volume} {99}},\ \bibinfo {pages} {165137} (\bibinfo {year}
  {2019})}\BibitemShut {NoStop}%
\bibitem [{\citenamefont {Tabanelli}\ \emph {et~al.}(2024)\citenamefont
  {Tabanelli}, \citenamefont {Castelnovo},\ and\ \citenamefont
  {\ifmmode~\check{S}\else \v{S}\fi{}trkalj}}]{PhysRevB.110.184208}%
  \BibitemOpen
  \bibfield  {author} {\bibinfo {author} {\bibfnamefont {H.}~\bibnamefont
  {Tabanelli}}, \bibinfo {author} {\bibfnamefont {C.}~\bibnamefont
  {Castelnovo}},\ and\ \bibinfo {author} {\bibfnamefont {A.}~\bibnamefont
  {\ifmmode~\check{S}\else \v{S}\fi{}trkalj}},\ }\bibfield  {title} {\bibinfo
  {title} {{Reentrant localization transitions and anomalous spectral
  properties in off-diagonal quasiperiodic systems}},\ }\href
  {https://doi.org/10.1103/PhysRevB.110.184208} {\bibfield  {journal} {\bibinfo
   {journal} {Phys. Rev. B}\ }\textbf {\bibinfo {volume} {110}},\ \bibinfo
  {pages} {184208} (\bibinfo {year} {2024})}\BibitemShut {NoStop}%
\end{thebibliography}%


\begin{thebibliography}{3}%
	\makeatletter
	\providecommand \@ifxundefined [1]{%
		\@ifx{#1\undefined}
	}%
	\providecommand \@ifnum [1]{%
		\ifnum #1\expandafter \@firstoftwo
		\else \expandafter \@secondoftwo
		\fi
	}%
	\providecommand \@ifx [1]{%
	\ifx #1\expandafter \@firstoftwo
		\else \expandafter \@secondoftwo
		\fi
	}%
	\providecommand \natexlab [1]{#1}%
	\providecommand \enquote  [1]{``#1''}%
	\providecommand \bibnamefont  [1]{#1}%
	\providecommand \bibfnamefont [1]{#1}%
	\providecommand \citenamefont [1]{#1}%
	\providecommand \href@noop [0]{\@secondoftwo}%
	\providecommand \href [0]{\begingroup \@sanitize@url \@href}%
	\providecommand \@href[1]{\@@startlink{#1}\@@href}%
	\providecommand \@@href[1]{\endgroup#1\@@endlink}%
	\providecommand \@sanitize@url [0]{\catcode `\\12\catcode `\$12\catcode
		`\&12\catcode `\#12\catcode `\^12\catcode `\_12\catcode `\%12\relax}%
	\providecommand \@@startlink[1]{}%
	\providecommand \@@endlink[0]{}%
	\providecommand \url  [0]{\begingroup\@sanitize@url \@url }%
	\providecommand \@url [1]{\endgroup\@href {#1}{\urlprefix }}%
	\providecommand \urlprefix  [0]{URL }%
	\providecommand \Eprint [0]{\href }%
	\providecommand \doibase [0]{https://doi.org/}%
	\providecommand \selectlanguage [0]{\@gobble}%
	\providecommand \bibinfo  [0]{\@secondoftwo}%
	\providecommand \bibfield  [0]{\@secondoftwo}%
	\providecommand \translation [1]{[#1]}%
	\providecommand \BibitemOpen [0]{}%
	\providecommand \bibitemStop [0]{}%
	\providecommand \bibitemNoStop [0]{.\EOS\space}%
	\providecommand \EOS [0]{\spacefactor3000\relax}%
	\providecommand \BibitemShut  [1]{\csname bibitem#1\endcsname}%
	\let\auto@bib@innerbib\@empty
	
	\bibitem [{\citenamefont {Wang}\ \emph {et~al.}(2016)\citenamefont {Wang},
	\citenamefont {Liu}, \citenamefont {Xianlong},\ and\ \citenamefont
	{Hu}}]{PhysRevB.93.104504}%
\BibitemOpen
\bibfield  {author} {\bibinfo {author} {\bibfnamefont {J.}~\bibnamefont
		{Wang}}, \bibinfo {author} {\bibfnamefont {X.-J.}\ \bibnamefont {Liu}},
	\bibinfo {author} {\bibfnamefont {G.}~\bibnamefont {Xianlong}},\ and\
	\bibinfo {author} {\bibfnamefont {H.}~\bibnamefont {Hu}},\ }\bibfield
{title} {\bibinfo {title} {{Phase diagram of a non-Abelian
			Aubry-Andr\'e-Harper model with $p$-wave superfluidity}},\ }\href
{https://doi.org/10.1103/PhysRevB.93.104504} {\bibfield  {journal} {\bibinfo
		{journal} {Phys. Rev. B}\ }\textbf {\bibinfo {volume} {93}},\ \bibinfo
	{pages} {104504} (\bibinfo {year} {2016})}\BibitemShut {NoStop}%
	\bibitem [{\citenamefont {Pietracaprina}\ \emph {et~al.}(2018)\citenamefont
		{Pietracaprina}, \citenamefont {Mac{\'e}}, \citenamefont {Luitz},\ and\
		\citenamefont {Alet}}]{pietracaprina2018shift}%
	\BibitemOpen
	\bibfield  {author} {\bibinfo {author} {\bibfnamefont {F.}~\bibnamefont
			{Pietracaprina}}, \bibinfo {author} {\bibfnamefont {N.}~\bibnamefont
			{Mac{\'e}}}, \bibinfo {author} {\bibfnamefont {D.~J.}\ \bibnamefont
			{Luitz}},\ and\ \bibinfo {author} {\bibfnamefont {F.}~\bibnamefont {Alet}},\
	}\bibfield  {title} {\bibinfo {title} {{Shift-invert diagonalization of large
				many-body localizing spin chains}},\ }\href@noop {} {\bibfield  {journal}
		{\bibinfo  {journal} {SciPost Physics}\ }\textbf {\bibinfo {volume} {5}},\
		\bibinfo {pages} {045} (\bibinfo {year} {2018})}\BibitemShut {NoStop}%
	\bibitem [{\citenamefont {Wei}\ \emph {et~al.}(2019)\citenamefont {Wei},
		\citenamefont {Cheng}, \citenamefont {Xianlong},\ and\ \citenamefont
		{Mondaini}}]{PhysRevB.99.165137}%
	\BibitemOpen
	\bibfield  {author} {\bibinfo {author} {\bibfnamefont {X.}~\bibnamefont
			{Wei}}, \bibinfo {author} {\bibfnamefont {C.}~\bibnamefont {Cheng}}, \bibinfo
		{author} {\bibfnamefont {G.}~\bibnamefont {Xianlong}},\ and\ \bibinfo
		{author} {\bibfnamefont {R.}~\bibnamefont {Mondaini}},\ }\bibfield  {title}
	{\bibinfo {title} {{Investigating many-body mobility edges in isolated
				quantum systems}},\ }\href {https://doi.org/10.1103/PhysRevB.99.165137}
	{\bibfield  {journal} {\bibinfo  {journal} {Phys. Rev. B}\ }\textbf {\bibinfo
			{volume} {99}},\ \bibinfo {pages} {165137} (\bibinfo {year}
		{2019})}\BibitemShut {NoStop}%
	\bibitem [{\citenamefont {Zhou}\ \emph {et~al.}(2025)\citenamefont {Zhou},
		\citenamefont {Yao}, \citenamefont {Wang}, \citenamefont {Wang},
		\citenamefont {Wei}, \citenamefont {Zhou},\ and\ \citenamefont
		{Liu}}]{zhou2025fundamentallocalizationphasesquasiperiodic}%
	\BibitemOpen
	\bibfield  {author} {\bibinfo {author} {\bibfnamefont {X.-C.}\ \bibnamefont
			{Zhou}}, \bibinfo {author} {\bibfnamefont {B.-C.}\ \bibnamefont {Yao}},
		\bibinfo {author} {\bibfnamefont {Y.}~\bibnamefont {Wang}}, \bibinfo {author}
		{\bibfnamefont {Y.}~\bibnamefont {Wang}}, \bibinfo {author} {\bibfnamefont
			{Y.}~\bibnamefont {Wei}}, \bibinfo {author} {\bibfnamefont {Q.}~\bibnamefont
			{Zhou}},\ and\ \bibinfo {author} {\bibfnamefont {X.-J.}\ \bibnamefont
			{Liu}},\ }\href {https://arxiv.org/abs/2503.24380} {\bibinfo {title} {The
			fundamental localization phases in quasiperiodic systems: A unified framework
			and exact results}} (\bibinfo {year} {2025}),\ \Eprint
	{https://arxiv.org/abs/2503.24380} {arXiv:2503.24380 [cond-mat.dis-nn]}
	\BibitemShut {NoStop}%
	\bibitem [{\citenamefont {Goblot}\ \emph {et~al.}(2020)\citenamefont {Goblot},
		\citenamefont {{\v{S}}trkalj}, \citenamefont {Pernet}, \citenamefont {Lado},
		\citenamefont {Dorow}, \citenamefont {Lema{\^\i}tre}, \citenamefont
		{Le~Gratiet}, \citenamefont {Harouri}, \citenamefont {Sagnes}, \citenamefont
		{Ravets} \emph {et~al.}}]{goblot2020emergence}%
	\BibitemOpen
	\bibfield  {author} {\bibinfo {author} {\bibfnamefont {V.}~\bibnamefont
			{Goblot}}, \bibinfo {author} {\bibfnamefont {A.}~\bibnamefont
			{{\v{S}}trkalj}}, \bibinfo {author} {\bibfnamefont {N.}~\bibnamefont
			{Pernet}}, \bibinfo {author} {\bibfnamefont {J.~L.}\ \bibnamefont {Lado}},
		\bibinfo {author} {\bibfnamefont {C.}~\bibnamefont {Dorow}}, \bibinfo
		{author} {\bibfnamefont {A.}~\bibnamefont {Lema{\^\i}tre}}, \bibinfo {author}
		{\bibfnamefont {L.}~\bibnamefont {Le~Gratiet}}, \bibinfo {author}
		{\bibfnamefont {A.}~\bibnamefont {Harouri}}, \bibinfo {author} {\bibfnamefont
			{I.}~\bibnamefont {Sagnes}}, \bibinfo {author} {\bibfnamefont
			{S.}~\bibnamefont {Ravets}}, \emph {et~al.},\ }\bibfield  {title} {\bibinfo
		{title} {{Emergence of criticality through a cascade of delocalization
				transitions in quasiperiodic chains}},\ }\href@noop {} {\bibfield  {journal}
		{\bibinfo  {journal} {Nature Physics}\ }\textbf {\bibinfo {volume} {16}},\
		\bibinfo {pages} {832} (\bibinfo {year} {2020})}\BibitemShut {NoStop}%
	\bibitem [{\citenamefont {Tabanelli}\ \emph {et~al.}(2024)\citenamefont
		{Tabanelli}, \citenamefont {Castelnovo},\ and\ \citenamefont
		{\ifmmode~\check{S}\else \v{S}\fi{}trkalj}}]{PhysRevB.110.184208}%
	\BibitemOpen
	\bibfield  {author} {\bibinfo {author} {\bibfnamefont {H.}~\bibnamefont
			{Tabanelli}}, \bibinfo {author} {\bibfnamefont {C.}~\bibnamefont
			{Castelnovo}},\ and\ \bibinfo {author} {\bibfnamefont {A.}~\bibnamefont
			{\ifmmode~\check{S}\else \v{S}\fi{}trkalj}},\ }\bibfield  {title} {\bibinfo
		{title} {{Reentrant localization transitions and anomalous spectral
				properties in off-diagonal quasiperiodic systems}},\ }\href
	{https://doi.org/10.1103/PhysRevB.110.184208} {\bibfield  {journal} {\bibinfo
			{journal} {Phys. Rev. B}\ }\textbf {\bibinfo {volume} {110}},\ \bibinfo
		{pages} {184208} (\bibinfo {year} {2024})}\BibitemShut {NoStop}%
	\bibitem [{\citenamefont {L\"uschen}\ \emph {et~al.}(2018)\citenamefont
		{L\"uschen}, \citenamefont {Scherg}, \citenamefont {Kohlert}, \citenamefont
		{Schreiber}, \citenamefont {Bordia}, \citenamefont {Li}, \citenamefont
		{Das~Sarma},\ and\ \citenamefont {Bloch}}]{PhysRevLett.120.160404}%
	\BibitemOpen
	\bibfield  {author} {\bibinfo {author} {\bibfnamefont {H.~P.}\ \bibnamefont
			{L\"uschen}}, \bibinfo {author} {\bibfnamefont {S.}~\bibnamefont {Scherg}},
		\bibinfo {author} {\bibfnamefont {T.}~\bibnamefont {Kohlert}}, \bibinfo
		{author} {\bibfnamefont {M.}~\bibnamefont {Schreiber}}, \bibinfo {author}
		{\bibfnamefont {P.}~\bibnamefont {Bordia}}, \bibinfo {author} {\bibfnamefont
			{X.}~\bibnamefont {Li}}, \bibinfo {author} {\bibfnamefont {S.}~\bibnamefont
			{Das~Sarma}},\ and\ \bibinfo {author} {\bibfnamefont {I.}~\bibnamefont
			{Bloch}},\ }\bibfield  {title} {\bibinfo {title} {{Single-Particle Mobility
				Edge in a One-Dimensional Quasiperiodic Optical Lattice}},\ }\href
	{https://doi.org/10.1103/PhysRevLett.120.160404} {\bibfield  {journal}
		{\bibinfo  {journal} {Phys. Rev. Lett.}\ }\textbf {\bibinfo {volume} {120}},\
		\bibinfo {pages} {160404} (\bibinfo {year} {2018})}\BibitemShut {NoStop}%
	\bibitem [{\citenamefont {Wang}\ \emph
		{et~al.}(2020{\natexlab{b}})\citenamefont {Wang}, \citenamefont {Zhang},
		\citenamefont {Niu}, \citenamefont {Yu},\ and\ \citenamefont
		{Liu}}]{PhysRevLett.125.073204}%
	\BibitemOpen
	\bibfield  {author} {\bibinfo {author} {\bibfnamefont {Y.}~\bibnamefont
			{Wang}}, \bibinfo {author} {\bibfnamefont {L.}~\bibnamefont {Zhang}},
		\bibinfo {author} {\bibfnamefont {S.}~\bibnamefont {Niu}}, \bibinfo {author}
		{\bibfnamefont {D.}~\bibnamefont {Yu}},\ and\ \bibinfo {author}
		{\bibfnamefont {X.-J.}\ \bibnamefont {Liu}},\ }\bibfield  {title} {\bibinfo
		{title} {{Realization and Detection of Nonergodic Critical Phases in an
				Optical Raman Lattice}},\ }\href
	{https://doi.org/10.1103/PhysRevLett.125.073204} {\bibfield  {journal}
		{\bibinfo  {journal} {Phys. Rev. Lett.}\ }\textbf {\bibinfo {volume} {125}},\
		\bibinfo {pages} {073204} (\bibinfo {year} {2020}{\natexlab{b}})}\BibitemShut
	{NoStop}%
\end{thebibliography}

\clearpage

\onecolumngrid

\begin{center}
	
	{\large \bf Supplementary Materials:
		\\ Distinct reentrant transitions in a quasi-periodic Raman lattice}\\
	
	\vspace{0.3cm}
	
\end{center}

\vspace{0.6cm}

\twocolumngrid

\beginsupplement

\paragraph{Matrix form of the Hamiltonian.---}Eq. \eqref{h1} in the main text can be written in the matrix form 
\begin{equation}\label{matrix}
	\hat{H}=\left(\begin{array}{cc}
		\hat{h} & \hat{\Delta} \\
		\hat{\Delta} & -\hat{h}
	\end{array}\right),
\end{equation}
where $\hat{h}$ and $-\hat{h}$ correspond to the spin-$\uparrow$ and spin-$\downarrow$ parts respectively. $\hat{\Delta}$ is the spin-flip term caused by SOC. The matrix form of $\hat{h}$ is
\begin{equation}
	\hat{h}=\left(\begin{array}{cccccc}
		V_0 & -t & & & & -t\\
		-t & V_1 & -t & & & \\
		& -t & V_2 & -t & & \\
		& & \ddots & \ddots & \ddots & \\
		& & & -t & V_{L-2} & -t \\
		-t & & & & -t & V_{L-1}
	\end{array}\right),
\end{equation}
and that of $\hat{\Delta}$  can be written as
\begin{equation}
	\hat{\Delta}=\left(\begin{array}{cccccc}
		0 & \Delta & & & & -\Delta\\
		-\Delta & 0 & \Delta & & & \\
		& -\Delta & 0 & \Delta & & \\
		& & \ddots & \ddots & \ddots & \\
		& & & -\Delta & 0 & \Delta \\
		\Delta & & & & -\Delta & 0
	\end{array}\right).
\end{equation}

\begin{figure}[t]
	\includegraphics[width=0.95\columnwidth,height=0.8\columnwidth]{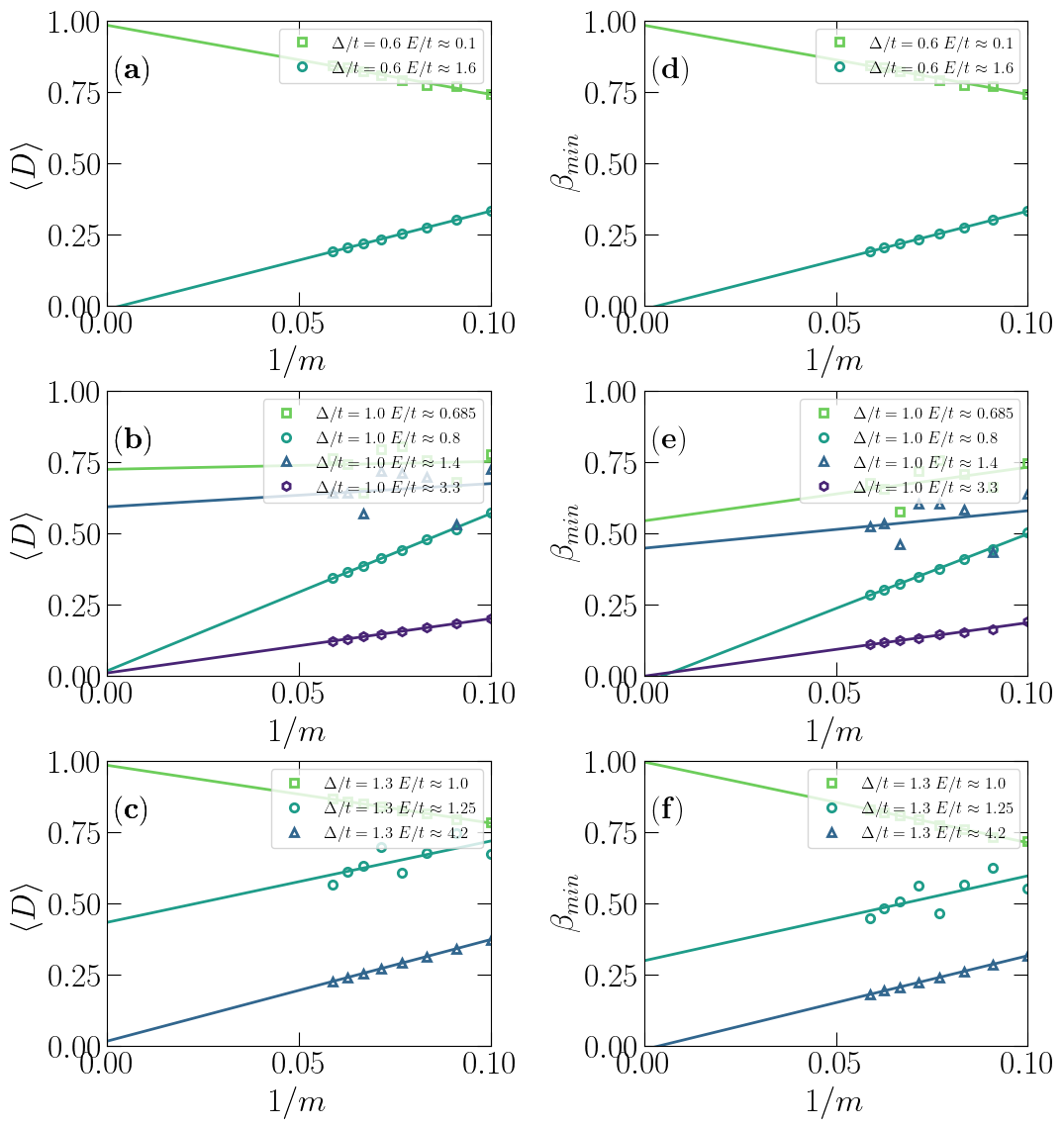}
	\vspace{-0.4cm}
	\caption{The $\langle D\rangle$ (left panels) and $\beta_{min}$ (right panels) as a function of $1/m$. The parameters of (a)\textendash(c) $\left[(d)\textendash(f)\right]$ correspond to those in Figs.~\ref{fig2} (b)\textendash(d) one by one. Here, we employ the shift-invert algorithm~\cite{pietracaprina2018shift,PhysRevB.99.165137} to calculate the fractal dimension, where the energy specified in the legend denotes the target energy of the algorithm.
	}
	\label{figadd7}
\end{figure}

\paragraph{Multifractal analysis of different states.---}
In Figs.~\ref{fig2} (b)\textendash(d), we present how $\langle D\rangle$ varies with size for different states. Here, we still fix $L=F_m$ as in the main text and extend our analysis to a more detailed multifractal study. The Fibonacci sequence is defined as $F_0=1$, $F_1=1$, and $F_m=F_{m-1}+F_{m-2}$. In the left panels of Fig.~\ref{figadd7}, $\langle D\rangle$ approaches zero (unity) for localized (extended) states in the thermodynamic limit for $m\to\infty$, whereas it is in the interval $0<\langle D\rangle<1$ for critical states.
Furthermore, we also calculate the minimum fractal dimension $\beta_{min}$ in the right panels of Fig.~\ref{figadd7}, following the method introduced in Ref.~\cite{PhysRevB.93.104504}. Specifically, we first calculate the occupation probability 
$p_j=u_j^2+v_j^2$ at each lattice site. The local fractal dimension at site $j$ is then defined as
$\beta_j = -\ln(p_j)/\ln(L) = -\ln(p_j)/\ln(F_m)$, from which we extract the minimum value $\beta_{\min}$. For extended states, $\beta_{min}$ tends to unity as $1/m$ decreases. Conversely, for localized states, $\beta_{min}$ approaches zero in the limit for $m\to\infty$. When $\beta_{min}$ is within the interval $0<\beta_{min}<1$, it indicates that the system is in a critical state. The scaling behavior of $\beta_{min}$ is consistent with that of $\langle D\rangle$, further confirming the properties of different states and supporting the existence of conventional and anomalous mobility edges.

\begin{figure}[t]
	\centering
	\includegraphics[width=1\columnwidth,height=0.8\columnwidth]{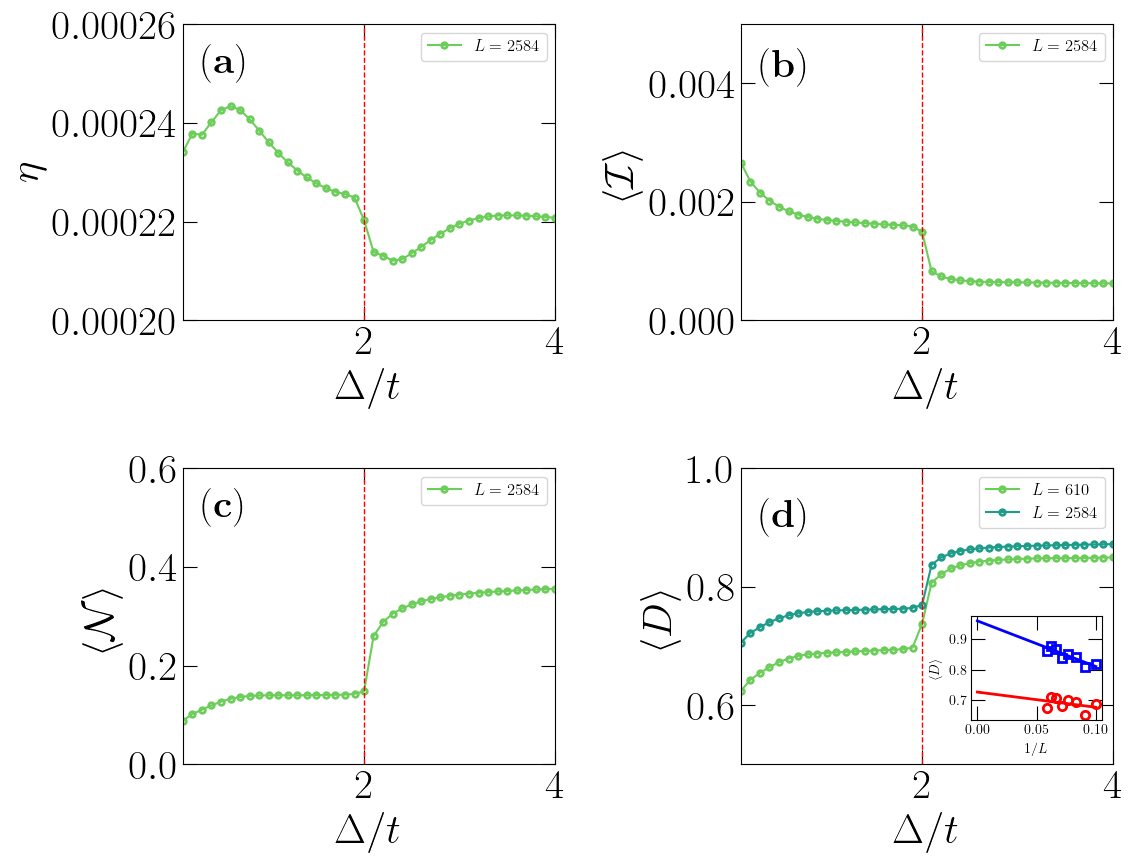}
	\vspace{-0.4cm}
	\caption{ (a) Composite participation ratio $\eta$, (b) Mean inverse participation ratio $\langle \mathcal{I} \rangle$, (c) Mean normalized participation ratio $\langle \mathcal{N} \rangle$, and (d) Mean fractal dimension $\langle D \rangle$ as functions of $\Delta/t$. $W/t=0$ and $V/t=2$. The inset in (d) shows the scaling analysis of $\langle D \rangle$ at $\Delta/t=0.5$ (red circles) and $\Delta/t=2.5$ (blue squares).
	}
	\label{figS0}
\end{figure}

\paragraph{Transition points in the phase diagram.---}In our work, it is important to clearly distinguish the different phases in the phase diagram. In Fig.~\ref{figS0}, we employ several diagnostics to characterize the transition from the purely critical phase to the purely extended phase. As shown in Fig.~\ref{figS0}(a), $\eta$ takes nearly the same value in both the purely extended and purely critical phases. It can also be observed that $\eta$ exhibits a pronounced change at the transition point for finite system sizes. This behavior originates from the sharp variation of the $\langle \mathcal{I} \rangle$ and $\langle \mathcal{N} \rangle$ at the transition point, as shown in Figs.~\ref{figS0} (b) and (c). In our work, the core role of $\eta$ is to distinguish purely phases from mixed (non-purely) phases. The distinction between the purely extended phase and the purely critical phase is instead established through the fractal dimension and its scaling behavior, as illustrated in Fig.~\ref{figS0} (d). Specifically, in the thermodynamic limit, $\langle D \rangle=1$ characterizes the purely extended phase, while $0<\langle D \rangle<1$ signals a purely critical phase, as shown in the inset of Fig.~\ref{figS0} (d). 

\begin{figure}[h!]
	\centering
	\includegraphics[width=1\columnwidth,height=1.2\columnwidth]{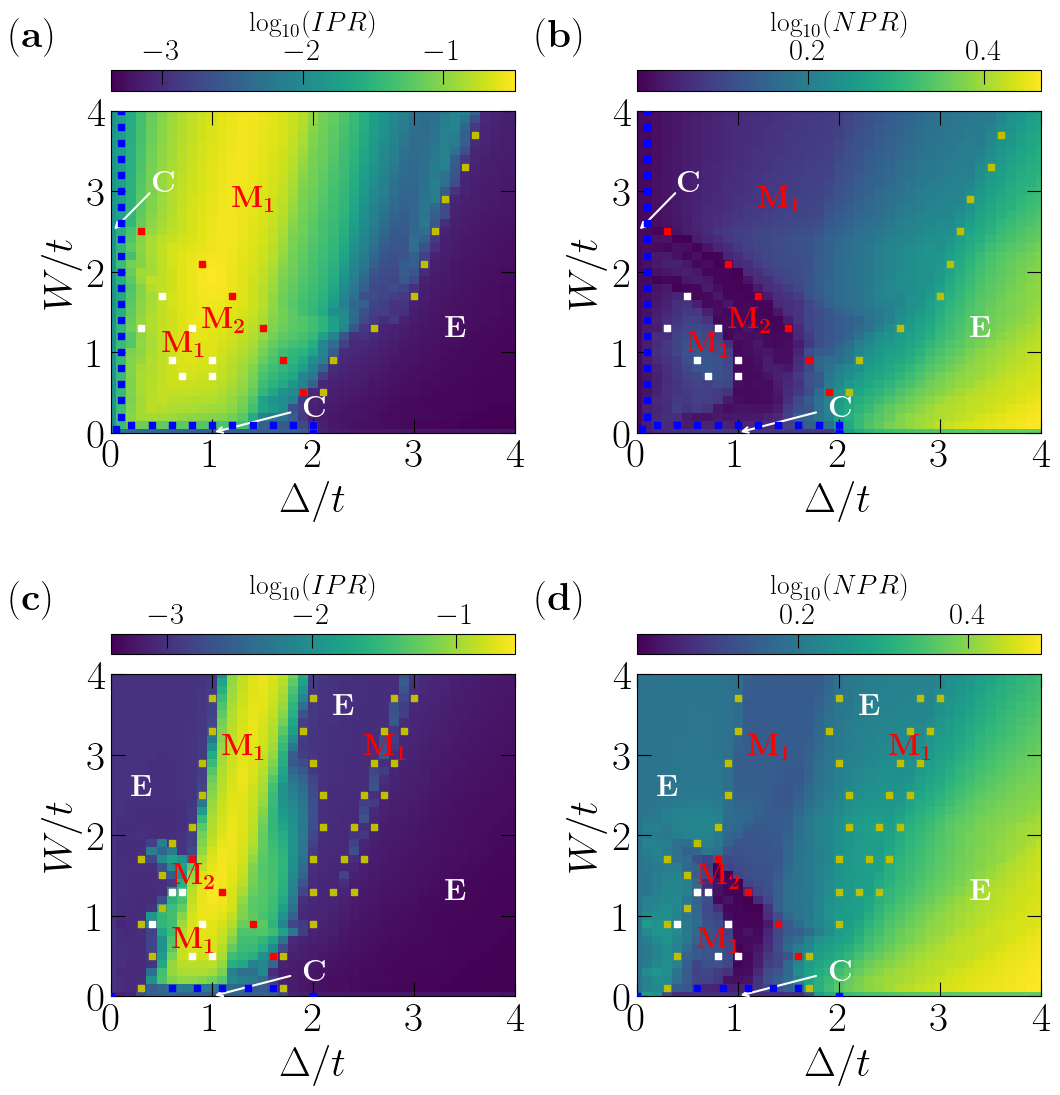}
	\vspace{-0.4cm}
	\caption{Phase diagrams in the $W$–$\Delta$ plane for $V/t=2.0$ (upper panel) and $V/t=1.4$ (lower panel). All other parameters are the same as those used in Fig. 1 of the main text. Colored markers denote transition points extracted from finite-size scaling analysis, following the procedure in Fig.~\ref{figadd7} and Fig.~\ref{figS0} (d). Blue markers enclose the purely critical phase, white markers enclose the $M_1$ phase, red markers separate the $M_1$ and $M_2$ phases, and yellow markers separate the purely extended phase from other phases. 
	}
	\label{figS01}
\end{figure}

In Fig.~\ref{figS01}, we present phase diagrams in the $W$–$\Delta$ plane for $V/t=2.0$ (upper panel) and $V/t=1.4$ (lower panel). The color scale indicates the inverse participation ratio for the left panel and the normalized participation ratio for the right panel. The transition points between different phases are determined by calculating the inverse participation ratio, the normalized participation ratio, and the composite participation ratio, extracted from the finite-size scaling analysis, following the procedure in Fig.~\ref{figadd7} and Fig.~\ref{figS0} (d). Blue markers enclose the purely critical phases, white markers enclose the $M_1$ phase, red markers separate the $M_1$ and $M_2$ phases, and yellow markers separate the E phase from other phases.

\paragraph{In the limiting cases.---}~\label{add1}
For SOC $\Delta/t\rightarrow\infty$, the Hamiltonian effectively becomes $\hat{H}_{eff}=\Delta\sum_{j}\left(\hat{c}_{j,\uparrow}^{\dagger}\hat{c}_{j+1,\downarrow}-\hat{c}_{j,\uparrow}^{\dagger}\hat{c}_{j-1,\downarrow} +\text{h.c.}\right)$. Then we perform the Fourier transform
\begin{equation}
	\begin{aligned}
		\hat{c}_{j,\sigma} = \frac{1}{\sqrt{L}} \sum_{k} e^{i k j} \hat{c}_{k,\sigma},
	\end{aligned}
\end{equation}
where $k$ is the momentum. Thus, the term $\sum_{j} \hat{c}_{j,\uparrow}^{\dagger} \hat{c}_{j+1,\downarrow}$ becomes
\begin{equation}
	\begin{aligned}
\sum_{j} \hat{c}_{j,\uparrow}^{\dagger} \hat{c}_{j+1,\downarrow} = &\sum_{j} \frac{1}{L} \sum_{k,k'} e^{-i k j} e^{i k' (j+1)} \hat{c}_{k,\uparrow}^{\dagger} \hat{c}_{k',\downarrow} \\
= &\sum_{k} e^{i k} \hat{c}_{k,\uparrow}^{\dagger} \hat{c}_{k,\downarrow}.
	\end{aligned}
\end{equation}
Similarly, $\sum_{j} \hat{c}_{j,\uparrow}^{\dagger} \hat{c}_{j-1,\downarrow}$ becomes
\begin{equation}
	\begin{aligned}
		\sum_{j} \hat{c}_{j,\uparrow}^{\dagger} \hat{c}_{j-1,\downarrow} = \sum_{k} e^{-i k} \hat{c}_{k,\uparrow}^{\dagger} \hat{c}_{k,\downarrow}.
	\end{aligned}
\end{equation}
Combining these and considering the Hermitian conjugate terms, the Hamiltonian in momentum space reads
\begin{equation}
	\begin{aligned}
\hat{H}^{'}_{eff} = \Delta \sum_{k} 2i \sin k \left( \hat{c}_{k,\uparrow}^{\dagger} \hat{c}_{k,\downarrow} - \hat{c}_{k,\downarrow}^{\dagger} \hat{c}_{k,\uparrow} \right).
	\end{aligned}
\end{equation}
This can be written as a matrix form for each momentum by using the two-component operator $\Psi_k = (\hat{c}_{k,\uparrow} \ \hat{c}_{k,\downarrow})^T$
\begin{equation}
	\begin{aligned}	
		H(k) =&2 \Delta \sin k \begin{pmatrix} 0 & i \\ -i & 0 \end{pmatrix}\\
 		=&-2\Delta \sin k \, \sigma_y,
	\end{aligned}
\end{equation}
where $\sigma_y$ is the Pauli matrix. Thus the momentum $k$ is a good quantum number and the eigenstates of the Hamiltonian are known as Bloch waves, corresponding to extended states. The spectrum can be obtained by diagonalizing $H(k)$ and eigenvalues are
\begin{equation}
	\begin{aligned}
E(k) = \pm 2\Delta \sin k.
\end{aligned}
\end{equation}

\begin{figure}[htbp]
	\includegraphics[width=1.0\columnwidth,height=0.8\columnwidth]{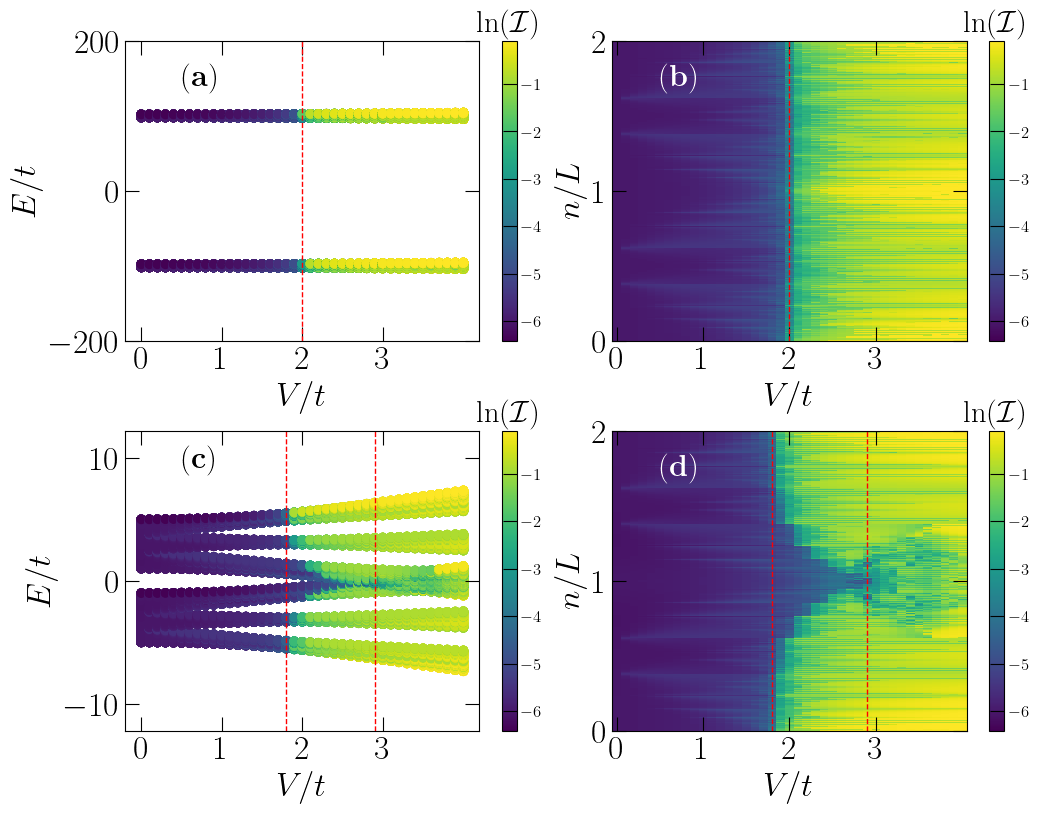}
	\vspace{-0.4cm}
	\caption{(a) and (b) Spectra as a function of $V/t$ for $W/t=100$. (c) and (d) Spectra as a function of $V/t$ for $W/t=3$. $L=610$, $\alpha=377/610$, $\Delta/t=0.5$, and $\phi=0$. The color indicates the value of $\ln(\mathcal{I})$. Red dotted lines mark the transition points.
	}
	\label{figadd1}
\end{figure}

In another limit $W/t\rightarrow\infty$, the diagonal elements in Eq.~\eqref{matrix}  dominate over the off-diagonal elements. This allows us to decompose the full Hamiltonian $\hat{H}$ into a dominant diagonal part $\hat{H_0}$ and a perturbative off-diagonal part $\hat{H}'$, such that
\begin{equation}
	\hat{H_0}=\left(\begin{array}{cc}
		\hat{h} & 0 \\
		0 & -\hat{h}
	\end{array}\right),
\end{equation}
\begin{equation}
	\hat{H}'=\left(\begin{array}{cc}
		0 & \hat{\Delta} \\
		-\hat{\Delta} & 0
	\end{array}\right).
\end{equation}
We now consider the first-order perturbation theory to study the effect of $\hat{H}'$ on the eigenstates of $\hat{H_0}$. According to standard non-degenerate perturbation theory, the first-order correction to the eigenstate $\left|\psi\right\rangle$ of $\hat{H}$ is
\begin{equation}
	\left|\psi\right\rangle=\left|\psi_{m}^{(0)}\right\rangle+\sum_{k\neq m} \frac{H_{k m}^{\prime}}{E_{m}^{(0)}-E_{k}^{(0)}}\left|\psi_{k}^{(0)}\right\rangle,
\end{equation}
where $\left|\psi_{m}^{(0)}\right\rangle$ and $\left|\psi_{k}^{(0)}\right\rangle$ are unperturbed eigenstates of $\hat{H_0}$, corresponding to different blocks $\hat{h}$ and $-\hat{h}$, with eigenvalues $E_{m}^{(0)}$ and $E_{k}^{(0)}$, respectively. $H_{k m}^{\prime}$ describes the coupling between these states due to the perturbation $\hat{H}'$. Since $\hat{H_0}$ has a block-diagonal form, its spectrum consists of two bands: one associated with $\hat{h}$ and the other with $-\hat{h}$. In the limit $W/t\rightarrow\infty$, these bands are well-separated, with an approximate energy gap of $2W/t$. As a result, the denominator becomes $(E_{m}^{(0)}-E_{k}^{(0)})\rightarrow\infty$, when $\left|\psi_{m}^{(0)}\right\rangle$ and $\left|\psi_{k}^{(0)}\right\rangle$ belong to different blocks. This suppresses the perturbative corrections, yielding $\left|\psi\right\rangle\approx\left|\psi_{m}^{(0)}\right\rangle$. Since both $\hat{h}$ and $-\hat{h}$ correspond to the AA model, we conclude that Eq. \eqref{h1} reduces to the AA model for $W/t\rightarrow\infty$. In Figs.~\ref{figadd1} (a) and (b), we show the spectra as a function of $V/t$ for $W/t=100$. The localization transition point is at $W/t\approx2.0$, which is consistent with that of the AA model. When $W/t$ decreases and the perturbation can no longer be neglected, the sharp transition point broadens into a transition region. For instance, when $W/t=3$ and $\Delta/t=0.5$ in Figs.~\ref{figadd1} (c) and (d), the transition occurs within the range $1.8<V/t<2.9$, where mobility edges emerge. In Fig.~\ref{fig1} (b) in the main text, we fix $V/t=1.4$, thus the system is in the purely extended phase for $W/t=3$ and $\Delta/t=0.5$.

\paragraph{Reentrant phenomenon and pure phase criterion.---}
In spinor notation $\Psi_j = (c_{j,\uparrow}, c_{j,\downarrow})^T$, the Hamiltonian in the main text can be written as
\begin{equation}
	H = \sum_j \Psi_j^\dagger M_j \Psi_j + \sum_j \left( \Psi_{j+1}^\dagger T \Psi_j + \Psi_j^\dagger T^\dagger \Psi_{j+1} \right),
\end{equation}
where the on-site matrix $M_j = V_j \sigma_z$ and the hopping matrix $T = -t \sigma_z - i\Delta \sigma_y$. This model possesses chiral symmetry: the operator $\Gamma = \sigma_x$ satisfies $\Gamma H \Gamma^{-1} = -H$.
According to Theorem I of the unified framework~\cite{zhou2025fundamentallocalizationphasesquasiperiodic}, a spinful quasi-periodic system exhibits pure phases (no mobility edges) if the following conditions are simultaneously met:
\begin{enumerate}
	\item Chiral symmetry is present.
	\item The on-site matrix $M_j$ is purely quasiperiodic (i.e., no uniform part, $W=0$).
	\item The hopping matrix $\Pi_j$ (here $T$) is either uniform or purely quasiperiodic.
\end{enumerate}
In our model, when $W\neq0$, $M_j$ is no longer purely quasiperiodic, thereby violating this condition. Consequently, although chiral symmetry is preserved, the uniform term $W$ is expected to induce mobility edges, rendering the transition point energy dependent.
The appearance of purely extended phases in Fig.~5 of the main text arises from the lack of overlap between the mobility edge and the energy band, which remains within the framework of Ref.~\cite{zhou2025fundamentallocalizationphasesquasiperiodic}.

\begin{figure}[t]
	\centering
	\includegraphics[width=1\columnwidth,height=0.8\columnwidth]{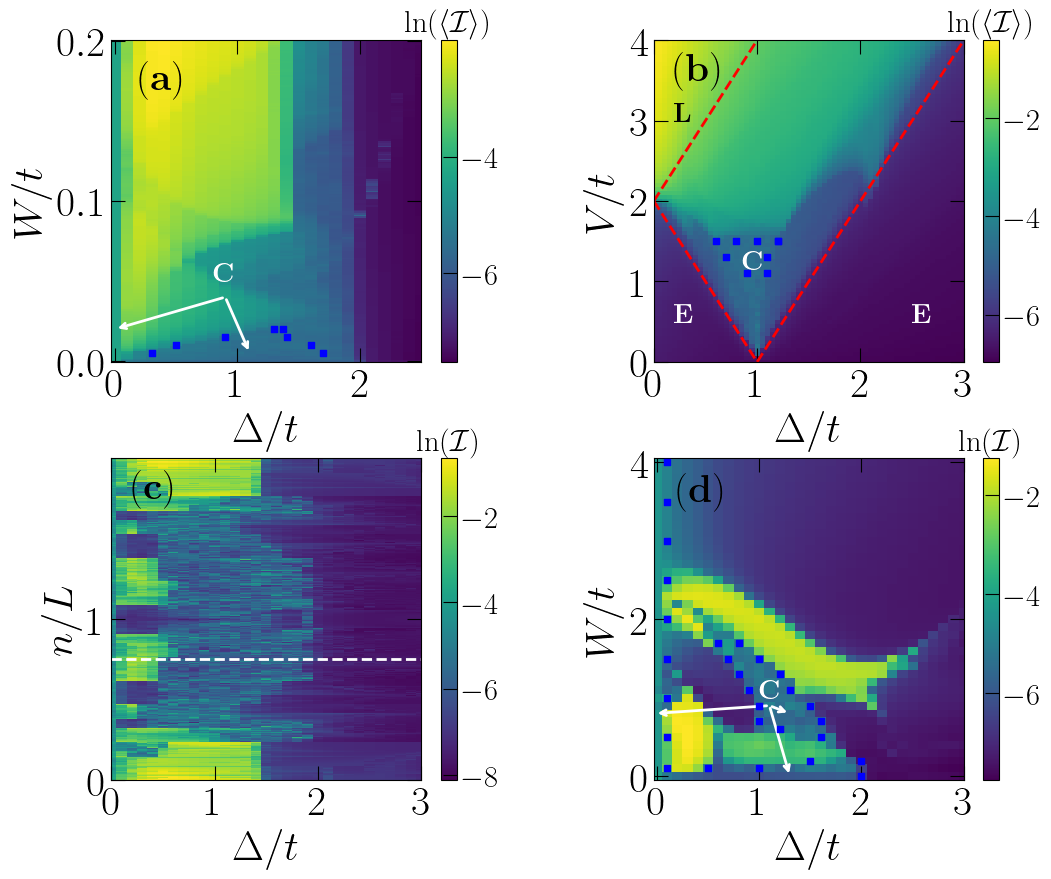}
	\vspace{-0.4cm}
	\caption{(a) The W-$\Delta$ phase diagram for $V/t=2$.  (b) The V-$\Delta$ phase diagram for $W/t=0.1$. (c) Spectrum as a function of $\Delta$ for $V/t=2$ and $W/t=0.1$. (d) The W-$\Delta$ phase diagram at $n/L=0.75$ (the corresponding position is marked by the white dashed line in (c)) for $V/t=2$. $L=2584$ and blue markers indicate the boundaries between the critical phase and other phases. In (b), the red dashed lines enclose the purely critical phase for $W/t=0$.
	}
	\label{figS1}
\end{figure}

\begin{figure}[htbp]
	\centering
	\includegraphics[width=1\columnwidth,height=0.8\columnwidth]{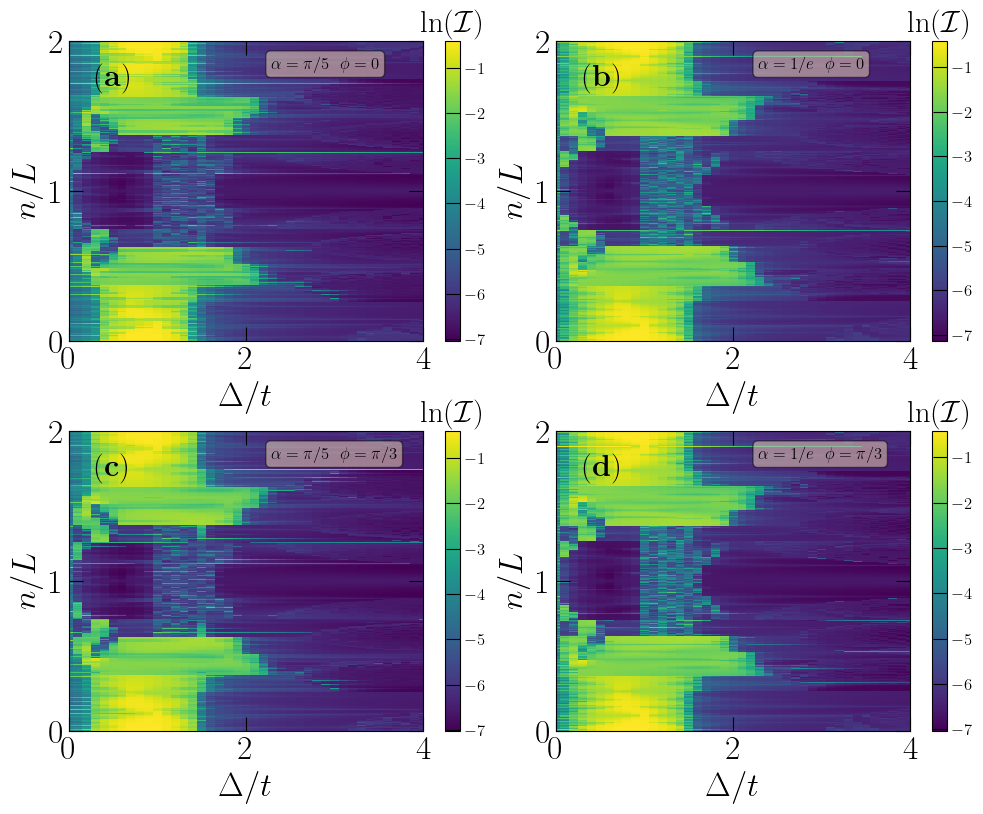}
	\vspace{-0.4cm}
	\caption{ Spectrum as a function of $\Delta$ for (a) $\alpha=\pi/5$, $\phi=0$; (b) $\alpha=1/e$, $\phi=0$; (c) $\alpha=\pi/5$, $\phi=\pi/3$; and (d) $\alpha=1/e$, $\phi=\pi/3$. $L=1000$ and other parameters are identical to those used in Fig.~2 in the main text.
	}
	\label{figS3}
\end{figure}

\paragraph{The effect of $W$ on the purely critical phase.---}
In Fig.~4 of the main text, we adopt a weak uniform Zeeman potential strength $W$ to ensure that the purely critical phase (where all eigenstates are critical) in the region $\Delta/t>0$ is not completely destroyed for $V/t=2$. 
In Fig.~\ref{figS1} (a), we present the evolution of the phase diagram with increasing $W$ for $V/t=2$. This indicates that the purely critical phase surrounded by blue dots survives only for sufficiently weak $W$ and is rapidly replaced by mixed phases as $W$ increases in the region $\Delta/t>0$. Although the values of $\ln(\langle \mathcal{I} \rangle)$ in certain regions for $W/t > 0.04$ are close to those observed in the purely critical phase, a detailed inspection of the spectrum confirms that the purely critical phase is absent in this regime.
In Fig.~\ref{figS1} (b), we fix $W/t=0.1$ and present the $V$-$\Delta$ phase diagram. The region enclosed by the blue dots corresponds to the purely critical phase. Compared with the $W/t=0$ case, where the purely critical phase is bounded by the red dashed lines, the uniform Zeeman potential significantly suppresses this phase. Although the purely critical phase disappears for generic finite values of $W$, partial critical states are still preserved in the spectrum, as shown in Fig.~\ref{figS1} (c), where we present the spectrum as a function of $\Delta$ for $W/t=0.1$. It can be found that eigenstates near the spectral edges evolve into localized or extended states, whereas the critical states at the central part of the spectrum remain robust. Thus, a generalized reentrant criticality transition can still be observed. Here, the term ``generalized reentrant criticality transition'' refers to a situation where a given band or energy level enters the critical regime multiple times as parameters vary. It does not require all eigenstates to be critical, analogous to the generalized reentrant localization transition discussed in previous studies~\cite{PhysRevB.110.184208,goblot2020emergence}. In Fig.~\ref{figS1} (d), we show the phase diagram at the energy level $n/L=0.75$ (the corresponding position is marked by the white dashed line in Fig.~\ref{figS1} (c)), where the system repeatedly enters the critical regime as $\Delta$ decreases from 3 to 0 for generic finite values of $W$.

\paragraph{The influence of disorder and interactions on phase diagrams.---}
To further demonstrate the influence of disorder and interactions on phase diagrams, we consider the following Hamiltonian
\begin{equation}
	\begin{aligned}
		\hat{H}=&-t\sum_{j}\left(\hat{c}_{j,\uparrow}^{\dagger}\hat{c}_{j+1,\uparrow}-\hat{c}_{j,\downarrow}^{\dagger} \hat{c}_{j+1,\downarrow}+\text{h.c.}\right) \\
		&+\Delta\sum_{j}\left(\hat{c}_{j,\uparrow}^{\dagger}\hat{c}_{j+1,\downarrow}-\hat{c}_{j+1,\uparrow}^{\dagger}\hat{c}_{j,\downarrow} +\text{h.c.}\right) \\
		&+ \sum_{j} (V\cos(2\pi\alpha j+\phi)+W+\delta_j) \left(\hat{n}_{j,\uparrow}-\hat{n}_{j,\downarrow}\right)\\
		&+ \sum_{j} U \hat{n}_{j,\uparrow}\hat{n}_{j,\downarrow},
	\end{aligned}
\end{equation}
where $\delta_j$ is a white-noise-type random disorder potential drawn uniformly from $[-\delta, \delta]$ and $U$ represents the on-site interaction. All other symbols and operators are defined as in the main text.

\begin{figure}[t]
	\centering
	\includegraphics[width=1\columnwidth,height=0.8\columnwidth]{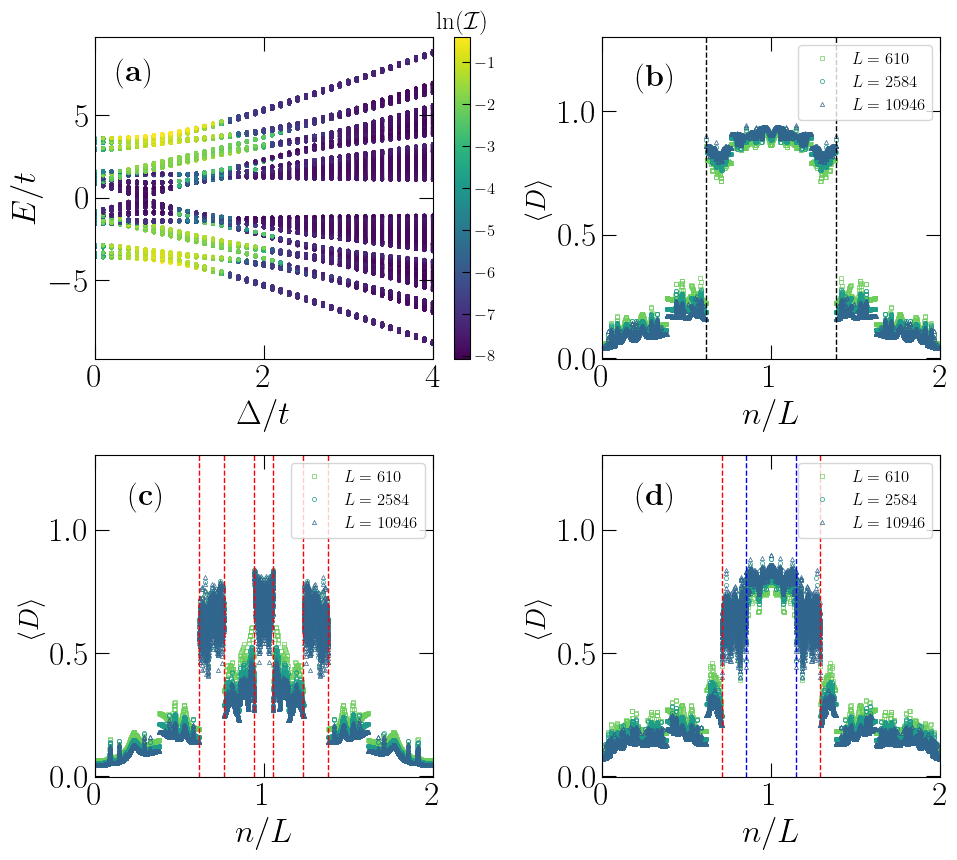}
	\vspace{-0.4cm}
	\caption{(a) Spectrum as a function of $\Delta$ in the presence of perturbative white-noise-type random disorder with $\delta=10^{-7}$. (b)–(d) Mean fractal dimension $\langle D\rangle$ at different energy levels for $\Delta/t=0.6$, $1.0$, and $1.3$, extracted from panel (a). Other parameters and symbols are the same as those in Fig.~2 of the main text.
	}
	\label{figS2}
\end{figure}

We first examine the robustness against disorder. In the main text, we focus on a quasi-periodic system, also referred to as a quasi-disordered system. Quasi-disorder is realized by tuning $\alpha$ to an irrational number, while the phase factor $\phi$ generates different disorder configurations. In Fig.~\ref{figS3}, we present the spectrum as a function of $\Delta$ for several choices of irrational $\alpha$ and $\phi$. By comparison, the transitions observed in Fig.~\ref{figS3} are consistent with those shown in Fig.~2 (a) of the main text (here we use $n/L$ as the ordinate), indicating that variations in the irrational parameter $\alpha$ and the phase factor $\phi$ do not significantly alter the phase-diagram structure, thereby demonstrating its robustness.
Moreover, we investigate the effect of random disorder on the phase diagram. In Fig.~\ref{figS2}, the perturbative white-noise-type random disorder potential is added to the quasi-disordered system. Compared with Fig.~2 of the main text, the introduction of such disorder does not alter the evolution of the system as $\Delta$ increases. The mobility edges and anomalous mobility edges shown in Figs.~2 (b)–(d) of the main text also remain robust in Fig.~\ref{figS2} (b)-(d). These results indicate that the phase diagram in Fig.~1 of the main text remains stable against weak random disorder perturbations.

\begin{figure}[t]
	\centering
	\includegraphics[width=1\columnwidth,height=0.5\columnwidth]{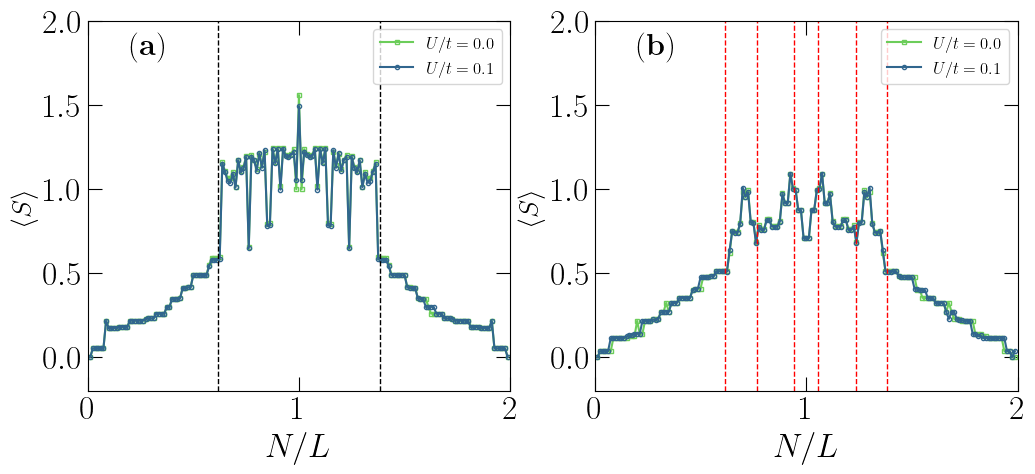}
	\vspace{-0.4cm}
	\caption{ (a) and (b) Mean entanglement entropy as a function of filling $N/L$ for $\Delta/t=0.6$ and $\Delta/t=1.0$, corresponding to Fig.~2 (b) and (c) of the main text. $L = 80$, other parameters and symbols are the same as those in Fig.~2 of the main text. In the DMRG calculations, we maintain $800$ states and perform 80 sweeps, yielding a truncation error below $10^{-9}$. To prevent convergence to local minima, a two-site optimization algorithm is employed.
	}
	\label{figSU}
\end{figure}

\begin{figure}[htbp]
	\centering
	\includegraphics[width=1\columnwidth,height=0.8\columnwidth]{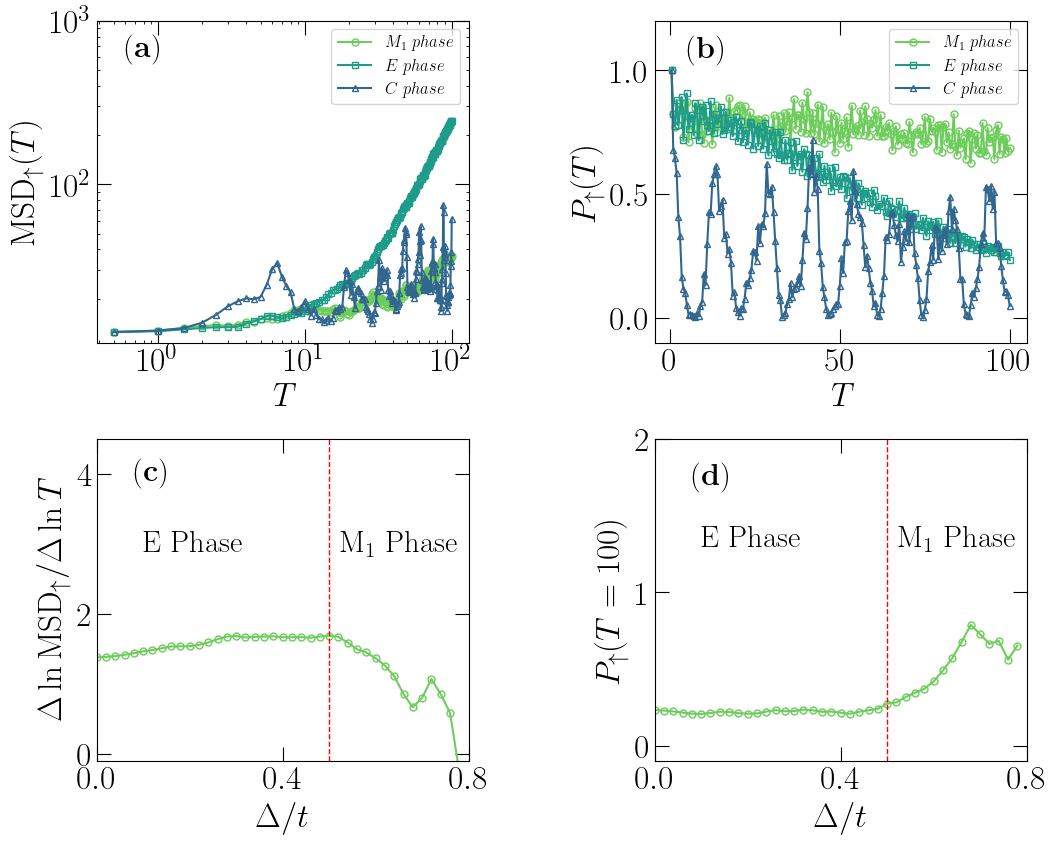}
	\vspace{-0.4cm}
	\caption{(a) $\mathrm{MSD}_{\uparrow}(T)$ and (b) $P_{\uparrow}(T)$ as functions of time for three representative cases: a M$_1$ phase with mobility edges, a purely extended (E) phase without mobility edges, and a purely critical (C) phase. (c) $\Delta \ln \mathrm{MSD}_{\uparrow}/ \Delta \ln T$ and (d) $P_{\uparrow}(T=100)$ as functions of $\Delta$. In (a) and (b), for the M$_1$ phase, $\Delta/t=0.7$, while for the purely extended phase, $\Delta/t=0.5$, all other parameters in these two cases are identical to those used in Fig.~5 of the main text. The parameters of the purely critical phase are the same as those in Fig.~4 of the main text with $\Delta/t=0.8$. $\sigma=5$ in the initial state. The red dotted lines in (c) and (d) mark the transition point at $\Delta/t=0.5$ between the $E$ and M$_1$ phases for $W/t=1.3$. 
	}
	\label{figStime}
\end{figure}

Next, we turn to the effect of interactions. In Fig.~\ref{figSU}, we employ the density-matrix renormalization group (DMRG) method to investigate the ground-state properties of the many-body system. The core quantity we calculated is entanglement entropy, defined as $S=-\sum_i\lambda_i\ln\lambda_i$, where $\lambda_i$ are the eigenvalues of the reduced density matrix obtained by tracing out half of the system.
In the absence of interactions ($U/t=0$), the Fermi energy increases as the particle number $N$ grows. When the Fermi energy lies in the localized-state region of Fig.~2 (b) in the main text, the particles are localized, resulting in low entanglement entropy. By contrast, when the Fermi energy lies in the extended-state region of Fig.~2 (b), some particles can extend across the chain, leading to a significant increase in the entanglement entropy. This behavior is illustrated in Fig.~\ref{figSU} (a), where an abrupt change of the entanglement entropy is observed as the Fermi energy crosses the mobility edges (black dotted lines).
Upon introducing weak perturbative interactions, the entanglement entropy does not change obviously compared with the noninteracting case ($U=0$), indicating the stability of the mobility edge against weak interactions. Similarly, in Fig.~\ref{figSU} (b), the behavior of the entanglement entropy is consistent with the evolution of the fractal dimension $\langle D \rangle$ shown in Fig.~2 (c) of the main text. Notably, the entanglement entropy decreases near $N/L=1$ in Fig.~\ref{figSU} (b). This occurs because the particles fill all energy levels with $E/t<0$, and the large band gap for $\Delta/t=1$ (see Fig.~2 (a) of the main text) drives the system into an insulating phase.
The inclusion of weak interactions again does not produce a noticeable change in the entanglement entropy, suggesting the robustness of the anomalous mobility edges (red dotted lines).
These results collectively demonstrate the stability of the phase diagram shown in Fig.~1 in the presence of weak perturbative interactions.

\begin{figure}[htbp]
	\includegraphics[width=1.0\columnwidth,height=0.8\columnwidth]{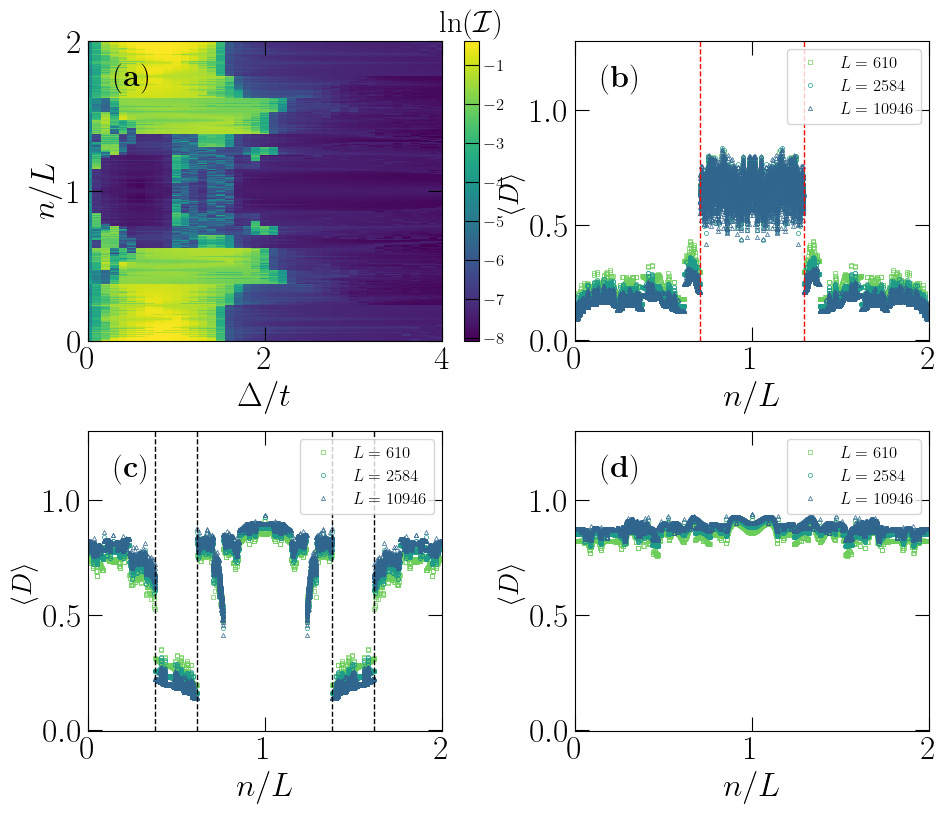}
	\vspace{-0.4cm}
	\caption{(a) Replotting Fig.~\ref{fig2} (a) using the energy level index as the ordinate. (b)\textendash(d) Mean fractal dimension $\langle D\rangle$ at various energy levels for $\Delta/t=1.4$, $1.8$, and $3.0$, respectively. Other parameters and dotted lines are consistent with Fig.~\ref{fig2} in the main text.
	}
	\label{figadd2}
\end{figure}

\paragraph{Experimental detection.---}
In our work, the reentrant delocalization shown in Fig.~5 of the main text originates from the shift of the mobility edge across the energy spectrum. Consequently, an unambiguous experimental verification of our proposed mechanism requires identifying the presence or absence of mobility edges. To this end, we follow established protocols based on real-time dynamical measurements, which have been widely used to detect mobility edges in quasiperiodic and disordered systems. The system under consideration contains two spin components, $\sigma=\uparrow,\downarrow$, and the single-particle state is described by a spinor wave function
\begin{equation}
	\Psi(t)=\bigl(\psi_{1\uparrow}(t),\dots,\psi_{L\uparrow}(t),
	\psi_{1\downarrow}(t),\dots,\psi_{L\downarrow}(t)\bigr)^{\mathrm T}.
\end{equation}
Experimentally, the initial state can be prepared such that a spin-up particle is localized near the center of the lattice, while the spin-down component is initially empty, namely,
\begin{equation}
	\psi_{j\uparrow}(0)=\mathcal{N}\exp \left[-\frac{(j-x_0)^2}{2\sigma^2}\right],\qquad
	\psi_{j\downarrow}(0)=0,
\end{equation}
where $x_0=L/2$ and $\mathcal{N}$ ensures normalization. In the following, we focus on the dynamics of the spin-up component. The time evolution is governed by the evolution operator $U=e^{-iHT}$, and all observables are evaluated using the spin-up probability density $\rho_{j \uparrow }(T)=|\psi_{j\uparrow}(T)|^2$. From this quantity, we extract the mean position and the mean-square displacement of the spin-up component, defined as $\langle x(T)\rangle_{\uparrow}=\sum_{j=1}^L j\,\rho_{j\uparrow}(T)$, and $\mathrm{MSD}_{\uparrow}(T)=\sum_{j=1}^L \bigl[j-\langle x(T)\rangle_{\uparrow}\bigr]^2\,\rho_{j \uparrow}(T)$, which characterize the spreading dynamics of the wave packet. In addition, the survival probability of the spin-up component, $P_{\uparrow}(T)=\left|\sum_{j=1}^L \psi^{*}_{j\uparrow}(0)\psi_{j\uparrow}(T)\right|^2$, provides a complementary dynamical diagnostic of localization properties.
Figs.~\ref{figStime} (a) and \ref{figStime} (b) show $\mathrm{MSD}_{\uparrow}(T)$ and $P_{\uparrow}(T)$ for three representative cases: a M$_1$ phase with mobility edges, a purely extended (E) phase without mobility edges, and a purely critical (C) phase. When the system hosts mobility edges, $\mathrm{MSD}_{\uparrow}(T)$ increases with time, indicating wave-packet expansion, while $P_{\uparrow}(T)$ tends to a finite value at long times, reflecting the persistence of localized components. The coexistence of these two behaviors provides clear dynamical evidence for mobility edges~\cite{PhysRevLett.120.160404}. In contrast, when mobility edges are absent, $\mathrm{MSD}_{\uparrow}(T)$ exhibits linear growth and $P_{\uparrow}(T)$ decays toward zero as time increases. In the purely critical phase, the growth rate of $\mathrm{MSD}_{\uparrow}(T)$ is lower than that in the purely extended phase and displays pronounced fluctuations, while $P_{\uparrow}(T)$ repeatedly oscillates between finite and vanishing values. This dynamical behavior is distinct from other phases. To verify our mechanism, we fix $W/t=1.3$ and present the rate of change of $\mathrm{MSD}_{\uparrow}(T)$, defined as $\Delta \ln \mathrm{MSD}_{\uparrow}/ \Delta \ln T$, together with $P_{\uparrow}(T=100)$, as functions of $\Delta$ in Figs.~\ref{figStime} (c) and \ref{figStime} (d). By calculating the quantities in the main text, we confirm that the transition point between the E phase and the M$_1$ phase for $W/t=1.3$ is at $\Delta/t=0.5$ (red dotted lines in Figs.~\ref{figStime} (c) and \ref{figStime} (d)).
As $\Delta$ decreases from 0.8 to 0.5, $\Delta \ln \mathrm{MSD}{\uparrow}/ \Delta \ln T$ increases, signaling an enhancement of the diffusion rate, whereas $P{\uparrow}(T=100)$ correspondingly decreases. These trends indicate an increasing fraction of extended states and a concomitant reduction of localized states in the spectrum. This behavior suggests that the mobility edge shifts with $\Delta$, thereby redistributing the relative weight between localized and extended states.

\paragraph{Destruction of the re-emerged extended states.---}
In Fig.\ref{fig2} (a) in the main text, we use energy as the ordinate to clearly show the band structure. However, due to the close proximity of many energy levels, some states are obscured in this representation. To address this, we replot the spectrum in Fig.\ref{figadd2} (a), where we use the energy level as the ordinate. It is observed that extended states re-emerge near the middle of the spectrum within the interval $1.1 < \Delta \leq 1.3$. This phenomenon is analyzed in detail in the main text. However, as $\Delta$ continues to increase, the re-emerged extended states are disrupted and transition into critical states. Specific details for $\Delta=1.4$ are illustrated in Fig.\ref{figadd2} (b), where only critical and localized states remain in the spectrum, with no extended states present. Combined with Fig.\ref{fig2} (a) in the main text, further increasing $\Delta$ leads to a decrease in the density of states near the middle of the spectrum, which is not conducive to the overlap of critical states, thereby inhibiting the transition of critical states into extended states. In Fig.\ref{figadd2} (c) and Fig.\ref{figadd2} (d), we present another two cases for $\Delta=1.8$ and $\Delta=3.0$, respectively. These figures demonstrate conventional mobility edges and purely extended phase, further illustrating the $\Delta$ induced transitions among different phases for $W/t=1$, as shown in Fig.\ref{fig1} (a) in the main text. 

\begin{figure}[htbp]
	\includegraphics[width=1.0\columnwidth,height=0.8\columnwidth]{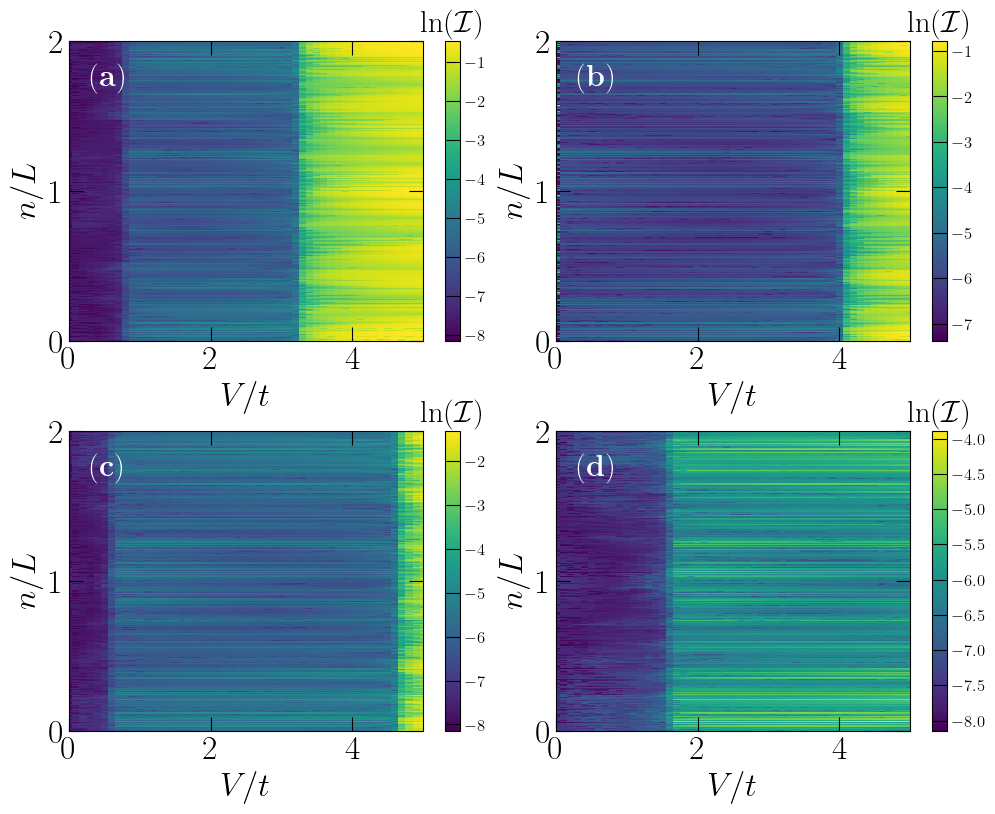}
	\vspace{-0.4cm}
	\caption{(a)\textendash(d) The IPR associated with the energy level index as a function of $V/t$ for $W/t=0$ and $\Delta/t=0.6$, $1.0$, $1.3$, and $1.8$, respectively. Other parameters are the same as those in Fig.~\ref{fig3}.
	}
	\label{figadd3}
\end{figure}

\begin{figure}[t]
	\includegraphics[width=0.95\columnwidth,height=0.6\columnwidth]{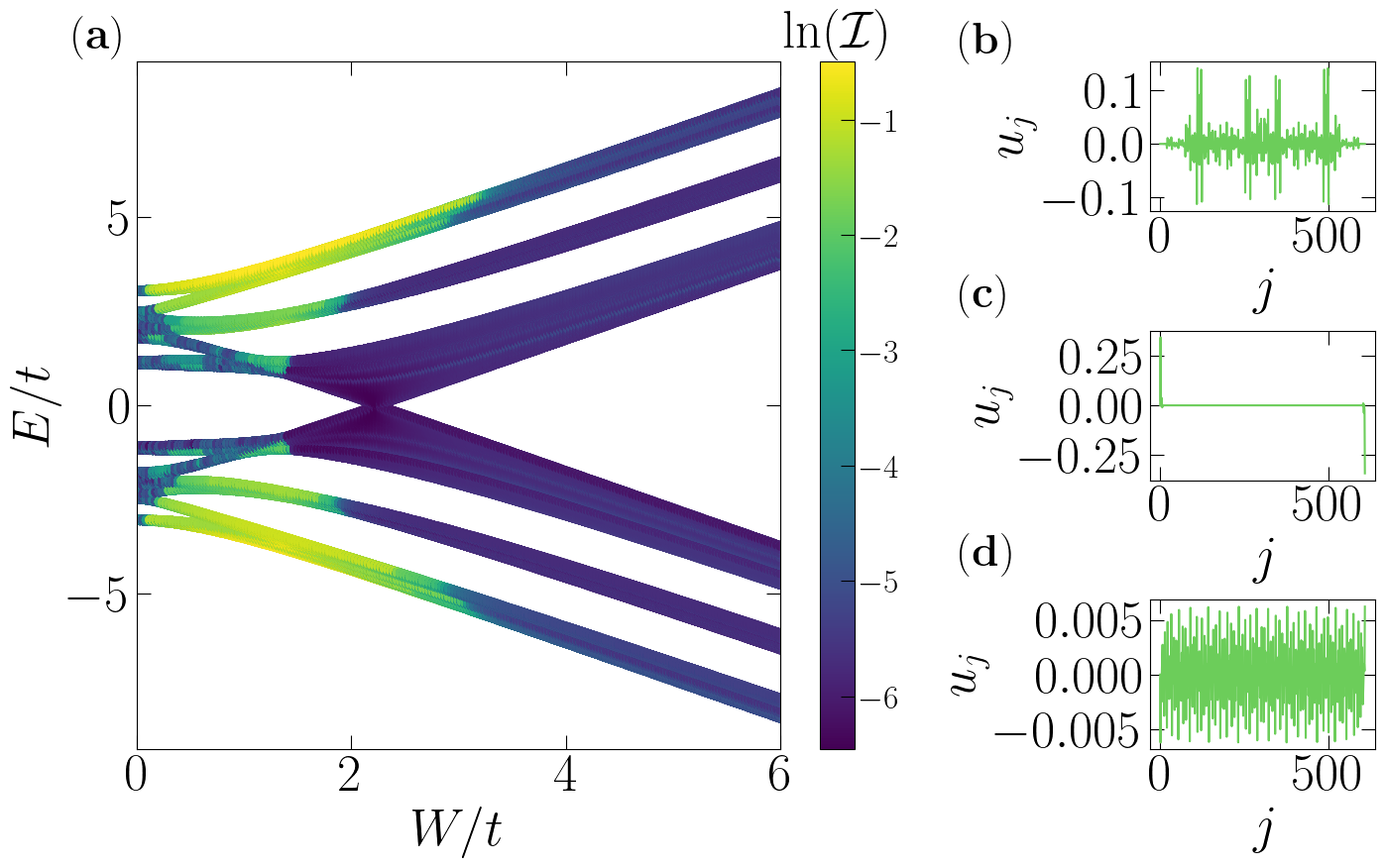}
	\vspace{-0.4cm}
	\caption{(a) Spectrum as a function of $W/t$. (b)\textendash(d) Typical critical, localized, and extended states, which are all taken from the ground state. $L=610$, $\alpha=377/610$, $\Delta/t=1$, $V/t=1.4$ and $\phi=0$. The color indicates the value of $\ln(\mathcal{I})$ in (a). The values of $W/t$ are $0.1$, $0.5$, and $5.0$ in (b)\textendash(d), respectively.
	}
	\label{figadd4}
\end{figure}

\paragraph{The transition in the case of $W/t=0$.---}
Fig.\ref{figadd3} (a) shows that the system undergoes a purely extended-critical-localized transition without conventional and anomalous mobility edges for $\Delta/t=0.6$. The transition
points are at $V/t=2-2\Delta/t=0.8$ and $V/t=2+2\Delta/t=3.2$, consistent with Ref.~\cite{PhysRevLett.125.073204,PhysRevB.93.104504}. As $\Delta$ increases, it first suppresses the purely extended phase. Typically, when $\Delta/t=1$ in Fig.\ref{figadd3} (b), the purely extended phase disappears completely. Continuing to increase $\Delta$, the purely extended phase reappears within the region $V/t<2\Delta/t-2$, as shown in Fig.\ref{figadd3} (c) and Fig.\ref{figadd3} (d). When $\Delta/t>1$, the increase of $\Delta$ is beneficial to the purely extended phase, and the transition point from the purely extended phase to the purely critical phase increases with $\Delta$.

\begin{figure}[t]
	\includegraphics[width=0.95\columnwidth,height=0.6\columnwidth]{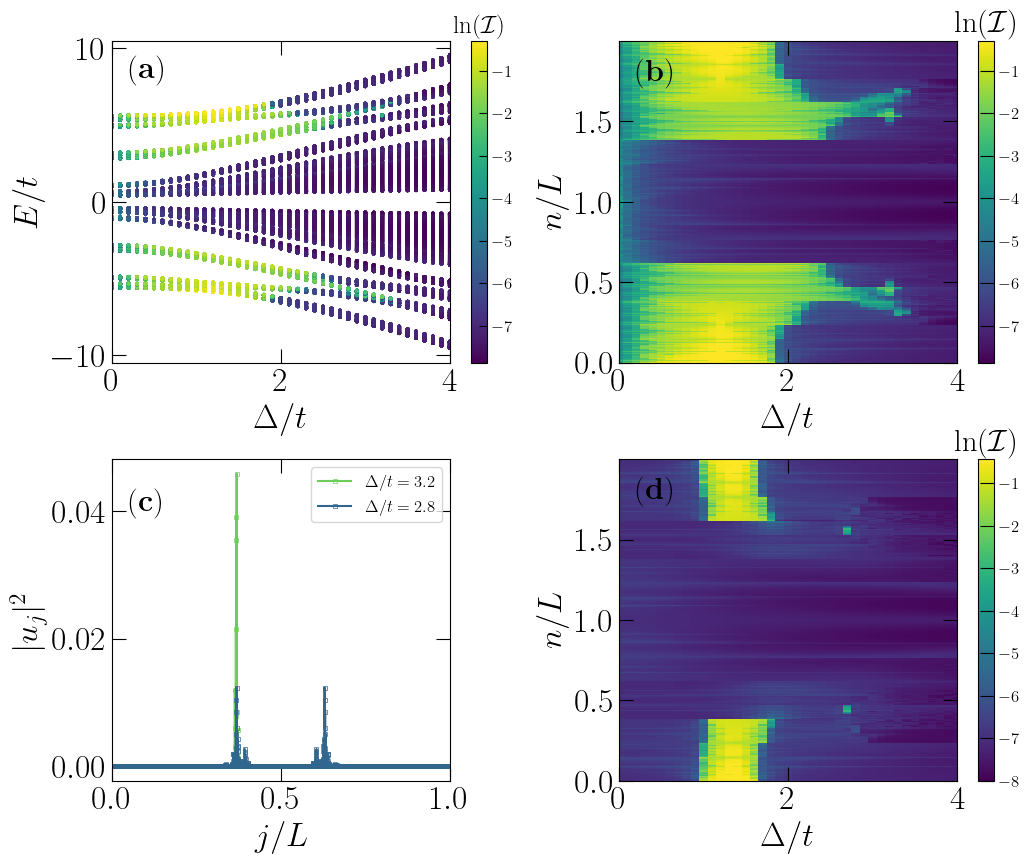}
	\vspace{-0.4cm}
	\caption{(a) and (b) Spectra with energy and energy level index on the ordinate as a function of $\Delta$ for $W/t=3$ and $V/t=2$, respectively. (c) wave function at $n=1151$ for $\Delta/t=3.2$ and $\Delta/t=2.8$ extracted from (a). (d) Spectrum as a function of $\Delta$ for $W/t=3$ and $V/t=1.4$. $L=2584$, $\alpha=1597/2584$, and $\phi=0$.
	}
	\label{figadd5}
\end{figure}

\begin{figure}[t]
	\includegraphics[width=0.95\columnwidth,height=0.8\columnwidth]{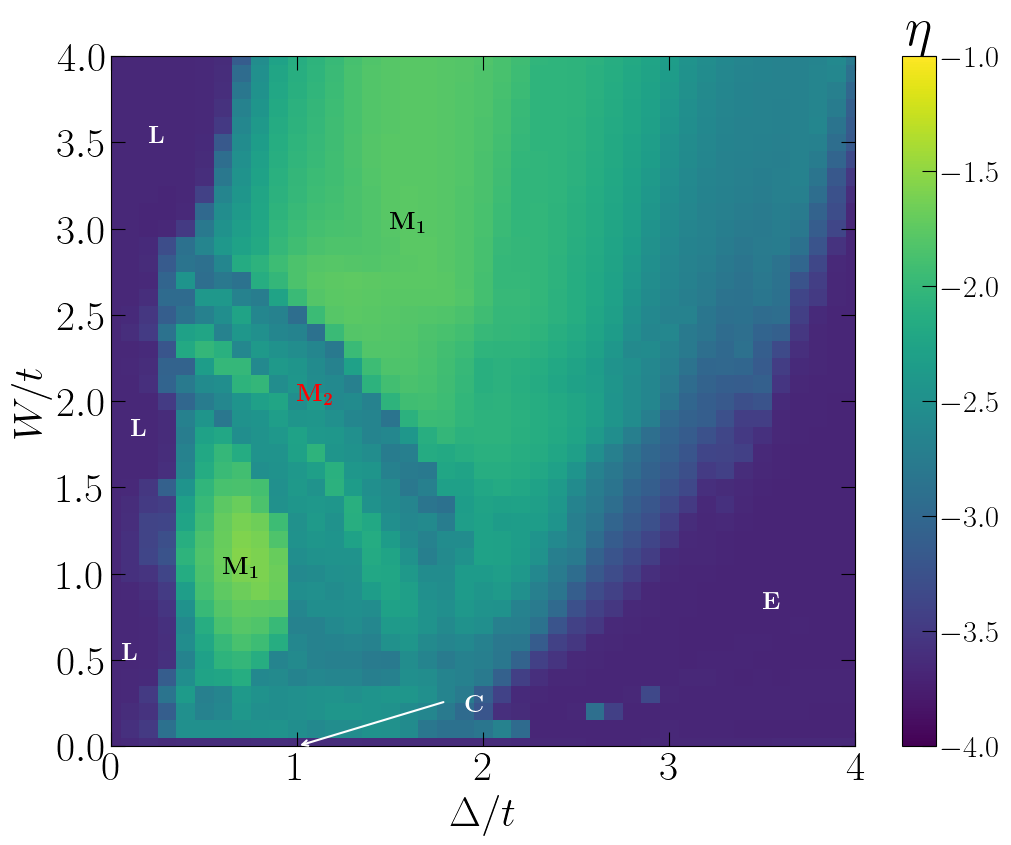}
	\vspace{-0.4cm}
	\caption{The phase diagram in the $W$ and $\Delta$ plane for $V/t=2.5$. The letter L indicates pure localized phases, and other parameters and labels are the same as those in Fig.~\ref{fig1}.
	}
	\label{figadd6}
\end{figure}

\paragraph{Spectrum for $\Delta/t=1$ and typical states.---}
Fig.~\ref{figadd4} (a) shows the spectrum as a function of $W$ for $V/t=1.4$ and $\Delta/t=1$. In the limiting case of $W/t=0$, the system is in the purely critical phase, where all states are critical states~\cite{PhysRevLett.125.073204, PhysRevB.93.104504}. In another limiting case of $W/t\rightarrow\infty$, Eq.~\eqref{h1} reduces to the AA model, details are provided above. Since we set $V/t=1.4$, the system lies in the purely extended phase for large $W$.  As $W/t$ increases from zero, localized states emerge and coexist with critical states in the spectrum, resulting in anomalous mobility edges in the range $0.11<W/t<1$. 
With further increase in $W/t$, critical states vanish, giving way to extended states, which coexist with localized states and form conventional mobility edges for $1<W/t<3.3$.
In Figs.~\ref{figadd4} (b)\textendash(d), we show representative examples of critical, localized, and extended states, respectively. They exhibit distinct spatial characteristics: the critical state manifests a fractal structure, the localized state displays exponential decay as the distance increases, and the extended state occupies the entire lattice~\cite{PhysRevLett.104.070601}.

\paragraph{$\Delta$ induced transition for large $W$.---} Figs.~\ref{figadd5} (a) and (b) show the same spectrum, but the ordinates represent energy and energy level index, respectively. It can be observed that with increasing $\Delta$, the system transitions from a purely critical phase to a purely extended phase, accompanied by the emergence of mobility edges. When we fix the $1151^{st}$ energy level ($n/L\approx 0.445$), the IPR at $\Delta/t=3.2$ is slightly larger than that at smaller $\Delta$, indicating that the wave function is more localized at this value, and the representative wave functions for $\Delta/t=3.2$ and $\Delta/t=2.8$ are displayed in Fig.~\ref{figadd5} (c). In Fig.~\ref{fig1} in the main text, we show that as $V$ decreases, the purely extended phase encroaches on the space of mixed phases. Specifically, states with small IPR are affected first, leading to the splitting of the M$_1$ phase into two regions for large $W$ as $\Delta$ increases in Fig.~\ref{fig1} (b). It should be noted that the reappearance of localized states for $V/t=1.4$ in Fig.~\ref{figadd5} (d) occurs at $\Delta/t=2.7$. which does not coincide with the position of the large-IPR states at $\Delta/t=3.2$ in Fig.~\ref{figadd5} (b). This discrepancy results from the change in $V$, which alters the spectrum and shifts the positions of states with large IPR.

\begin{figure}[t]
	\includegraphics[width=1.0\columnwidth,height=0.8\columnwidth]{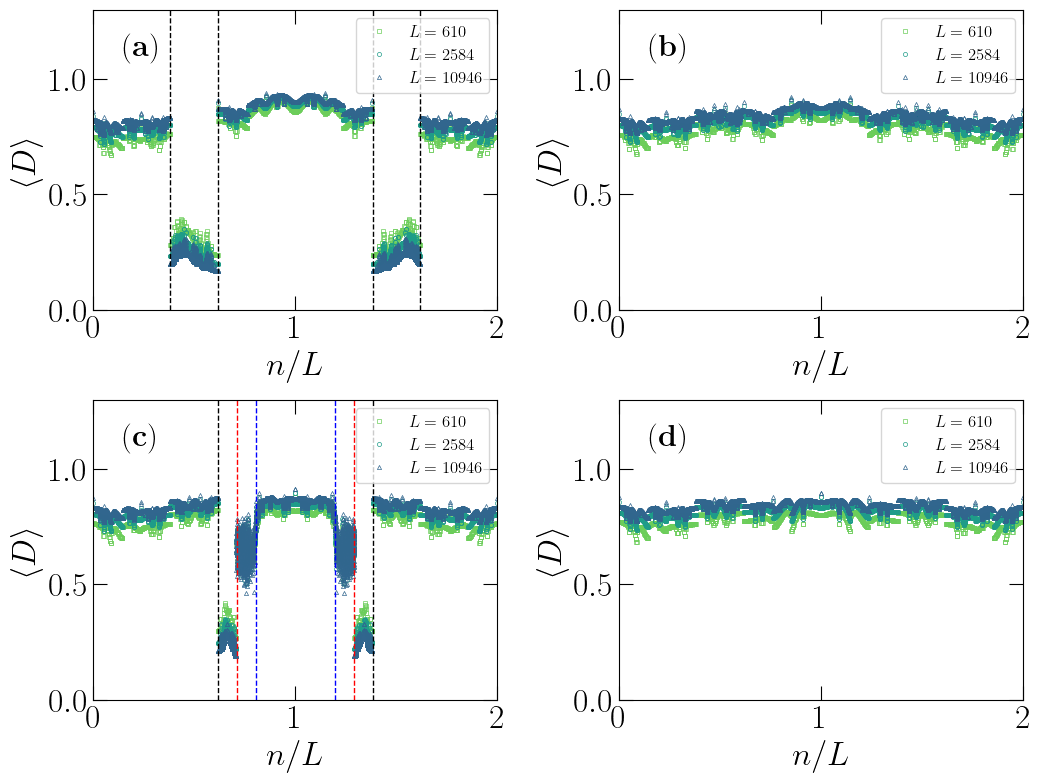}
	\vspace{-0.4cm}
	\caption{(a)\textendash(d) Mean fractal dimension $\langle D\rangle$ at various energy levels for $W/t=1.0$, $1.4$, $1.7$, and $2.5$, corresponding to orange markers \protect\solidSquare[orange]{1.6ex}, \protect\solidCircle[orange]{1.6ex}, \protect\solidTriangle[orange]{1.9ex}, and \protect\solidStar[orange]{1.9ex} in Fig.~\ref{fig1} (b). The dotted lines of different colors are consistent with those in Figs.~\ref{fig2} (b)\textendash(d).
	}
	\label{figadd8}
\end{figure}

{The phase diagram for $V/t=2.5$.---}	
In Fig.~\ref{figadd6}, we show the phase diagram for $V/t=2.5$. Compared to the cases of $V/t=2$ and $V/t=1.4$ in Fig.~\ref{fig1}, the most significant difference is the appearance of purely localized phases. For $\Delta/t=0$, Eq.~\eqref{h1} degenerates into two separated AA models for each spin. When $V/t=2.5$, the system is in purely localized phases. Thus, the ordinate of the phase diagram in Fig.~\ref{figadd6} corresponds to purely localized phases. Similar to Fig.~\ref{fig1} (b), the purely localized phases encroach upon the space of mixed phases.

{Reentrant delocalization transitions.---}
To further present the details of Fig.~\ref{fig5} (d), we study the mean fractal dimension $\langle D\rangle$ as a function of energy levels for $W/t = 1.0$, $1.4$, $1.7$, and $2.5$ in Fig.~\ref{figadd8}. The scaling behavior of $\langle D\rangle$ in Fig.~\ref{figadd8} (a) reveals the coexistence of extended and localized states in the spectrum for $W/t = 1.0$. The two types of states are separated by four conventional mobility edges (black dotted lines). As $W/t$ increases to $1.4$, all eigenstates in Fig.~\ref{figadd8} (b) exhibit that $\langle D\rangle$ increases with system size and approaches unity asymptotically, signaling the system is in E. Upon further increasing the amplitude to $W/t = 1.7$ in Fig.~\ref{figadd8} (c), the system enters a mixed phase where extended, critical, and localized states coexist. In this case, both conventional and anomalous mobility edges emerge. Finally, for $W/t = 2.5$ in Fig.~\ref{figadd8} (d), the system reenters E, with the scaling behavior of $\langle D\rangle$ resembling that observed for $W/t = 1.4$ in Fig.~\ref{figadd8} (b).

\end{document}